%% file: collas2025_conference.tex
\documentclass{article} 
\usepackage{collas2025_conference,times}
\usepackage{easyReview}

\input{math_commands.tex}

\usepackage{hyperref}
\hypersetup{
    colorlinks=true,
    linkcolor=red,
    filecolor=magenta,
    urlcolor=blue,
    citecolor=purple,
    pdftitle={Overleaf Example},
    pdfpagemode=FullScreen,
    }

\usepackage[utf8]{inputenc} 
\usepackage[T1]{fontenc}    
\usepackage{kotex}          
\usepackage{url}            
\usepackage{booktabs}       
\usepackage{amsfonts}       
\usepackage{nicefrac}       
\usepackage{microtype}      
\usepackage{xcolor}         

\definecolor{Blue9}{rgb}{0.098,0.3,0.9}
\usepackage{microtype}
\usepackage{graphicx}
\usepackage{subfigure}
\usepackage{subcaption}
\usepackage{booktabs} 
\usepackage[capitalize,noabbrev]{cleveref}

\usepackage{tcolorbox}
\usepackage{tabularray}
\usepackage{amsmath}
\usepackage{array}
\usepackage{color}
\usepackage{enumitem}
\usepackage{multirow}
\usepackage{bbm}
\usepackage{duckuments}
\usepackage{stfloats}
\usepackage{xspace}
\usepackage{caption}
\usepackage{verbatim}
\usepackage{wrapfig}
\usepackage{multicol}

\usepackage{amsmath}
\usepackage{amssymb}
\usepackage{mathtools}
\usepackage{amsthm}
\usepackage{bbm}

\newcommand{\multilinecell}[2][c]{%
  \begin{tabular}[#1]{@{}c@{}}#2\end{tabular}}
\newcommand{\stdv}[1]{\scalebox{.80}{~$\pm$~#1}}

\newcommand{\metabbr}{B-MoCA\xspace}
\newcommand{\tasknum}{131\xspace}
\newcommand{\Custom}{Custom\xspace}
\newcommand{\custom}{custom\xspace}

\title{Benchmarking Mobile Device Control Agents\\across Diverse Configurations}

\author{Juyong Lee$^1$ $\;$ Taywon Min$^1$ $\;$ Minyong An$^2$ $\;$ Dongyoon Hahm$^1$ 
\\
\textbf{Haeone Lee}$^1$ $\;$ \textbf{Changyeon Kim}$^1$ $\;$ \textbf{Kimin Lee}$^1$
\\
$^1$KAIST $\;$ $^2$Yonsei University \\
\texttt{agi.is@kaist.ac.kr}
}

\collasfinalcopy 

\begin{document}

\maketitle

\begin{abstract}
Mobile device control agents can largely enhance user interactions and productivity by automating daily tasks. 
However, despite growing interest in developing practical agents, 
the absence of a commonly adopted benchmark in this area makes it challenging to quantify scientific progress.
In this work, we introduce \metabbr, a novel benchmark with interactive environments for evaluating and developing mobile device control agents.
To create a realistic benchmark, we develop \metabbr based on the Android operating system and define \tasknum common daily tasks.
Importantly, we incorporate a randomization feature that changes the configurations of mobile devices, including user interface layouts and language settings, to assess generalization performance. 
We benchmark diverse agents, including agents employing large language models (LLMs) or multi-modal LLMs as well as agents trained with imitation learning using human expert demonstrations. 
While these agents demonstrate proficiency in executing straightforward tasks, their poor performance on complex tasks highlights significant opportunities for future research to improve effectiveness. 
Our source code is publicly available at \href{https://b-moca.github.io}{https://b-moca.github.io}.
\end{abstract}

\input{sections/intro}

\input{sections/method}
\input{sections/baselines}
\input{sections/experiments}
\input{sections/analyses}
\input{sections/relatedwork}

\input{sections/conclusion}

\section*{Acknowledgments}
We thank Dongjun Lee, Kyuyoung Kim, and Ahjeong Seo for providing sincere suggestions for improving our work. 
This work was supported by Institute for Information \& communications Technology Planning \& Evaluation(IITP) grant funded by the Korea government(MSIT) (RS-2019-II190075, Artificial Intelligence Graduate School Program(KAIST)) 
and Artificial intelligence industrial convergence cluster development project funded by the Ministry of Science and ICT(MSIT, Korea)\&Gwangju Metropolitan City.

\bibliography{collas2025_conference}
\bibliographystyle{collas2025_conference}

\clearpage
\appendix
\input{sections/appendix}

\end{document}

%% file: math_commands.tex
\usepackage{amsmath,amsfonts,bm}

\def\eqref#1{equation~\ref{#1}}

\def\1{\bm{1}}

\DeclareMathAlphabet{\mathsfit}{\encodingdefault}{\sfdefault}{m}{sl}
\SetMathAlphabet{\mathsfit}{bold}{\encodingdefault}{\sfdefault}{bx}{n}

\DeclareMathOperator*{\argmax}{arg\,max}

%% file: sections/intro.tex
\section{Introduction} \label{sec:intro}

Autonomous agents controlling mobile devices have great potential benefits.
For example, these agents can improve the accessibility of user interactions, especially for users with physical disabilities or those facing challenges in operating devices.
Additionally, they can boost productivity by automating daily tasks.
Such advantages have led to increased interest in developing practical agents for {\em mobile} device control.
Various approaches have been introduced, including agents based on large language models (LLMs) \citep{wen2023empowering,yan2023gpt} and agents trained with human demonstrations~\citep{sun2022meta,li2023uinav,rawles2023android}.
These innovations aim to create assistive agents capable of understanding device screen layouts and manipulating user interfaces (UI) to execute human commands.

Despite recent progress in developing mobile device control agents based on real systems, such as Android emulators~\citep{toyama2021androidenv,shvoEtAl2021appbuddy,zhang2023mobile}, prior works often overlook several important properties.
One primary aspect is testing the generalization ability across diverse device configurations, which is crucial in deploying agents in real devices.
Moreover, practical tasks essential for life (such as setting an alarm or making emergency calls) are often neglected because of the challenges in defining a wide range of such tasks with robust success criteria in various device settings.
The lack of a unified benchmark encompassing these important properties has impeded scientific progress in this field.

\begin{figure}[t!]
\vspace{-20pt}
    \centering
    \includegraphics[width = 0.77\textwidth]{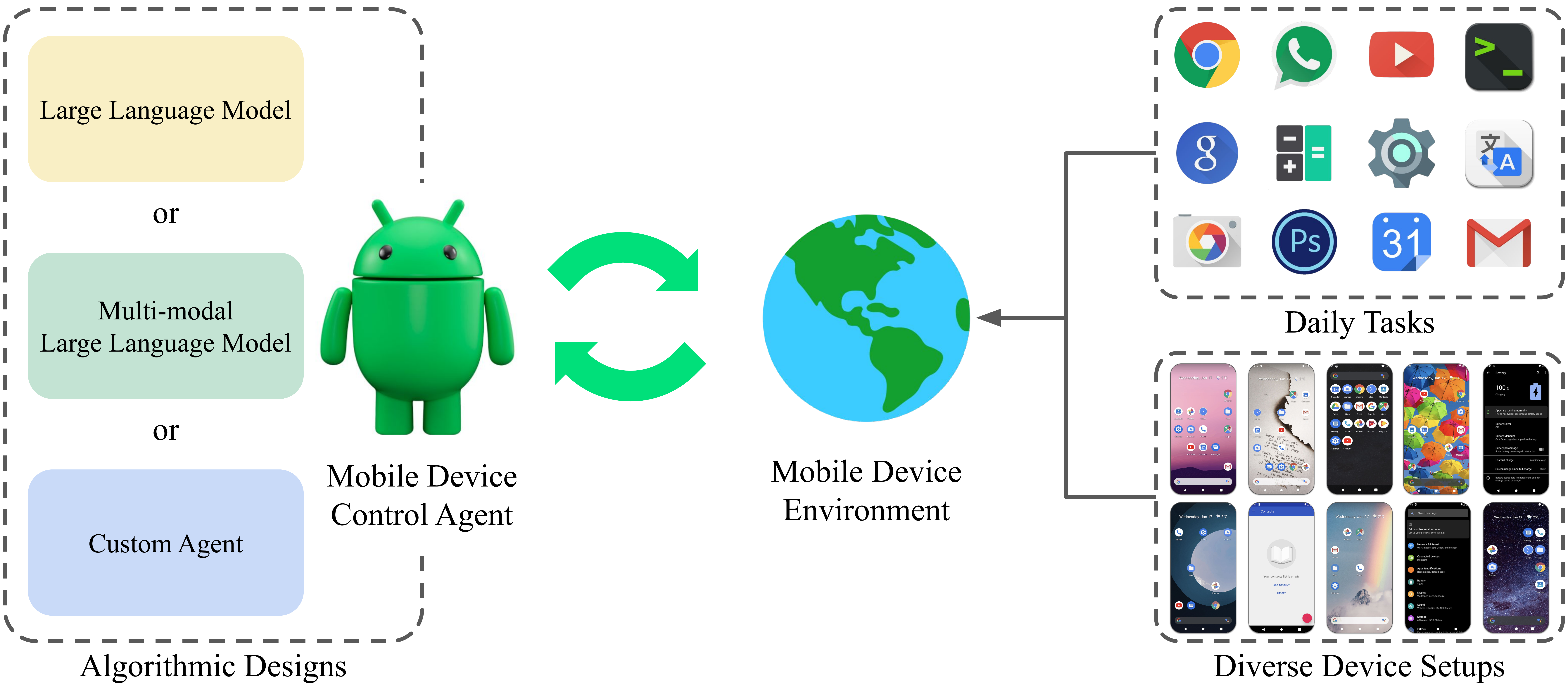}
\vspace{-5pt}
    \caption{Illustration of \metabbr. 
    We present a realistic benchmark for assessing the performances of mobile device control agents in executing everyday tasks. 
    A key feature of \metabbr is supporting randomization that changes various device attributes to analyze generalization ability.
    We benchmark agents leveraging LLMs or MLLMs as well as custom agents trained using human demonstrations.}
    \label{fig:overview}
\vspace{-15pt}
\end{figure}

In this work, we introduce \metabbr: a \textbf{B}enchmark designed for evaluating \textbf{Mo}bile device \textbf{C}ontrol \textbf{A}gents across diverse configurations (see \autoref{fig:overview}).
For real-system interactive evaluation, \metabbr is developed based on the Android operating system.
A key feature of \metabbr is supporting numerous customization, designed to mirror diverse device configurations in real-use cases,
including variations in icon placements, sizes, wallpapers, languages, and device types.
Utilizing this feature, one can easily create diverse environments with various configurations to evaluate the agents' generalization ability. 
Notably, our environment simulates an open-ended environment by, for example, incorporating access to the web (inducing dynamicity in the tasks using Chrome, Wikipedia, Walmart, and Instagram), challenging the agent to face continuously changing contexts.
This also allows \metabbr to be a useful testbed for open-ended learning and continual learning, such as task incremental learning and skill discovery~\citep{li2017learning, liu2024skillact}.

We define \tasknum practical tasks grounded in realistic scenarios, such as opening specific applications, initializing searches over the web, and adjusting device settings. 
To ensure reliable evaluation across diverse configurations, \metabbr provides rule-based success detectors that automatically signal task completion during the agents' interactions over the environments.
Within the open digital world, we examine the adaptabilities of agents in environments with different configurations, including the transferability of knowledge of agents trained with multiple tasks.
To be specific, the baselines include agents employing foundation models, such as large language models (LLMs) or multi-modal LLMs (MLLMs), which benefit from extensive knowledge obtained through pre-training.
We evaluate closed-source models, such as GPT-4o~\citep{gpt4o} and Gemini-1.5-pro~\citep{team2023gemini}, and open-source models, such as Llama-3~\citep{llama3}.
Additionally, we consider building agents by training policies using behavior cloning (BC;~\citealt{pomerleau1988alvinn,schaal1997learning}).

In our experiments, we find that the tested agents demonstrate capabilities in solving straightforward tasks.
However, agents employing foundation models (like LLMs) show limitations in more challenging scenarios that require multiple interactions.
Custom agents trained with BC successfully mimic expert behaviors but lack the ability to generalize to unseen configurations.
Our extensive experiments reveal the limitations of existing methods, calling for future research.

Our contributions are as follows:
\begin{itemize}[leftmargin=7mm, itemsep=0mm]
\item We propose \metabbr, a new benchmark designed to measure progress in developing device control agents, including various features such as environment randomization.
\item We evaluate several baseline agents for mobile device control, identifying their shortcomings, such as their limited generalization ability in UI elements understanding and manipulation.
\item We explore different design choices for leveraging foundation models, and analyze the impact of data diversity on the effectiveness of agents trained using human expert demonstrations.
\item We open-source all the source codes and relevant materials for easy reproduction of our environments and experiments. 
\end{itemize}
We hope \metabbr helps future researchers identify challenges in building assistive agents and easily compare the efficacy of their methods over the prior work.

%% file: sections/method.tex
\vspace{-5pt}

\section{B-MoCA} \label{sec:method}

\vspace{-5pt}

In this section, we introduce \metabbr, a benchmark designed to develop agents capable of executing common daily tasks on mobile devices with diverse configurations.

\vspace{-5pt}

\subsection{Design factors}

\vspace{-5pt}

Designing a meaningful benchmark poses significant challenges, particularly in developing a realistic platform that incorporates practical tasks.
Our benchmark is built on Android, a widely used open-source operating system, ensuring authentic environments.
To reflect the multi-step nature of real interactions, we model the device control task as a sequential decision-making problem (Section~\ref{sec:problem_formulation}). 
The benchmark includes \tasknum tasks involving both default applications like Chrome and Calendar, and third-party applications such as Instagram and Wikipedia, selected for their prevalence and utility in everyday life.\footnote{
We highlight that we designed the environments and tasks to support topics related to continual learning naturally. We include detailed guidelines for using \metabbr for continual learning in Appendix~\ref{app:continual_learning}.}
Each task is equipped with a success detector to evaluate the agent's performance in accurately completing the task (Section~\ref{sec:daily_task}).

Given the diverse nature of user mobile device setups, such as variations in icon placements, wallpaper choices, languages, and device types,
it is important to test the generalization abilities of device control agents across diverse setups.
To assess generalization performance, we incorporate a randomization feature in our benchmark.
This feature is designed to simulate various real-world scenarios by changing the settings of mobile devices (Section~\ref{sec:environment}). 

\begin{figure}[t!]
  \centering
  \includegraphics[width=0.9\textwidth]{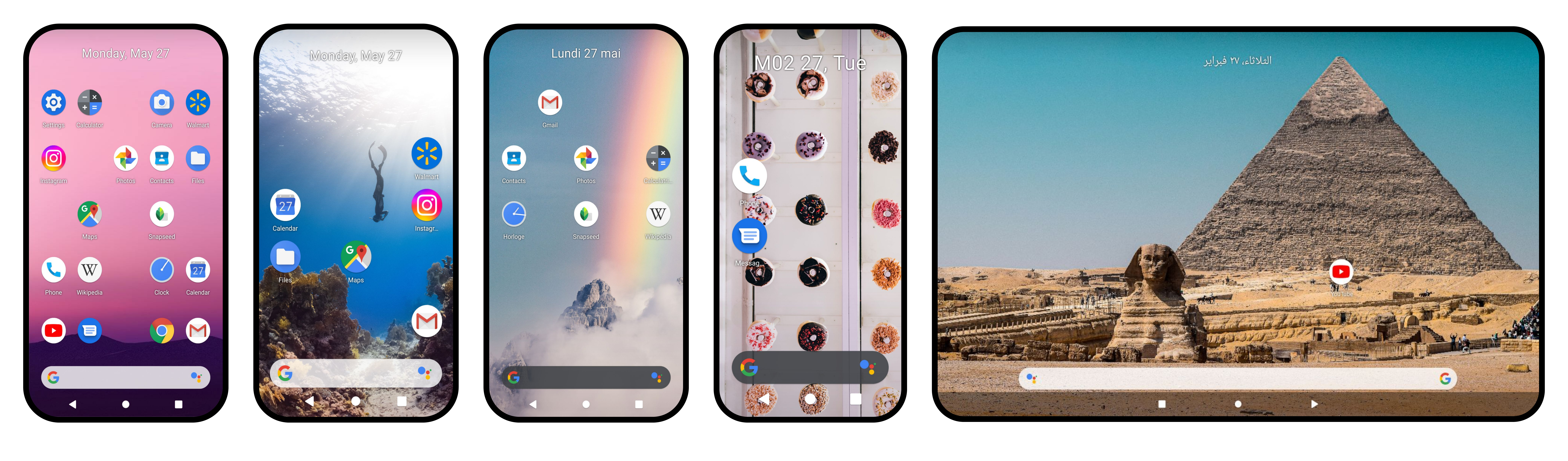}
  \caption{Examples of home screen images from environments in \metabbr.
  The randomized features span icon location, font size, wallpaper, language, and device type and challenge the generalization ability of agents.
  }
  \label{fig:environment_randomization}
\end{figure}

\subsection{Problem formulation}\label{sec:problem_formulation}

We formulate the device control task as a sequential decision-making problem, where an agent interacts with an environment (i.e., an Android emulator).
Formally, given a task instruction $c$, the agent receives an observation $o_t$ and takes an action $a_t$ based on its policy at each timestep $t$.
The environment returns a success signal $r_t$ and then transitions to the next observation $o_{t+1}$.

Observations, which capture the UI elements, can be represented as either screen pixels, screen descriptions in text derived from the Android view hierarchy, or a combination of both.
The action space includes a set of screen-touching actions.
In \metabbr, we support both continuous and discrete actions.
Continuous actions are defined as dual-gesture actions, similar to \citet{rawles2023android}. 
Each dual-gesture action comprises a pair of $(x, y)$ screen locations.
A dual-gesture action is identified as to {\tt tap} when the two locations are identical within a specified threshold, and as to {\tt swipe} when the distance between the two locations exceeds this threshold.
Also, agents can perform to {\tt press} navigation buttons (i.e., back, home, and overview) by touching the corresponding locations on the screen.
Discrete actions are defined as direct interactions with specific screen locations (such as the center of a UI element's bounding box or predefined locations), swiping in specified directions, or pressing individual buttons.
We note that our benchmark supports text-based actions,
enabling the utilization of both LLMs and MLLMs (see Section~\ref{sec:llm_agent} for details).

We refer readers for further details on the environment implementation to Appendix~\ref{app:environment_implementation}.

\subsection{Daily tasks}\label{sec:daily_task}

\begin{figure}[t!]
\centering  
    \includegraphics[width=0.9\textwidth]{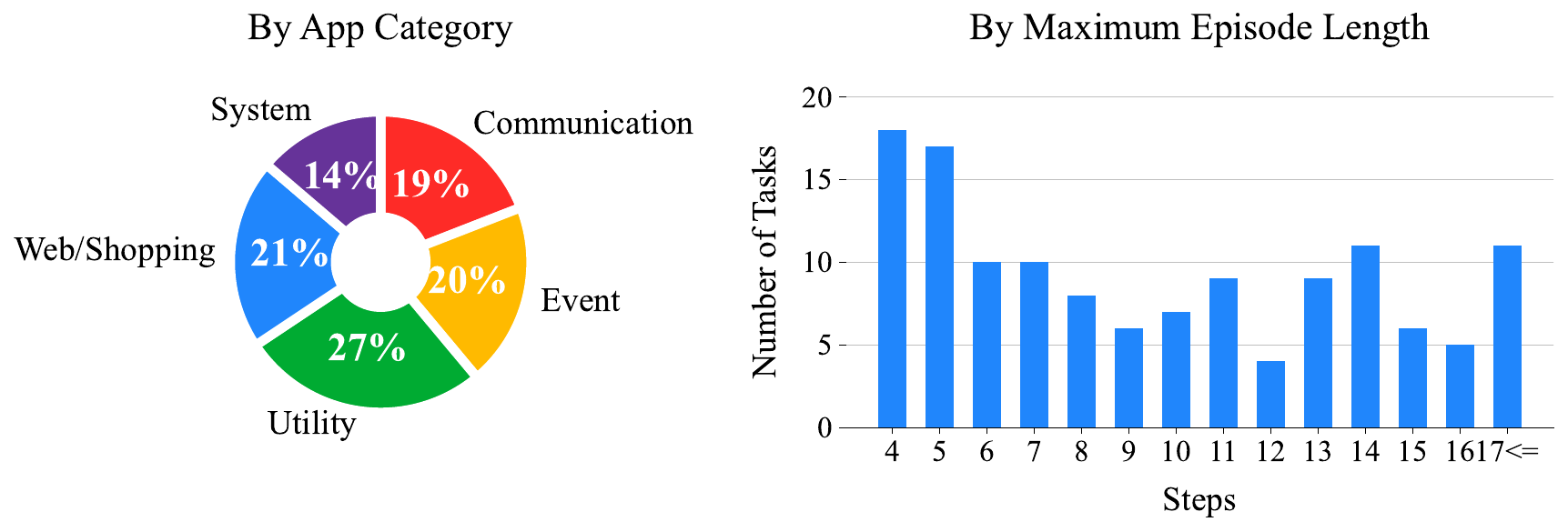}
    \caption{Distribution of tasks by app category and the maximum episode length.}
    \label{fig:task_statistics}
\end{figure}

\metabbr includes \tasknum tasks that can be used to assess essential skills for mobile device management.
Each task is designed to be grounded in realistic situations to provide functionalities useful in daily routines, such as setting the alarm and enabling airplane mode. 
These tasks require agents to navigate the device among diverse screens and manipulate various UI elements. 
\autoref{fig:task_statistics} shows the statistics of the tasks.
For a comprehensive list of tasks, we refer readers to Appendix~\ref{app:task_list}. 

Task completion is determined by a rule-based success detector that relies on three sources:
the system information, app data, and attributes of UI elements in the Android view hierarchy.
To monitor these sources,
we have developed a set of interfaces using tools like Android Debug Bridge (ADB) and Appium.
The success detector identifies the successful completion based on pre-defined criteria within these information sources. 
For example, it automatically detects the matching regular expression in the system logs or checks whether the attributes in specific UI elements (e.g., the `checked' attribute of the checkbox) are arranged correctly.
The success signal has the value of $+1$ when the task is completed, and $0$ otherwise.
An episode terminates as a success if the success detector signals completion, or as a failure if the agent exceeds a maximum number of steps without meeting the criteria.

\subsection{Environment randomization} \label{sec:environment}

In mobile device control, developing agents that can generalize across various device setups is crucial.
To evaluate their generalization ability, \metabbr incorporates a randomization feature that changes icon placements and sizes, wallpapers, languages, and device types.
Users can select a device type from a device list that includes popular models like Pixel 3, Pixel 4, Pixel 6, and WGXA Tablet.
They can also specify the locales to set the language and region, choose wallpapers from a selection of custom images, and activate dark mode for further environmental variation. 
Moreover, the sizes and locations of application icons can be customized to simulate real-world usage patterns.

Using randomization features, we create 45 unique environments in \metabbr, with examples shown in \autoref{fig:environment_randomization}.
To assess the generalization ability, we divide these environments into two sets: 35 for training and 10 for testing.
We employ domain randomization~\citep{tobin2017domain} to train agents, enabling them to perform tasks robustly across diverse device configurations.
We then evaluate the performance on test environments, which include unseen device setups. 
A detailed list of environment device configurations we prepare is available in Appendix~\ref{app:environment_configurations}.

%% file: sections/baselines.tex
\section{Baseline agents} \label{sec:baselines}

In this work, we benchmark various approaches for building mobile device control agents.
Section~\ref{sec:llm_agent} describes LLM agents and MLLM agents, where the agents are developed with frontier LLMs and MLLMs, respectively.
In Section~\ref{sec:custom_agent}, we introduce custom agents that are equipped with either fine-tuned open-source LLMs or encoders using vision-language models (VLM encoder) trained using human expert demonstrations.

\subsection{Closed-source models} \label{sec:llm_agent} 

Utilizing foundation models like LLMs and MLLMs, which contain extensive knowledge and possess emergent capabilities, has become a major approach in developing mobile device control agents~\citep{wen2023empowering,yan2023gpt}. 
We benchmark two types of agents that employ different foundation models: LLMs and MLLMs.
LLM agents utilize only the text descriptions of the screen layout to generate text actions, while MLLM agents leverage both text and visual inputs. 

\begin{figure}[t!]
\begin{minipage}[t]{\textwidth}
    \begin{minipage}[b]{0.45\textwidth}
        \centering
        \scalebox{0.9}{
        \small
        \begin{tabular}{l l}
        \toprule
            \textbf{Action option} & \textbf{Description} 
            \\ \hline
            \\[-0.6em]
                \multirow{2}{*}{{\tt dual-gesture(*)}} 
                & Operate a dual-gesture action 
            \\
                & with arguments {\tt(*)}.
            \\
            \\[-0.5em]
                \multirow{2}{*}{{\tt tap(numeric tag)}} 
                & Tap UI element labeled 
            \\
                & with {\tt numeric tag}. 
            \\ 
            \\[-0.5em]
                {\tt swipe(direction)} 
                & Swipe to {\tt direction}. 
            \\
            \\[-0.5em]
                {\tt press("HOME")} 
                & Press home button. 
            \\
            \\[-0.5em]
                {\tt press("BACK")} 
                & Press back button. 
            \\
            \\[-0.5em]
                {\tt press("OVERVIEW")} 
                & Press overview button. 
            \\ 
            \bottomrule
        \end{tabular}
        }
    \captionof{table}{A set of action options for agents generating text-based actions. 
    The options include both continuous and discrete actions.
    }
    \label{tab:llm_actions}
    \end{minipage}
\hfill
    \begin{minipage}[b]{0.5\textwidth}
        \centering
        \includegraphics[width=0.9\textwidth]{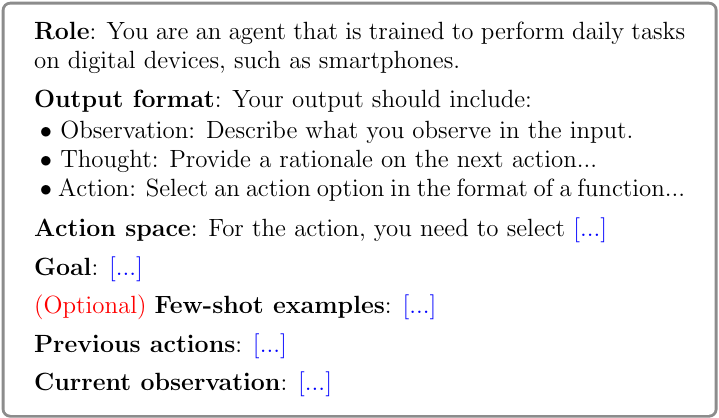}
        \captionof{figure}{An overview of prompts for text-based agents, with abbreviated relevant information as \textcolor{blue}{[...]}. The complete prompt is at Appendix~\ref{app:agent_llm_prompt}.}
        \label{fig:llm_prompt}
    \end{minipage}
\end{minipage}
\end{figure}

To facilitate the interactions of LLM and MLLM agents with an Android emulator, we define a screen translator that parses the observation from the Android view hierarchy~\citep{zhang2023mobile,yang2023appagent}. 
This translator converts screen layout information (i.e., the Android view hierarchy presented in XML format) into a text description of the UI elements.
Each description includes a numeric tag and details of each UI element, such as the class information specifying the type of UI.
Additionally, we define a set of possible action options that can be selected by agents in text format, as detailed in Table~\ref{tab:llm_actions}.

In prompts, we include the role of agents, action space definition, goal, (optional) few-shot examples, previous actions taken by the agent, and the current observation.
Our prompts, outlined in \autoref{fig:llm_prompt}, also incorporate the Chain-of-Thought technique~\citep{wei2022chain} to enhance the reasoning ability by enforcing a certain output format.
We illustrate an overview of LLM agents in Appendix~\ref{app:agent_llm_overview}.

\subsection{Open-source models} \label{sec:custom_agent} 

Despite the promising results of closed-sourced foundation models, leveraging them presents several challenges such as the difficulties in fine-tuning.
In our benchmark analysis, we also explore agents employing customized models, named \custom agents.
We consider two types of \custom agents: \custom agents using fine-tuned foundation models (i.e., Llama-3) and \custom agents equipped with VLM encoders.
Similar to agents using closed-source LLMs or MLLMs, the \custom agents using fine-tuned open-source LLMs process task instructions and screen layout descriptions in text form to produce text-based actions.
We fine-tune the open-source LLMs leveraging actions collected from human experts.
While evaluating, we use prompts that specify their general roles and relevant task-specific information.\footnote{We do not employ the CoT technique in \custom agents due to challenges in preparing the dataset with such thought processes.}

The \custom agents with VLM encoder, where the policy is denoted by $\pi_\theta$, take task instructions $c$ and screen images $o_t$ as inputs, and output discrete actions $a_t$.
Input embeddings, extracted using a pre-trained VLM encoder~\citep{yang2023auto}, are processed through a transformer module~\citep{vaswani2017attention} to generate actions. 
Each action is a vector of size 385, where the first 378 values correspond to tapping pre-defined locations (14$\times$27 bins) on the screen, four values to swiping directions (up, down, right, left), and the last three values to pressing buttons (back, home, overview).
For further details on the network architecture, we refer readers to Appendix~\ref{app:agent_bc_architecture}.
We train the policies of \custom agents with VLM encoder using BC.
The agents are optimized to imitate the human expert demonstrations $\mathcal{D}= \{(o_t,a_t^*,c)\}$ by minimizing $\sum_{(o_t,a_t^*,c) \sim \mathcal{D}} L_{\text{BC}(\pi_\theta(a_t|o_t,c), a_t^*)}$ with the cross entropy loss $L_{\text{BC}}$. 

%% file: sections/experiments.tex
\section{Experiments} \label{sec:exp}

\begin{figure}[t!]
    \centering  
    \subfigure{\includegraphics[width=0.8\linewidth]{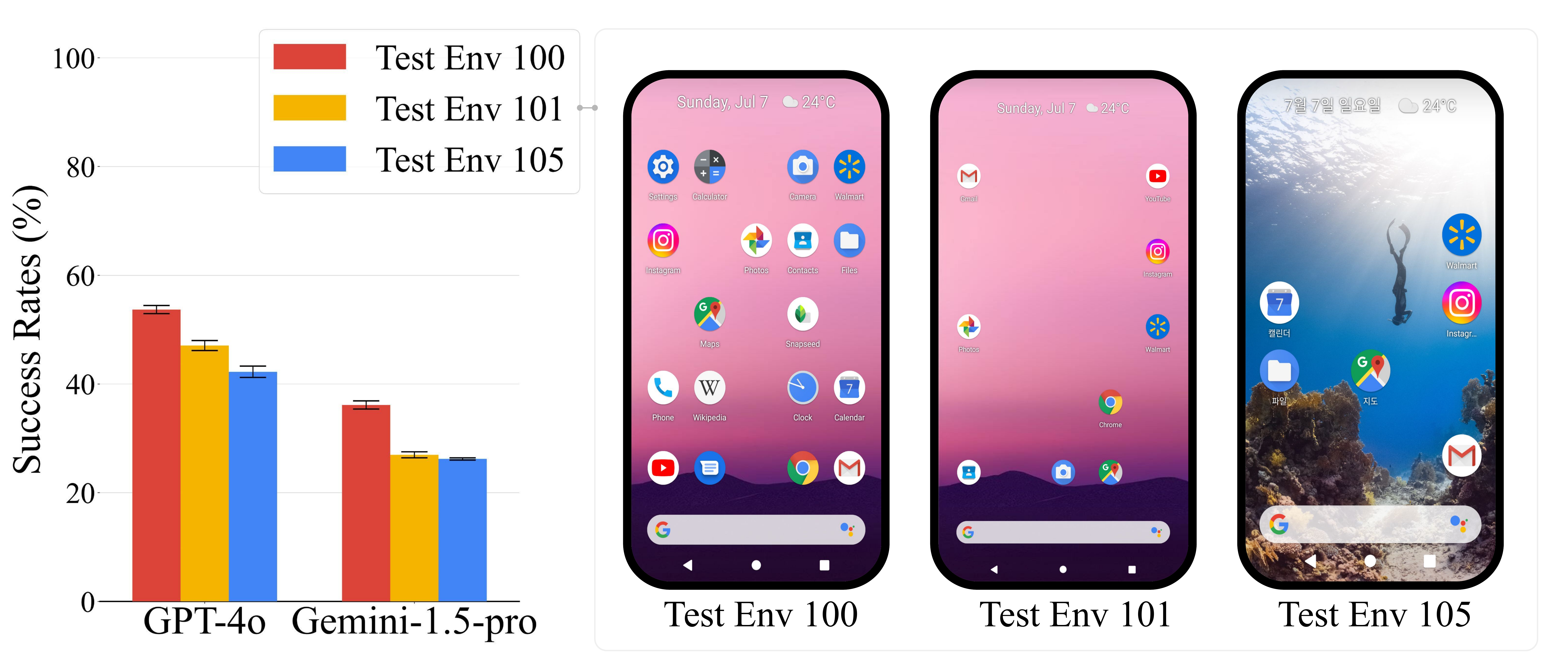} 
    }
    \caption{
    Average success rates of state-of-the-art LLM agents on all \tasknum tasks in three test environments. 
    We report the zero-shot performance of the agents with three runs.
    The different success rates among environments reveal the unique challenge of B-MoCA in assessing the generalization ability.
    }
\label{fig:all_results}
\end{figure}

\begin{figure}[t!]
    \centering  
    \subfigure{\includegraphics[width=0.8\linewidth]{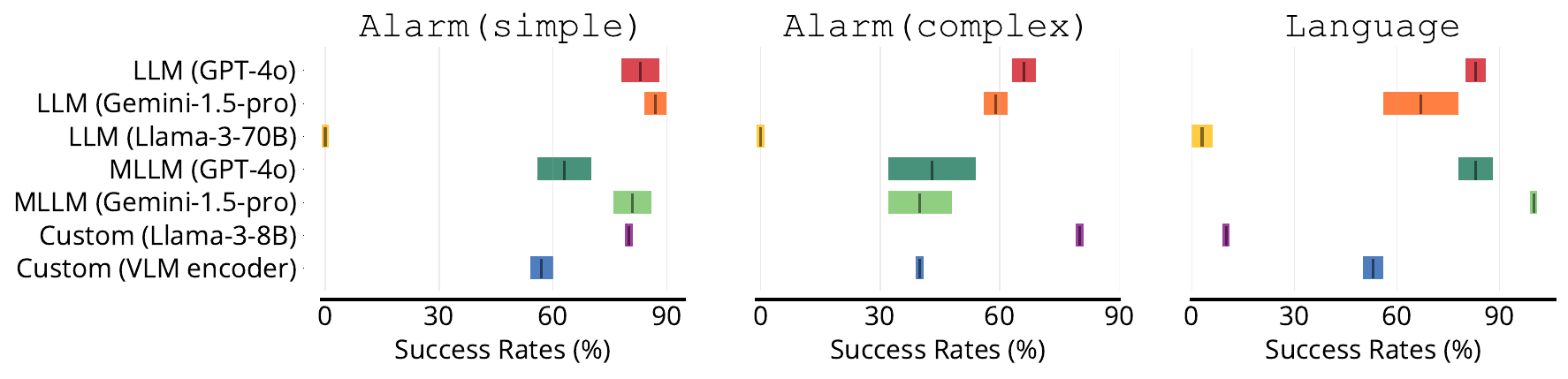} 
    }
    \subfigure{\includegraphics[width=0.8\linewidth]{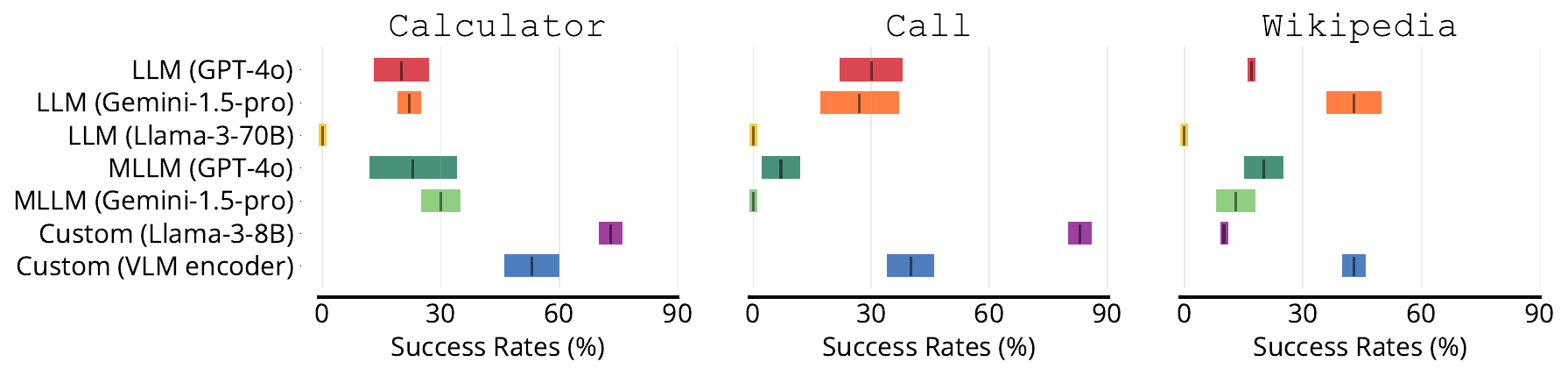} 
    }
    \caption{
    Average success rates of baseline agents in the test environments. 
    We report the mean and standard error across three runs.
    LLM agents and MLLM agents are evaluated using few-shot examples.
    \Custom agents are trained with BC using human demonstrations.
    }
\label{fig:main_results}
\end{figure}

We design our experiments to investigate the following research questions:
\begin{itemize}[leftmargin=6mm]
    \item How well do agents employing state-of-the-art LLMs perform daily tasks in \metabbr? (\autoref{fig:all_results}) 
    \item What are the different behaviors of the baseline agents shown in representative tasks? (\autoref{fig:main_results}) 
    \item How are the LLM agents affected by the randomization of each environmental factor? (\autoref{fig:exp_environmental_feature})
    \item How do different choices for employing LLMs affect performance? (\autoref{tab:exp_llm_few_shot})
    \item How crucial is data diversity when training \custom agents? (\autoref{fig:exp_data_diversity}) 
\end{itemize}

\subsection{Experimental setup}\label{sec:exp_setup}

We employ three state-of-the-art foundation models.
For LLM agents, we employ closed-source LLMs: GPT-4o~({\tt gpt-4o-2024-05-13};~\citealt{gpt4o}) and Gemini-1.5-pro ({\tt gemini-1.5-pro-001};~\citealt{team2023gemini}) with text-only input.
We also develop LLM agents using an open-source model Llama-3~({\tt meta-llama/Meta-Llama-3-70B-Instruct};~\citealt{llama3}) but without fine-tuning.
We evaluate LLM agents with and without few-shot examples.
For the few-shot examples, we select examples from a pool of 210 human expert demonstrations across 35 training environments (see Appendix~\ref{app:exp_dataset_collection} for dataset collection).
For MLLM agents, we leverage GPT-4o and Gemini-1.5-pro by providing additional image inputs.
We provide more details on the configurations used for the foundation models in Appendix~\ref{app:exp_llm_configuration}.
In addition, we experiment with two types of \custom agents.\footnote{Auxiliary studies on the agents trained with reinforcement learning is available in Appendix~\ref{app:exp_rl_discussions}.}
For the \custom agents using open-source LLMs, we fine-tune Llama-3~({\tt meta-llama/Meta-Llama-3-8B-Instruct}) using 210 human demonstrations from 35 training environments, adopting LoRA~\citep{hu2021lora}.
For \custom agents with VLM encoder, we fully fine-tune the cross-attention transformer module and visual encoder.
We refer the readers to Appendix~\ref{app:exp_custom_training_detail} for details on the training procedure.

We conduct two main experiments.
In the first main experiment, 
we examine state-of-the-art closed-source LLM agents (i.e., GPT-4o and Gemini-1.5-pro) on all \tasknum tasks.
The agents are evaluated in zero-shot on three test environments: a vanilla environment with all target applications in the home screen ("Test Env 100" with id 100, described in Appendix~\ref{app:environment_configurations}), another environment with randomized settings of icons in the home screen and size randomized ("Test Env 101" with id 101), and the other environment with randomized settings of icons in the home screen, size, wallpaper, and language ("Test Env 105" with id 105).
These varying configurations challenge the generalization ability of agents.
For example, in tasks for setting the alarm, the clock UI appears to be either circular in "Test Env 100" or rectangular in "Test Env 105", shown in Appendix~\ref{app:UI_elements_change}.

In the second main experiment, we study all the baseline agents on six representative challenging tasks: {\tt Alarm(simple)}, {\tt Alarm(complex)}, {\tt Calculator}, {\tt Call}, {\tt Language}, and {\tt Wikipedia}.
These tasks are selected as they require navigating multiple pages and manipulating diverse UI elements in a long horizon (i.e., episode length of at least 7).
In the {\tt Alarm} tasks, for example, the agents not only need to reach the alarm tab in the clock application but also operate the clock UI to set the alarms.
We provide exemplary expert demonstrations for these tasks in Appendix~\ref{app:task_demonstration_example}.
For each task, the goal instruction provided to the agents is as follows:
\begin{itemize}[leftmargin=6mm,itemsep=-0.1em]
\small
    \item {{\tt Alarm(simple)}: ``create alarm at 10:30 am''}
    \item {{\tt Alarm(complex)}: ``create alarm at 10:30 am on every weekday''}
    \item {{\tt Language}: ``go to the `add a language' page in setting''}
    \item {{\tt Calculator}: ``input `cos(180)' in Calculator''}
    \item {{\tt Call}: ``call the white house (202-456-1111)''}
    \item {{\tt Wikipedia}: ``disable the top 2 and `randomizer' topics at feed customize setting on Wikipedia and go back to the feed''}
\end{itemize}

For all experiments, we report the mean and standard error across three different runs.

\subsection{Main results}\label{sec:exp_main}

\autoref{fig:all_results} shows the success rates of LLM agents (without few-shot examples) on all \tasknum tasks, and \autoref{fig:main_results} displays the success rates of LLM agents (with few-shot examples), MLLM agents (with few-shot examples), and \custom agents on the six challenging representative tasks.
The agents employing the foundation model complete simple tasks with high performance by leveraging their pre-trained base knowledge.
For example, LLM agents using closed-source models achieve success rates higher than 80\% on the {\tt Alarm(simple)} task.
However, their performances significantly decrease as the tasks become complex (e.g., degradation on the {\tt Alarm(complex)} task).
On the other hand, \custom agents with VLM encoder imitate the behaviors of experts and exhibit average success rates greater than or equal to 40\% on all six challenging representative tasks. 
However, they still show limited generalization ability (less than 60\%).
Similarly, \custom agents using fine-tuned Llama-3 show an extreme discrepancy of proficiencies across the six challenging tasks.
The shortcomings of these baselines call for new algorithms for more efficient building mobile device control agents.

We provide more remarks on each agent type below. 

\paragraph{Challenges for agents using closed-source foundation models}
The LLM agents show remarkable performances in controlling mobile devices, even in zero-shot, as shown in \autoref{fig:all_results}.
However, we witness several limitations of agents employing the foundation models, across both LLM agents and MLLM agents.
First, LLM and MLLM agents frequently hallucinate task completion. 
For example, on the {\tt Calculator} task, the agents often conclude that a task is complete before entering all the requested strings (e.g., entering `cos(18)' instead of `cos(180)').
Second, these agents face difficulties with long-horizon tasks that require multiple interactions. 
For instance, they frequently make mistakes when typing the sequence of numbers on the {\tt Call} task.
The analysis with more number of failure cases of these agents is available in Appendix~\ref{app:exp_failure_cases}.

\paragraph{LLM agents across the randomized environments}
\autoref{fig:all_results} shows the different success rates of LLM agents across the three environments.
In the "Test Env 101" environment, compared to the "Test Env 100," we find that both LLM agents using GPT-4o and Gemini-1.5-pro make simple mistakes regarding the randomized icon locations.
For example, the agent employing GPT-4o taps the Walmart icon to open the Wikipedia application, if the Wikipedia icon is not available on the home screen.
In the "Test Env 105" environment, with fewer icons on the home screen, UI elements described in Korean, and a larger icon size setting, we observe similar performance degradation.
We include a more rigorous analysis of the effect of each environmental feature in Section~\ref{sec:exp_analysis}.

\paragraph{Comparison of LLM agents with MLLM agents}
In our experiments, gains from additional image inputs are marginal or even detrimental, as shown by comparing performances with and without additional screenshot inputs (see red vs. dark green and orange vs. light green). 
We expect that this is due to domain gaps in visual inputs. 
Additionally, we hypothesize that multi-modal agents suffer from the increased length of input sequence associated with additional image tokens. 
A similar observation was made in recent work involving agents that control desktop computers~\citep{xie2024osworld}.
These results indicate the remaining headroom in leveraging multi-modal inputs more effectively.

\paragraph{Differences between closed-source and open-source LLMs}
In the setting of agents using the LLMs with few-shot examples but without fine-tuning, frontier closed-source LLMs like GPT-4o (red) outperform open-sourced LLMs like Llama-3 (yellow) across all six representative challenging tasks, as shown in \autoref{fig:main_results}.
First, the shorter context length supported by Llama-3 largely restricts the number of few-shot examples that can be included.
We discuss the effect of few-shot examples for LLM agents in detail in Section~\ref{sec:exp_analysis}.
Also, we observe that the ability to understand current screen layouts differs between them.
For example, while describing what they observe, GPT-4o agents typically produce more detailed texts, such as a precise list of application icons, compared to Llama-3 agents.
More discussions and the evaluation results of LLM agents using Llama-3 across all \tasknum tasks are also present in Appendix~\ref{app:exp_llm_agent_all_tasks}.

\paragraph{On \custom agents using fine-tuned LLMs}
We observe that \custom agents leveraging fine-tuned Llama-3 models achieve superior performance on several complex tasks, exhibiting greater proficiency compared to agents using standard Llama-3 across all tasks.
These agents successfully complete the {\tt Alarm} tasks (both simple and complex), on the test environments demonstrating high similarities with the training environments.
Their proficiencies in many tasks reveal the benefits of fine-tuning LLMs to build agents.
However, their failure on the {\tt Language} and {\tt Wikipedia} tasks in most of the test environments demonstrates a clear limitation, with more analysis available in Appendix~\ref{app:exp_agent_llama}, necessitating more effective algorithms.

\paragraph{Generalization ability of \custom agents}
In our experiments, the \custom agents achieve reasonable performances in many complex tasks, even where agents using closed-source foundation models fail.
The \custom agents successfully imitate expert behaviors in navigating applications and manipulating diverse UI elements in training environments.
However, they still show limitations on generalization ability.
The agents using fine-tuned Llama-3, for example, fail to generalize their behaviors to environments having a language setting of Korean and a device setting of Tablet.
Also, the agents equipped with VLM encoder largely degrade in test environments, while achieving high success rates in training environments (e.g., higher than 90\%; see Appendix~\ref{app:exp_bc_training_env_performance} for training performances).
Specifically, they struggle with tasks involving severe visual changes induced by unseen device configurations.
These observations highlight the need to develop more efficient algorithms that improve generalization against changes in device configurations.

%% file: sections/analyses.tex
\begin{figure}[!t]
\centering
\small    
\begin{minipage}{0.47\textwidth}
    \centering
    \includegraphics[width=\textwidth]{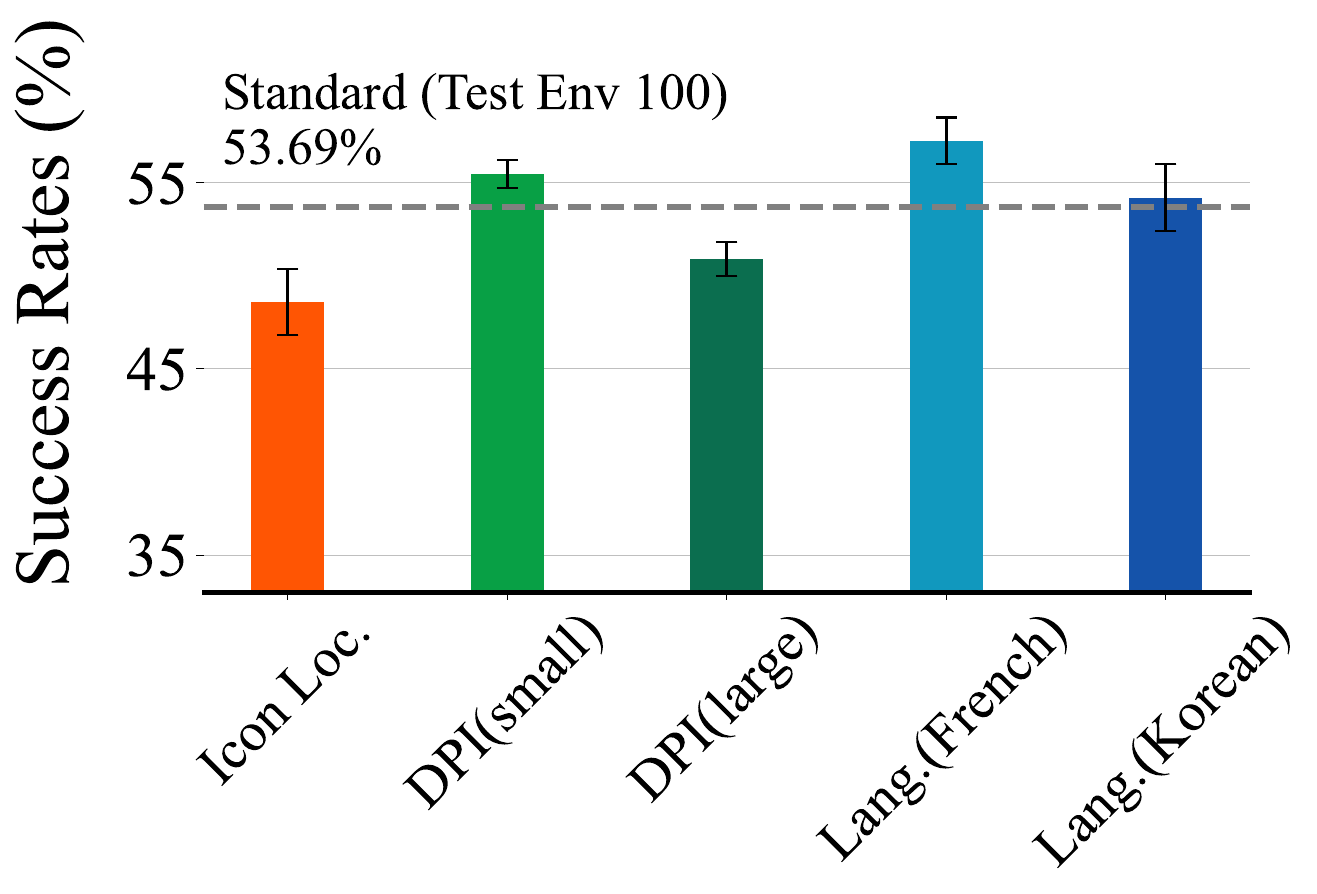}
    \caption{
        The success rates of LLM agents using GPT-4o across all tasks, on environments each feature randomized.
        Icon location affects the agents most significantly.
    }
    \label{fig:exp_environmental_feature}
\end{minipage}
\hfill
\begin{minipage}{0.5\textwidth}
    \centering
    \includegraphics[width=0.9\linewidth]{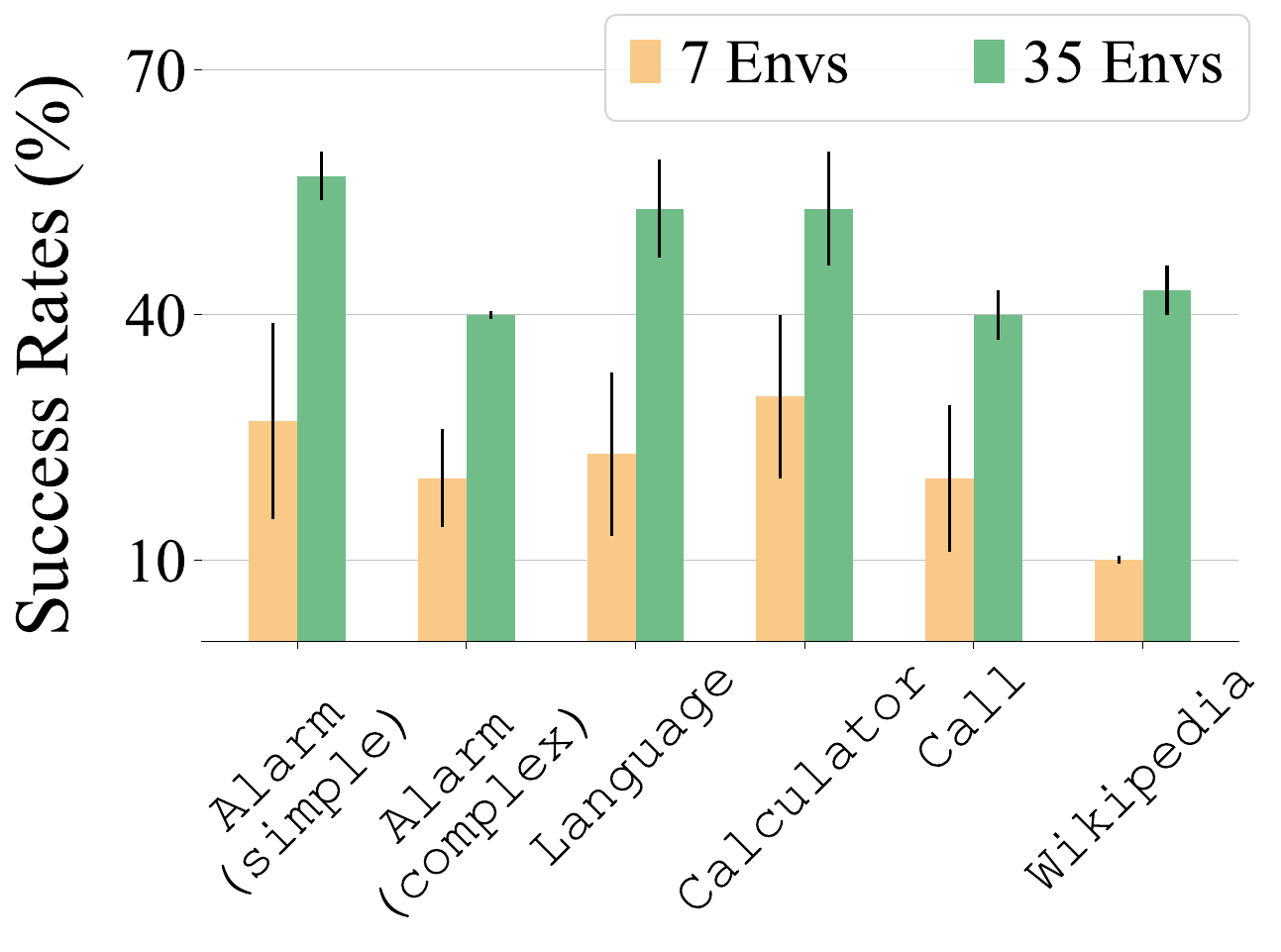}
    \captionof{figure}{
    Success rates of agents trained with BC on varying numbers of training environments. 
    A general trend of escalating success rates with more environments appears.
    }
    \label{fig:exp_data_diversity}
\end{minipage}
\vspace{-20pt}
\end{figure}

\subsection{Further analyses}\label{sec:exp_analysis}

In this section, we further analyze the experiments conducted in \metabbr.

\paragraph{Generalization ability of LLM agents on each environmental factor}
To examine the impact of randomizing each environment feature, 
we design five new environments by modifying individual components of a standard test environment (referred to as `Standard'). 
These modifications include changing the icon location on the home screen, adjusting the size of UI elements (options of small or large), and changing the language settings (French or Korean) while keeping all other settings constant.
As shown in \autoref{fig:exp_environmental_feature}, the impact of each environmental feature varies.
LLM agents using GPT-4o agents (without few-shot examples) exhibit significant performance decreases when icon locations are altered, as requiring the agents to explore in order to locate target applications.
On different DPI settings, the performances differ more with larger DPI settings compared to smaller DPI settings, as larger settings typically induce more abrupt changes in UIs on our environments.
The altered language settings have a limited impact on GPT-4o agents compared to other environmental factors, presumably as they have learned robust representations via large-scale multilingual training, but we anticipate that this will pose a greater challenge for models with less multilingual training.

\begin{wraptable}{r}{0.43\textwidth}
\centering
    \resizebox{0.94\linewidth}{!}{
        \begin{tabular}{c c c}
            \toprule
            & \multilinecell{LLM agents\\(zero-shot)} & \multilinecell{LLM agents\\(few-shot)} \\
            \midrule
            {\tt Alarm(simple)} & $87$\stdv{07} & $83$\stdv{05} \\
            {\tt Alarm(complex)} & $33$\stdv{05} & $67$\stdv{03} \\
            {\tt Language} & $97$\stdv{03} & $83$\stdv{03} \\
            {\tt Calculator} & $17$\stdv{03} & $17$\stdv{07} \\
            {\tt Call} & $00$\stdv{00} & $30$\stdv{08} \\
            {\tt Wikipedia} & $30$\stdv{22} & $20$\stdv{00} \\
            \midrule
            Average & $44$\stdv{03} & $50$\stdv{04} \\
            \bottomrule
        \end{tabular}
    }
    \captionof{table}{
    Success rates of LLM agents using GPT-4o with and without few-shot examples.
    While including few-shot examples enhances performance on the {\tt Alarm(complex)} task, it reduces performance on the {\tt Wikipedia} task.
    }
\vspace{-15pt}
\label{tab:exp_llm_few_shot}
\end{wraptable}

\paragraph{LLM agents with few-shot examples}

\autoref{tab:exp_llm_few_shot} compares the performance of LLM agents using GPT-4o with and without few-shot examples.
We observe that providing three examples to the agents improves the success rates on several tasks (e.g., {\tt Alarm(complex)} and {\tt Call}).
However, few-shot examples do not always enhance the performance and can sometimes even degrade the performance.
Specifically, the agents often fail to utilize these examples effectively.
For example, we observe that the agents frequently select the final action in the exemplary trajectory (see \autoref{fig:failure_phone} in Appendix~\ref{app:exp_failure_cases}).
This highlights a significant challenge in using few-shot examples for LLM agents; naively mimicking actions from human demonstrations may often lead to errors, particularly when the UI elements vary across different device settings.
We also find a similar trend in MLLM agents, where details are available in Appendix~\ref{app:exp_mllm_few_shot}.

\paragraph{Effect of training data diversity on custom agents}

We train \custom agents with varying numbers of training environments (see Appendix~\ref{app:exp_dataset_collection} for details of the experimental setup).
As shown in \autoref{fig:exp_data_diversity}, the performance of \custom agents escalates as the number of training environments increases.
Specifically, on the {\tt Language} task, for example, the agents exhibit success rates of 23\% and 53\% as trained with 7 and 35 training environments, respectively.
We believe this verifies the efficacy of the environment randomization incorporated in our benchmark for developing practical agents.

%% file: sections/relatedwork.tex
\section{Related work} \label{sec:related}

\paragraph{Open-ended digital environment for autonomous agents} 
Towards general-purpose agents, it is crucial to curate open-ended environments that can feature diverse aspects of learning, such as continual learning~\citep{rusu2016progressive,li2017learning,rolnick2019experience} and environment adaptation~\citep{cobbe2020leveraging}.
To this end, several researchers have focused on utilizing open-ended game environments~\citep{fan2022minedojo,tan2024towards}.
Recently, other types of open-ended digital environments are gaining interest, such as websites~\citep{yao2022webshop,zhou2024webarena}, operating systems~\citep{xie2024osworld}, and mobile devices~\citep{toyama2021androidenv,rawles2023android}.
These environments are characterized by diverse interaction modalities, continuously changing interfaces, and complex action spaces. 
Unlike static environments, such as board games~\citep{silver2016mastering,schrittwieser2020mastering} or fixed virtual worlds~\citep{bellemare13arcade,james2019rlbench}, digital interfaces evolve over time, requiring agents to continually adapt to diverse layouts, applications with different versions, and varying UIs. 
Specifically, we propose a novel open-world environment based on realistic Android emulators, with a focus on diverse device configurations.

\paragraph{Benchmark for decision-making agents}

There have been continuous efforts to build reliable benchmarks for sequential decision-making in video games~\citep{bellemare13arcade}, locomotion~\citep{brockman2016openai}, and robotic manipulation~\citep{james2019rlbench}. 
Lately, researchers have proposed benchmarks for solving device control tasks, viewing it as another decision-making problem.
For example, \citet{yao2022webshop} and \citet{zhou2024webarena} have presented benchmark simulating web platforms, while \citet{toyama2021androidenv}, \citet{shvoEtAl2021appbuddy}, and \citet{zhang2023mobile} have suggested RL environments adopting Android emulators.
In this work, inspired by special-purpose benchmarks quantifying the generalization ability of the agents~\citep {cobbe2020leveraging,stone2021distracting,mediratta2023generalization}, we newly propose a benchmark with a randomization feature.
We provide additional related works regarding foundation models for decision-making system and developing assistive agent for device control in Appendix~\ref{app:related_work}.

%% file: sections/conclusion.tex
\section{Conclusion} \label{sec:concl}

We present \metabbr, a new benchmark designed for evaluating mobile device control agents.
Our benchmark provides diverse tasks applicable to everyday routines and environments that simulate numerous device configurations.
We conduct extensive experiments and demonstrate that \metabbr can serve as a standardized platform for developing different types of agents in a unified setting.
We refer the readers to Appendix~\ref{app:limitation} for future directions and limitations.
Toward practical mobile device control agents, we hope that \metabbr stands as a valuable platform with helpful resources for innovative breakthroughs.








%% file: sections/appendix.tex
\begin{center}{\bf {\LARGE Appendix:}}\end{center}
\begin{center}{\bf {\large Benchmarking Mobile Device Control Agent across Diverse Configurations}}\end{center}

\input{sections/appendix/environment_details}
\input{sections/appendix/task_details}
\input{sections/appendix/continual_learning}
\input{sections/appendix/agent_details}
\input{sections/appendix/experiment_setup_details}
\input{sections/appendix/experiment_result_details}
\input{sections/appendix/rl_details}
\input{sections/appendix/related_work}
\input{sections/appendix/limitation}

%% file: sections/appendix/environment_details.tex
\section{Environment details}

\subsection{Environment implementation and interface} \label{app:environment_implementation}

\paragraph{Environment}
\metabbr is based on Android OS for real-system interactive evaluation.
The environment is simulated with Android virtual devices, containing the device hardware profile, system image, storage area, and other relevant properties.
The dynamics of the environments, such as the transition rules, are governed by Android OS and applications.

Each environment is represented as an Android device, running on top of the Android emulator.
To be specific, we define each environment as a snapshot, a stored image of the Android virtual device.
Each snapshot is built by saving an image of the target device after the configurations.
These configurations include installing third-party applications and randomizing the features of environments.
The features of the environments we randomize encompass placing icons in random locations, setting dots per inch (DPI), modifying wallpapers, and changing the language.
During the configuration, adjusting several device settings for accurate evaluation, such as changing the database of applications, is also conducted.

To facilitate interactions between the environment and agents, we develop a set of interfaces on top of the Python library AndroidEnv~\citep{toyama2021androidenv}.
The interfaces we develop encompass various functionalities: to provide the task descriptions in text to the agents, to capture screenshots of the virtual device, to provide the Android view hierarchy in XML format and parse the text description of the screen, to extract dual-gesture actions from text-based actions, and to deliver the dual-gesture action to the Android emulator.

\paragraph{Interaction frequency}
The Android emulators run asynchronously, independent of the agent that interacts with the environments.
However, this asynchronicity between the agent and the environment may cause several issues such as incomplete transition of the environments or delayed success signals.
To alleviate the issue, we adjust the interaction frequency between agents and environments. 
Specifically, this adjustment is operated by forcing the agent to wait a pre-defined time before fetching the screen information from the environment.
In our experiments, we fix the interaction frequency during evaluation to be $1/3\textrm{Hz}$ across all types of agents, except Instagram with longer latency (e.g., $1/30\textrm{Hz}$) as locale change requires additional loading.
We also allow users to adjust this interaction frequency.

\paragraph{Observation space}
The observation space is comprised of either a screen image, a text description of the screen in XML formats based on the Android view hierarchy, or both.
We provide an example of an observation, as a pair of a screen image and corresponding a text description in \autoref{tab:observation_example}.

The screen images are used for multi-modal large language model (MLLM) agents and \custom agents.
Each image is resized into a resolution of $256 \times 512$ for MLLM agents and $128 \times 256$ for \custom agents. 
The text descriptions are used for agents with LLMs and MLLMs.
To build the text description, the Android debug bridge (ADB) UI Automator is employed for acquiring the Android view hierarchy in XML format. 
A pre-defined screen translator, then, converts the information of UI elements in the XML file into a set of text descriptions of UI elements.
The description includes a numeric tag and details of the UI elements, including the class name or content descriptions.
Additionally, we provide attributes if the UI elements are checked or selected.
In our interface, the parser captures the descriptions of all the nodes in the Android view hierarchy, not specified to the leaf nodes.
Also, the users can choose either to include or exclude the bounding box x-y coordinates specifying the location of the elements, as a pair of x-y coordinates of top-left and bottom-right corners, where the x and y values range from 0.0 (top/left edge) to 1.0 (bottom/right edge), with the origin (0.0, 0.0) at the screen's top-left corner.
In our experiments, we exclude the bounding box information in the text-based observation, to avoid the input being excessively long, especially considering the few-shot examples.

\begin{table}[!ht]
\centering
\begin{minipage}{0.25\textwidth}
    \centering
    \includegraphics[width=0.95\textwidth]{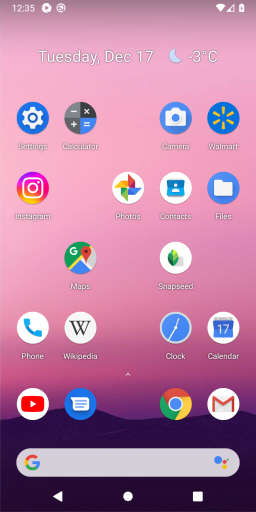} 
    \label{fig:observation_example_image}
\end{minipage}
\hfill 
\begin{minipage}{0.73\textwidth}
    \centering
    \resizebox{0.98\textwidth}{!}{
        \begin{tabular}[h]{@{}l@{}}
        \small
        \centering
        \tcbox[colback=white,boxrule=1pt,arc=3mm]{
            \begin{tblr}{colspec={@{}X@{}},rowsep=1pt}
            \texttt{[}
            \{`numeric\_tag': 0, `resource\_id': `', `class': `View', `content\_description': `Apps list', `text': `', `checked': `false', `bbox\_location': `((0.47, 0.71), (0.53, 0.74))'\}, 
            \{`numeric\_tag': 1, `resource\_id': `scrim\_view', `class': `View', `content\_description': `', `text': `', `checked': `false', `bbox\_location': `((0.00, 0.00), (1.00, 1.00))'\}, 
            \\
            $\cdots$
            \\
            \{`numeric\_tag': 9, `resource\_id': `', `class': `TextView', `content\_description': `Walmart', `text': `Walmart', `checked': `false', `bbox\_location': `((0.78, 0.18), (0.96, 0.31))'\}, 
            \{`numeric\_tag': 10, `resource\_id': `', `class': `TextView', `content\_description': `Instagram', `text': `Instagram', `checked': `false', `bbox\_location': `((0.04, 0.31), (0.22, 0.45))'\}, 
            \{`numeric\_tag': 11, `resource\_id': `', `class': `TextView', `content\_description': `Photos', `text': `Photos', `checked': `false', `bbox\_location': `((0.41, 0.31), (0.59, 0.45))'\}, 
            \{`numeric\_tag': 12, `resource\_id': `', `class': `TextView', `content\_description': `Contacts', `text': `Contacts', `checked': `false', `bbox\_location': '((0.59, 0.31), (0.78, 0.45))'\}, 
            \\
            $\cdots$
            \\
            \{`numeric\_tag': 39, `resource\_id': `', `class': `FrameLayout', `content\_description': `', `text': `', `checked': `false', `bbox\_location': `((0.00, 0.00), (1.00, 1.00))'\}
            \texttt{]}
            \end{tblr}}
        \end{tabular}
    }
    \label{tab:observation_example_text}
\end{minipage}
\caption{
An example of observation provided to agents, which is a pair of a screen image (left) and a text describing the screen layout (right).
The text-based observation consists of a list of descriptions of UI elements, parsed from Android view hierarchy in XML format.
The intermediate part of the list in the text is abbreviated.
To describe, the parts with `numeric\_tag' 9 correspond to an icon of the `Walmart' application, which is observable in the top-right part in the screen image.
}
\label{tab:observation_example}
\end{table}

\paragraph{Action space}
The action space of the agents is defined as either continuous or discrete actions.
The continuous action, here, refers to a dual-gesture action $\{a |\; a=(y_{\text{touch}}, x_{\text{touch}}, y_{\text{lift}}, x_{\text{lift}}) \in \mathbb{R}^4\}$, similar to \citet{rawles2023android}.
Each value of dual-gesture action $a$ is normalized to be in between $[-1, 1]$ with respect to the screen resolutions.
The former two values specify the location of the screen to touch, while the latter two values determine the location of the screen to lift.
This definition enables interpreting useful actions in digital device control, i.e., tapping or swiping the screens, in a precise and compressive manner.
Also, our interface allows pressing the navigator buttons available by touching the screen to support the essential actions for manipulating Android devices.

The discrete action, on the other hand, refers to the pre-defined set of actions for tapping, swiping, and button-pressing.
For agents generating text-based actions, the discrete action is defined to be each action option (i.e., callable function).
For agents trained from scratch, we define each action to be a vector with the size of 378, where the first 378 values correspond to tapping pre-defined locations (14×27 bins) on the screen, four values to swiping directions (up, down, right, left), and the last three values to pressing buttons (back, home, overview).

For dual-gesture actions, we implement an interface that determines whether the action is a tap, swipe, or pressing of navigation buttons i.e., back, home, and overview.
The action parsing interface converts the action into taps, swipes, or pressing buttons following the rule as follows:
\begin{itemize}[leftmargin=6mm]
    \item The action is {\tt tapping}, if $d((x_{\text{touch}}, y_{\text{touch}}), (x_{\text{lift}}, y_{\text{lift}})) < \text{threshold}$
        \begin{itemize}
        \item The {\tt tapping} is to press BACK button, if $(x_{\text{touch}}, y_{\text{touch}}) = (0.95, 0.22)$
        \item The {\tt tapping} is to press HOME button, if $(x_{\text{touch}}, y_{\text{touch}}) = (0.95, 0.50)$
        \item The {\tt tapping} is to press OVERVIEW button, if $(x_{\text{touch}}, y_{\text{touch}}) = (0.95, 0.78)$
        \end{itemize}
    \item The action is {\tt swiping}, if $d((x_{\text{touch}}, y_{\text{touch}}), (x_{\text{lift}}, y_{\text{lift}})) \geq \text{threshold}$,
\end{itemize}
where the threshold value is defined as 0.14.
This value is adjustable by users, while we find that the value of 0.14 ensures proper interactions over UI  elements, e.g., {\tt tapping} the target application icon, in all of our experiments.
These specific values are tested to be consistent across different device types, ensuring that the positions correspond to the correct buttons in all \metabbr environments.

For agents with foundation models, we further define an action converter that translates text-based actions into legal emulator actions.
Following the action space definition, the action options are designed to be either dual-gesture actions or discrete actions.
We prompt the LLM agents to output actions among six possible options:  dual-gesture action, tap, swipe, press(``HOME''), press(``BACK''), and press(``OVERVIEW''). 
The action converter translates the text-based actions into legal actions as below:
\begin{itemize}[leftmargin=6mm]
    \item For the dual-gesture action, it converts the text action into the four floating points by rounding each value into the second decimal point. 
    \item For tap actions, the agent outputs an integer value specifying the numeric tag assigned to the UI element. 
    Given the tapping action with a numeric tag, it converts the action into a tapping dual-gesture action with the bounding box information of the chosen UI element. 
    \item For swipe actions, a direction `up', `down', `left' and `right' is converted into a corresponding dual-gesture action with the value of $(0.8, 0.5, 0.2, 0.5)$, $(0.2, 0.5, 0.8, 0.5)$, $(0.5, 0.2, 0.5, 0.8)$, and $(0.5, 0.8, 0.5, 0.2)$, respectively. 
    \item For the action press(``HOME''), press(``BACK''), and press(``OVERVIEW''), it converts the actions to tap the corresponding screen location. 
\end{itemize}
During the evaluation, we ignore the action in the wrong format by skipping the transition of the environments but penalizing the agents by incrementing a step taken.

\subsection{Training and test environments configurations} \label{app:environment_configurations}

We construct 45 unique environments in \metabbr, where 35 environments are for training and 10 environments are for testing.
Each environment is provided with a unique identification (ID) number, to distinguish the environments easily.
From \autoref{tab:environment_details_0} to \autoref{tab:environment_details_2} show the list of the device configurations and the home screen images of exemplary environments.  

To construct environments, we use popular device types: Pixel 3, Pixel 4, Pixel 4 XL, Pixel 6, and WGXA Tablet. 
For training environments, only Pixel 3 is employed.
For evaluation environments, we use all device types Pixel 3, Pixel 4, Pixel 4 XL, Pixel 6, and WGXA Tablet. 
In these models, we alter the size of UI elements (including application icons) by changing the dots per inch (DPI) values and adjusting the font size of the devices.
For each device type, we prepare three different sizes that users can select.
We, then, change the wallpaper with 13 images collected from a free license image website. 
These wallpaper image files are shared in the open-source repository.
We also customize the background images with the dark theme mode.
If the dark theme mode is activated, the device provides screen images with light-dark color reversed.
For instance, the wallpaper of the application list page is white in the default setting, while it becomes black with dark theme mode activated.
Furthermore, we incorporate changes in locale, specifying the language and location of the devices.
12 different locales are used for 35 training environments, while we include three more locales for the test environments.

\clearpage
\begin{table}[h!]
\centering
    \begin{tabular}{@{}l >{\centering\arraybackslash}m{0.15\linewidth} >{\centering\arraybackslash}m{0.15\linewidth} >{\centering\arraybackslash}m{0.15\linewidth} >{\centering\arraybackslash}m{0.15\linewidth} >{\centering\arraybackslash}m{0.15\linewidth}@{}}
    \toprule
    & 
    \includegraphics[width=0.12\textwidth]{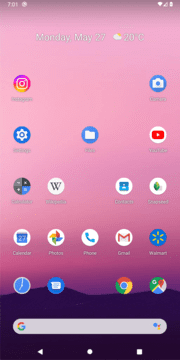} & 
    \includegraphics[width=0.12\textwidth]{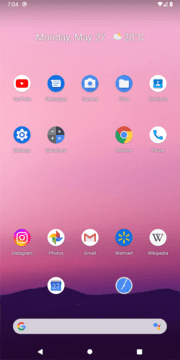} & 
    \includegraphics[width=0.12\textwidth]{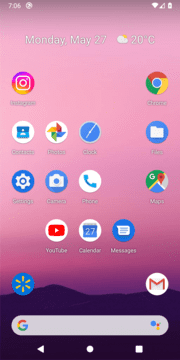} & 
    \includegraphics[width=0.12\textwidth]{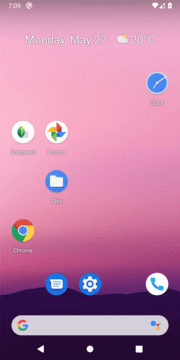} & 
    \includegraphics[width=0.12\textwidth]{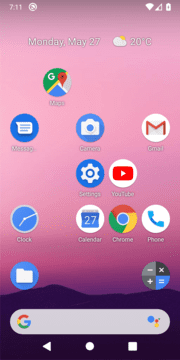} \\
    \midrule
    ID & 000 & 001 & 002 & 003 & 004 \\
    Device type & Pixel 3 & Pixel 3 & Pixel 3 & Pixel 3 & Pixel 3\\
    DPI & 330 & 330 & 440 & 440 & 550 \\
    Font size & 1.15 & 1.15 & 1.0 & 1.0 & 0.85 \\
    Locale & en-US & en-US & en-US & en-US & en-US \\
    Wallpaper & 00$\_$default & 00$\_$default & 00$\_$default & 00$\_$default & 00$\_$default \\
    Dark theme & - & - & - & - & - \\
    \midrule
    &
    \includegraphics[width=0.12\textwidth]{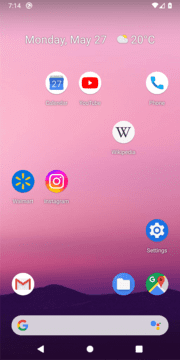} & 
    \includegraphics[width=0.12\textwidth]{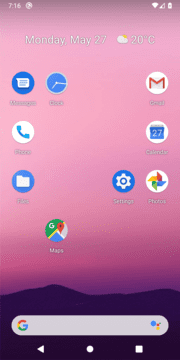} & 
    \includegraphics[width=0.12\textwidth]{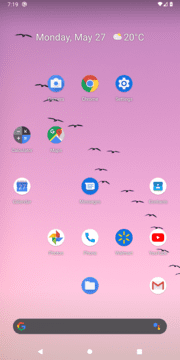} & 
    \includegraphics[width=0.12\textwidth]{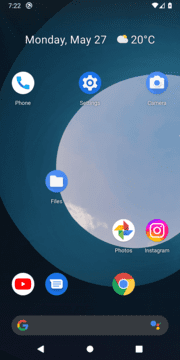} & 
    \includegraphics[width=0.12\textwidth]{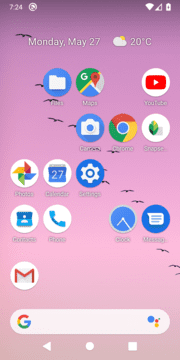} \\
    \midrule
    ID & 005 & 006 & 007 & 008 & 009 \\
    Device type & Pixel 3 & Pixel 3 & Pixel 3 & Pixel 3 & Pixel 3\\
    DPI & 440 & 440 & 330 & 440 & 550 \\
    Font size & 1.0 & 1.0 & 1.15 & 1.0 & 0.85 \\
    Locale & en-US & en-US & en-US & en-US & en-US \\
    Wallpaper & 00$\_$default & 00$\_$default & 01$\_$red & 02$\_$blue & 01$\_$red \\
    Dark theme & - & - & \checkmark & \checkmark & - \\
    \midrule
    &
    \includegraphics[width=0.12\textwidth]{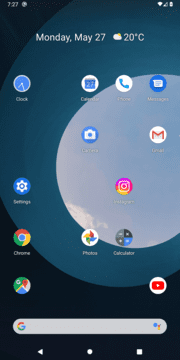} & 
    \includegraphics[width=0.12\textwidth]{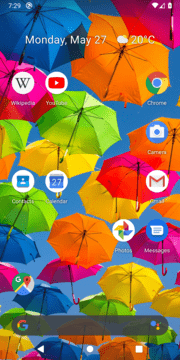} & 
    \includegraphics[width=0.12\textwidth]{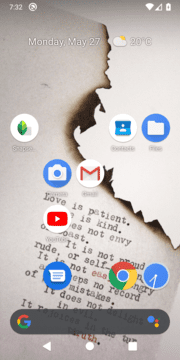} & 
    \includegraphics[width=0.12\textwidth]{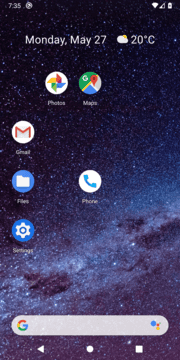} & 
    \includegraphics[width=0.12\textwidth]{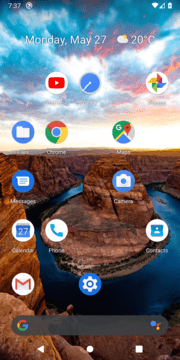} \\
    \midrule
    ID & 010 & 011 & 012 & 013 & 014 \\
    Device type & Pixel 3 & Pixel 3 & Pixel 3 & Pixel 3 & Pixel 3\\
    DPI & 330 & 440 & 550 & 440 & 440 \\
    Font size & 1.15 & 1.0 & 0.85 & 1.0 & 1.0 \\
    Locale & en-US & en-US & en-US & en-US & en-US \\
    Wallpaper & 02$\_$blue & 08$\_$colors & 03$\_$paper & 10$\_$galaxy & 13$\_$canyon \\
    Dark theme & - & \checkmark & \checkmark & - & \checkmark \\
    \toprule
    \end{tabular}
\caption{The device configuration of each environment with the home screen image.}
\label{tab:environment_details_0}
\end{table}

\clearpage

\begin{table}[h!]
\centering
    \begin{tabular}{@{}l >{\centering\arraybackslash}m{0.15\linewidth} >{\centering\arraybackslash}m{0.15\linewidth} >{\centering\arraybackslash}m{0.15\linewidth} >{\centering\arraybackslash}m{0.15\linewidth} >{\centering\arraybackslash}m{0.15\linewidth}@{}}
    \toprule
    & 
    \includegraphics[width=0.12\textwidth]{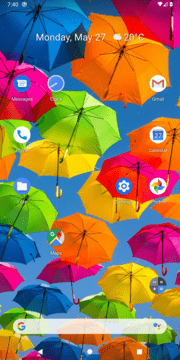} & 
    \includegraphics[width=0.12\textwidth]{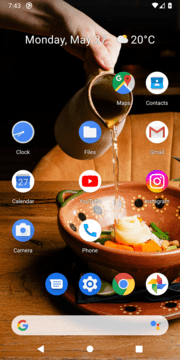} & 
    \includegraphics[width=0.12\textwidth]{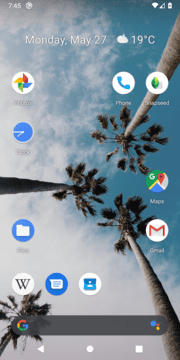} & 
    \includegraphics[width=0.12\textwidth]{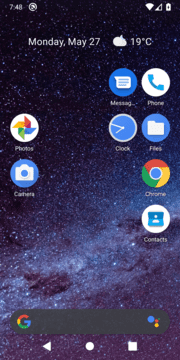} & 
    \includegraphics[width=0.12\textwidth]{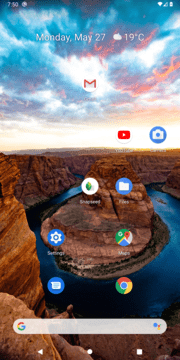} \\
    \midrule
    ID & 015 & 016 & 017 & 018 & 019 \\
    Device type & Pixel 3 & Pixel 3 & Pixel 3 & Pixel 3 & Pixel 3\\
    DPI & 330 & 440 & 440 & 550 & 330 \\
    Font size & 1.15 & 1.0 & 1.0 & 0.85 & 1.15 \\
    Locale & en-US & en-US & en-US & en-US & en-US \\
    Wallpaper & 08$\_$colors & 07$\_$food & 04$\_$sky & 10$\_$galaxy & 13$\_$canyon \\
    Dark theme & - & - & \checkmark & \checkmark & - \\
    \midrule
    & 
    \includegraphics[width=0.12\textwidth]{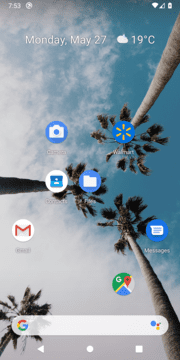} & 
    \includegraphics[width=0.12\textwidth]{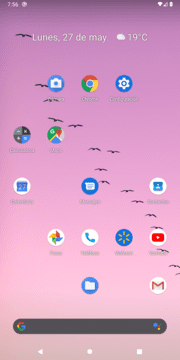} & 
    \includegraphics[width=0.12\textwidth]{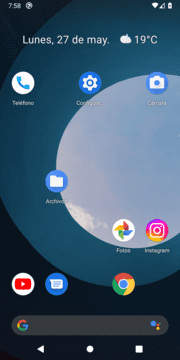} & 
    \includegraphics[width=0.12\textwidth]{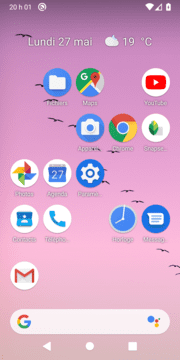} & 
    \includegraphics[width=0.12\textwidth]{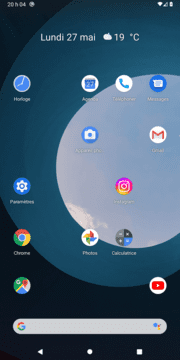} \\
    \midrule
    ID & 020 & 021 & 022 & 023 & 024 \\
    Device type & Pixel 3 & Pixel 3 & Pixel 3 & Pixel 3 & Pixel 3\\
    DPI & 440 & 330 & 440 & 550 & 330 \\
    Font size & 1.0 & 1.15 & 1.0 & 0.85 & 1.15 \\
    Locale & en-US & es-US & es-US & fr-CA & fr-CA \\
    Wallpaper & 04$\_$sky & 01$\_$red & 02$\_$blue & 01$\_$red & 02$\_$blue \\
    Dark theme & - & \checkmark & \checkmark & - & - \\
    \midrule
    & 
    \includegraphics[width=0.12\textwidth]{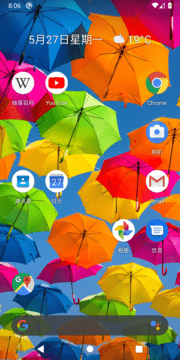} & 
    \includegraphics[width=0.12\textwidth]{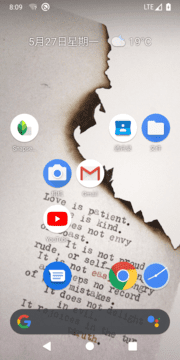} & 
    \includegraphics[width=0.12\textwidth]{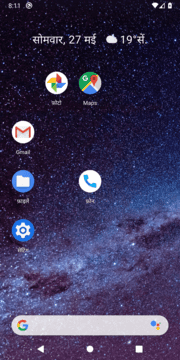} & 
    \includegraphics[width=0.12\textwidth]{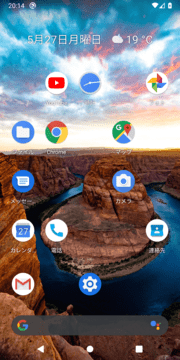} & 
    \includegraphics[width=0.12\textwidth]{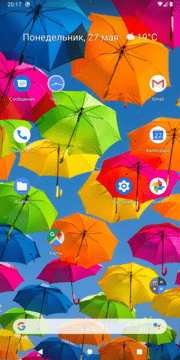} \\
    \midrule
    ID & 025 & 026 & 027 & 028 & 029 \\
    Device type & Pixel 3 & Pixel 3 & Pixel 3 & Pixel 3 & Pixel 3\\
    DPI & 440 & 550 & 440 & 440 & 330 \\
    Font size & 1.0 & 0.85 & 1.0 & 1.0 & 1.15 \\
    Locale & zh-hans-CN & zh-hans-CN & hi-IN & ja-JP & ru-MD \\
    Wallpaper &08$\_$colors & 03$\_$paper & 10$\_$galaxy & 13$\_$canyon & 08$\_$colors \\
    Dark theme & \checkmark & \checkmark & - & \checkmark & - \\
    \toprule
    \end{tabular}
\caption{The device configuration of each environment with the home screen image.}
\label{tab:environment_details_1}
\end{table}

\clearpage

\begin{table}[h!]
\centering
    \begin{tabular}{@{}l >{\centering\arraybackslash}m{0.15\linewidth} >{\centering\arraybackslash}m{0.15\linewidth} >{\centering\arraybackslash}m{0.15\linewidth} >{\centering\arraybackslash}m{0.15\linewidth} >{\centering\arraybackslash}m{0.15\linewidth}@{}}
    \toprule
    & 
    \includegraphics[width=0.12\textwidth]{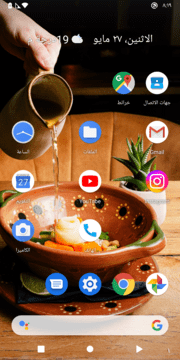} & 
    \includegraphics[width=0.12\textwidth]{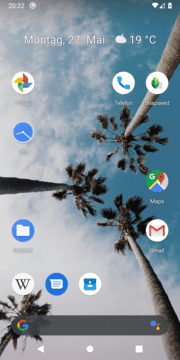} & 
    \includegraphics[width=0.12\textwidth]{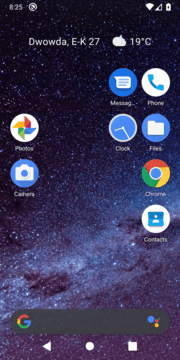} & 
    \includegraphics[width=0.12\textwidth]{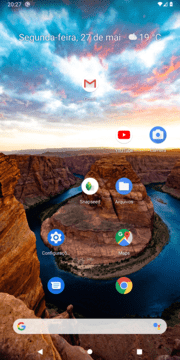} & 
    \includegraphics[width=0.12\textwidth]{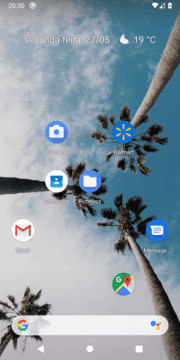} \\
    \midrule
    ID & 030 & 031 & 032 & 033 & 034 \\
    Device type & Pixel 3 & Pixel 3 & Pixel 3 & Pixel 3 & Pixel 3\\
    DPI & 440 & 440 & 550 & 330 & 440 \\
    Font size & 1.0 & 1.0 & 0.85 & 1.15 & 1.0 \\
    Locale & ar-AE & de-DE & ak-GH & pt-BR & pt-PT \\
    Wallpaper & 07$\_$food & 04$\_$sky & 10$\_$galaxy & 13$\_$canyon & 04$\_$sky \\
    Dark theme & - & \checkmark & \checkmark & - & - \\
    \midrule
    & 
    \includegraphics[width=0.12\textwidth]{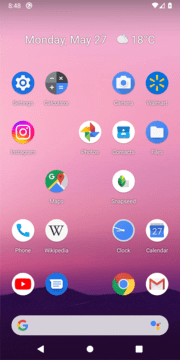} & 
    \includegraphics[width=0.12\textwidth]{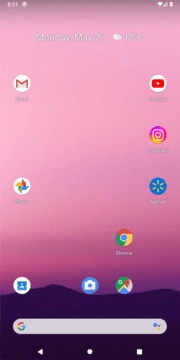} & 
    \includegraphics[width=0.12\textwidth]{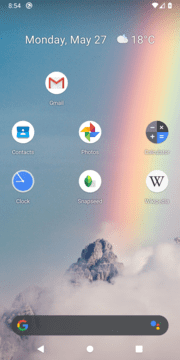} & 
    \includegraphics[width=0.12\textwidth]{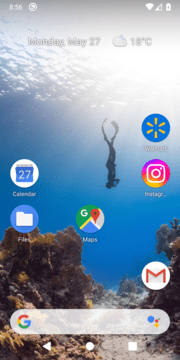} & 
    \includegraphics[width=0.12\textwidth]{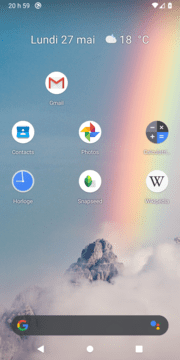} \\
    \midrule
    ID & 100 & 101 & 102 & 103 & 104 \\
    Device type & Pixel 3 & Pixel 3 & Pixel 3 & Pixel 3 & Pixel 3\\
    DPI & 440 & 330 & 440 & 550 & 440 \\
    Font size & 1.0 & 1.15 & 1.0 & 0.85 & 1.0 \\
    Locale & en-US & en-US & en-US & en-US & fr-CA \\
    Wallpaper & 00$\_$default & 00$\_$default & 09$\_$rainbow & 12$\_$ocean & 09$\_$rainbow \\
    Dark theme & - & - & \checkmark & - & \checkmark \\
    \midrule
    \multicolumn{2}{m{0.25\textwidth}}{\centering\includegraphics[width=0.25\textwidth]{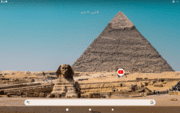}} & 
    \includegraphics[width=0.12\textwidth]{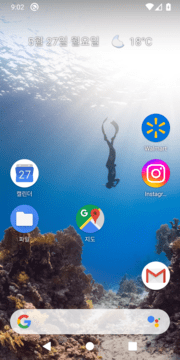} & 
    \includegraphics[width=0.12\textwidth]{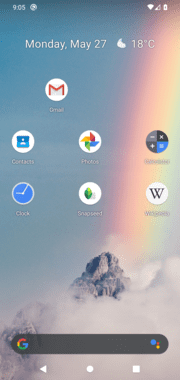} & 
    \includegraphics[width=0.12\textwidth]{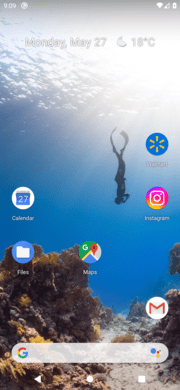} & 
    \includegraphics[width=0.12\textwidth]{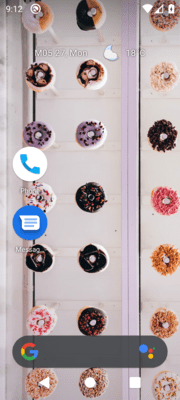} \\
    \midrule
    ID & 109 & 105 & 106 & 107 & 108 \\
    Device type & WXGA Tablet & Pixel 3 & Pixel 4 & Pixel 5 & Pixel 6 \\
    DPI & 160 & 550 & 440 & 440 & 700 \\
    Font size & 1.0 & 0.85 & 1.0 & 1.0 & 0.85 \\
    Locale & ar-EG & ko-KR & en-US & en-US & ur-PK \\
    Wallpaper & 12$\_$ocean & 09$\_$rainbow & 12$\_$ocean & 05$\_$doughnut & 11$\_$pyramid \\
    Dark theme & - & \checkmark & - & \checkmark & - \\
    \toprule
    \end{tabular}
\caption{The device configuration of each environment with the home screen image.}
\label{tab:environment_details_2}
\end{table}

\clearpage

\subsection{Exemples of UI elements changes in the randomized environments}\label{app:UI_elements_change}

\begin{figure}[ht!]
    \centering    \subfigure{\includegraphics[width=0.45\linewidth]{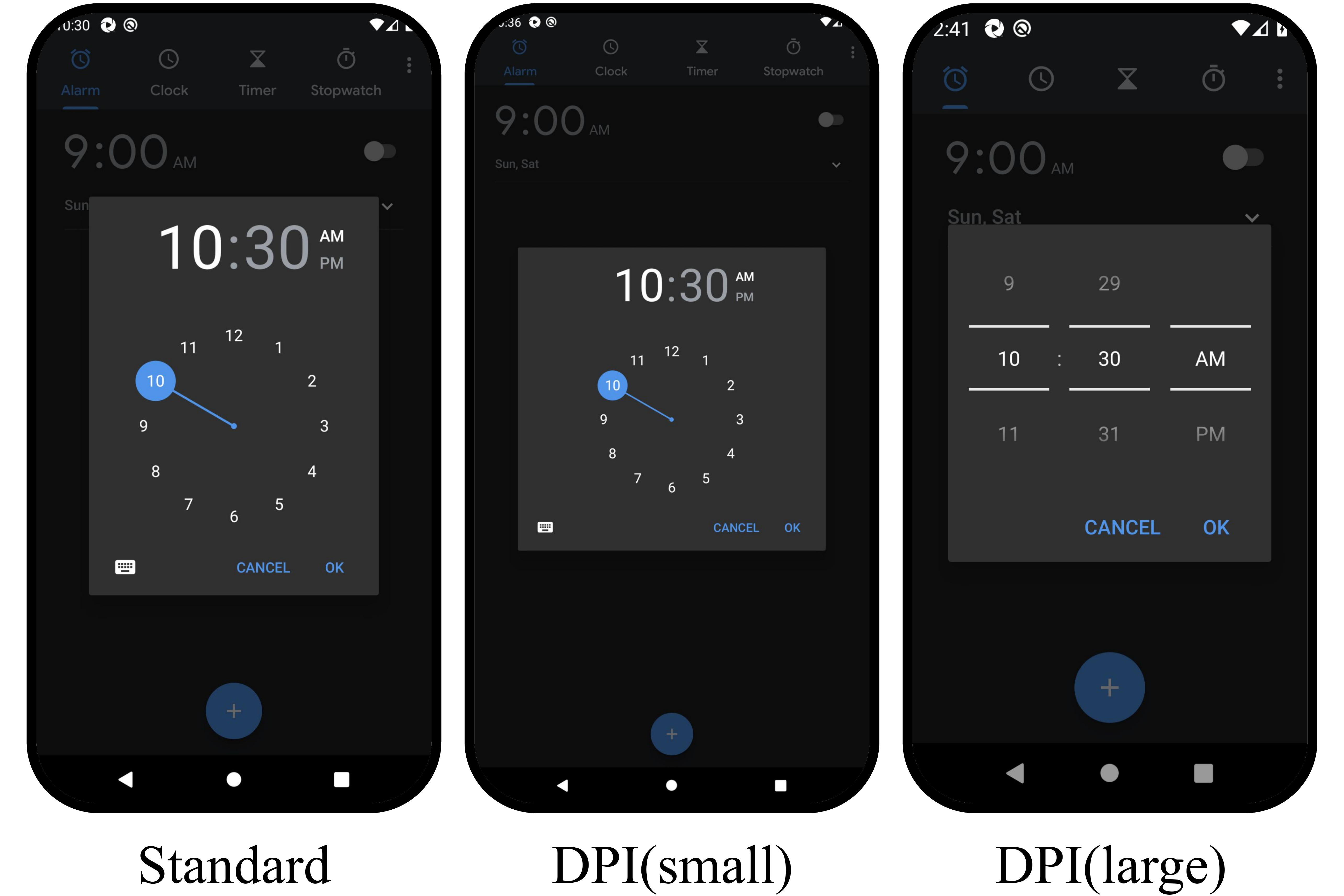}} 
    \subfigure{\includegraphics[width=0.45\linewidth]{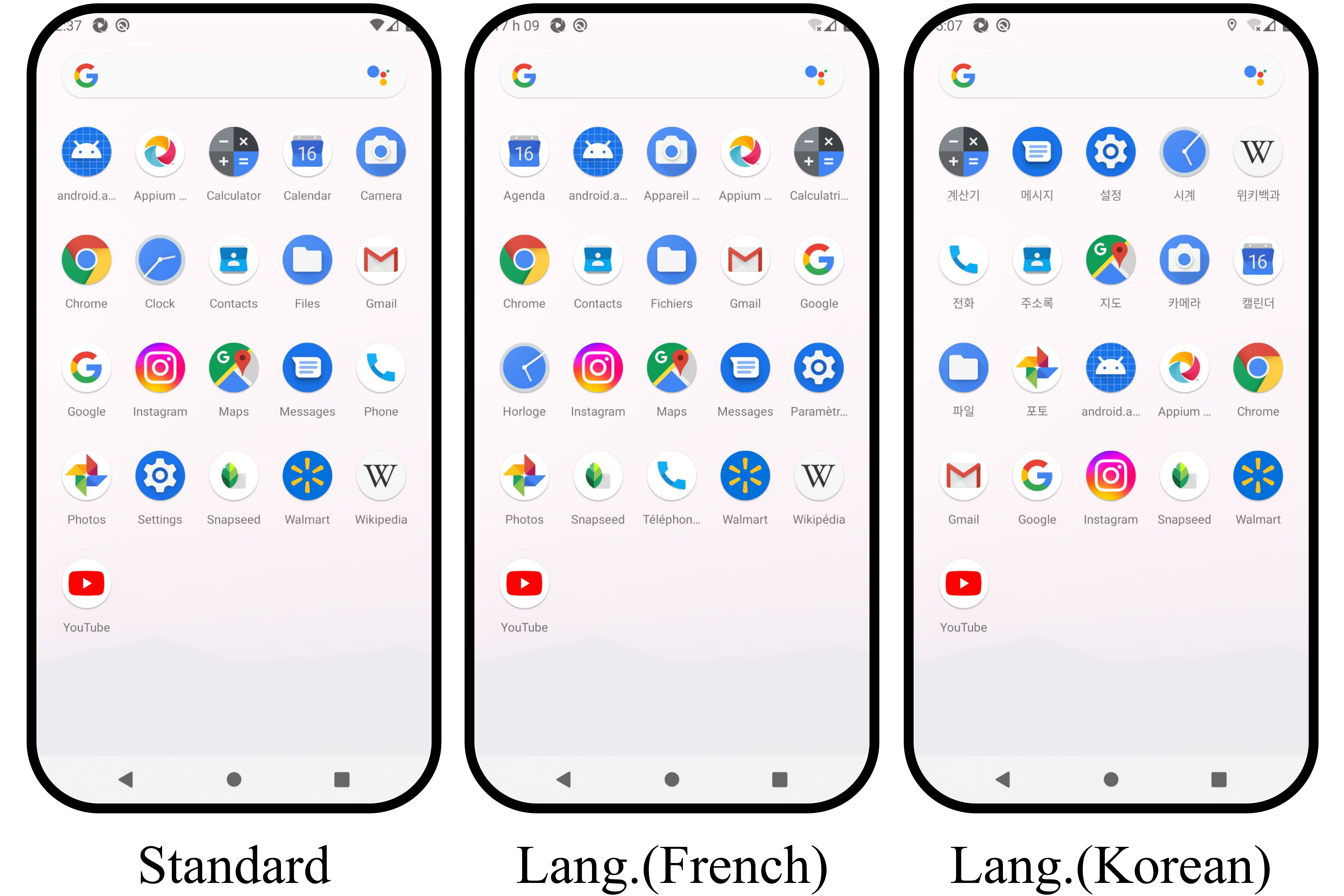}} 
    \caption{
    Visualization of the impact of randomizing each environmental feature on representative UI elements. 
    The agents are challenged to manipulate the differing UI elements appropriately as the environmental features are randomized.
    }
    \label{fig:UI_element_changes_image}
\end{figure}

We experiment with different environments where environmental features of icon location, DPI setting, and language settings are randomized.
For each setting, except the icon location, we display exemplary screens in \autoref{fig:UI_element_changes_image} that show the UI changes most intuitively, compared to "Standard" (i.e., "Test Env 100").
DPI changes affect the shape of UI (e.g., clock UI) and the way of interaction.
Language changes affect the description and positions of application icons in the app list screen (in alphabetical order in each language).
Language changes also affect the description of text-based description by changing the detailed description of UI elements, as shown in an example of changes in the text-based description when the language setting is randomized in \autoref{tab:UI_element_changes_text}.

\begin{table}[ht]
\centering
\resizebox{0.75\textwidth}{!}{
    \begin{tabular}[h]{@{}l@{}}
    \small
    \centering
    \tcbox[colback=white,boxrule=1pt,arc=3mm]{
        \begin{tblr}{colspec={@{}X@{}},rowsep=1pt}
        \texttt{[}
        \{`numeric\_tag': 0, `resource\_id': `', `class': `View', `description': `앱목록'\}, 
        \{`numeric\_tag': 1, `resource\_id': `id\_scrim\_view', `class': `View', `description': `'\}, 
        ...,
        \{`numeric\_tag': 7, `resource\_id': `', `class': `TextView', `description': `캘린더'\}, 
        \{`numeric\_tag': 8, `resource\_id': `', `class': `TextView', `description': `Instagram'\}, 
        \{`numeric\_tag': 9, `resource\_id': `', `class': `TextView', `description': `파일'\}, 
        \{`numeric\_tag': 10, `resource\_id': `', `class': `TextView', `description': `지도'\}, 
        ...,
        \{`numeric\_tag': 27, `resource\_id': `', `class': `FrameLayout', `description': `'\}
        \texttt{]}
        \end{tblr}}
    \end{tabular}
}
\caption{
Description of the impact of randomizing an environmental feature of the language setting on the text-based observation. 
The agents are challenged to understand the UI elements described in an altered language setting.
}
\label{tab:UI_element_changes_text}
\end{table}

%% file: sections/appendix/task_details.tex
\clearpage
\section{Task details}

\subsection{List of daily tasks} \label{app:task_list}

\metabbr presents \tasknum daily tasks that are common in everyday life.
The tasks are designed to operate in diverse environments seamlessly and cover commonly used applications.
Daily tasks effectively simulate a wide range of essential skills for mobile device control problems, such as manipulating UI elements (including application icons, checkboxes, and sliders), and can be employed for evaluating mobile device control agents' capabilities in performing tasks that mirror our daily activities.

\paragraph{Task category}
We categorize each task based on the application into five groups: System, Web/Shopping, Communication, Utility, and Event.
The group System includes tasks using applications of Files and Settings.
The group Web/Shopping includes tasks using applications of Chrome, Google, Walmart, and Wikipedia.
The group Communication includes tasks using applications of Contacts, Gmail, Instagram, Message, Phone, and Youtube.
The group Utility includes tasks using applications of Calculator, Camera, Maps, Photos, and Snapseed.
The group Event includes tasks using applications of Calendar and Clock.
Additionally, \metabbr presents several tasks requiring the agent to control the device over several applications (e.g., Clock and Settings).
For those tasks, we categorize the tasks to be in a group where the major sub-task belongs to and add a star mark (\*) on the list of tasks, presented in from \autoref{tab:task_list_0} to \autoref{tab:task_list_7}.

\paragraph{Score criteria}
We have defined success criteria for each task by analyzing significant changes in the environment (i.e., three key information sources) observed during task completion in human expert demonstrations.
We also set the maximum step limits, which are set for the rigorous evaluation of the agents’ proficiency in each task.
From \autoref{tab:task_list_0} to \autoref{tab:task_list_7}, we include the detailed list of tasks with the detailed success criteria.
For more precise score criteria, please refer to the supplementary code materials.

In detail, the source of success criteria are threefold: the system information, app data, and attributes of the UI elements in the Android view hierarchy.
\begin{itemize}[leftmargin=3mm]
    \item System information includes system log and system setting. 
    \begin{itemize}[leftmargin=3mm]
        \item For the system log, the success detector checks the system log matches with the criteria defined in a regular expression.
        The log that matches the criteria should satisfy two parts: a filter specifying the target application or activity (denoted as $\lbrack...\rbrack$ at the success criteria column in from \autoref{tab:task_list_0} to \autoref{tab:task_list_7}) and a regex specifying the detail of the log (denoted as "..." at the success criteria column in from \autoref{tab:task_list_0} to \autoref{tab:task_list_7}). 
        \item For the system setting, the success detector utilizes ADB commands to retrieve specific information related to fields. 
        The success detector, then, parses the results of these commands with a regular expression to extract the necessary system information and check it (denoted as \{"(adb command)": "(regular expression)"\} at the success criteria column in from \autoref{tab:task_list_0} to \autoref{tab:task_list_7}).
    \end{itemize}
    \item For the attributes of the UI elements, the success detector checks if certain attributes of specified UI elements satisfy the pre-defined conditions.
    We define the condition with UI elements having certain IDs (denoted as $\lbrack...\rbrack$ at the success criteria column in from \autoref{tab:task_list_0} to \autoref{tab:task_list_7}) and the status as the value of attributes (denoted as \{"(attribute key)": "(status value)"\} at the success criteria column in from \autoref{tab:task_list_0} to \autoref{tab:task_list_7}).
    \item For the app data, the success detector checks if the values of particular attributes in either the database or shared preferences (used to store simple data instead of a database as an XML file) meet the criteria.
    We define the condition as selected attributes from the database or shared preference (path denoted as $\lbrack...\rbrack$ at the success criteria column in from \autoref{tab:task_list_0} to \autoref{tab:task_list_7}) to be specific values (denoted as \{"(attribute key)": "(status value)"\} at the success criteria column in from \autoref{tab:task_list_0} to \autoref{tab:task_list_7}).
\end{itemize}

\paragraph{Accounts for log-in}
Performing several daily tasks in the Instagram application requires a log-in process.
We warn that exploiting private accounts on these tasks may cause a leak of personal information and recommend creating a dummy account.
To avoid unintended data loss or exposure, we set the tasks to be concise and the maximum step limits to be as small as possible.

\clearpage 

\begin{table}[!ht]
\centering
\resizebox{\textwidth}{!}{
\small
    \begin{tabular}{>{\centering}p{0.1\textwidth}p{0.25\textwidth}>{\centering}p{0.05\textwidth}>{\centering}p{0.15\textwidth}p{0.45\textwidth}}
    \toprule
    Application & Task instruction  & \multilinecell{Step\\limit} & \multilinecell{Success criteria\\(source)} & Success criteria (detail) \\
    \midrule
    Calculator & ``open Calculator'' & 4 & UI elements & \{``id'': ``com.google.android.calculator:id/clr'', ``enabled'': ``true''\} \\
    \midrule
    Calculator & ``input 1 in Calculator'' & 5 & UI elements & \{``id'': ``com.google.android.calculator:id/formula'', ``text'': ``1''\} \\
    \midrule
    Calculator & ``input factorial of 6 in Calculator'' & 7 & UI elements & \{``id'': ``com.google.android.calculator:id/formula'', ``text'': ``6!''\} \\
    \midrule
    Calculator & ``input `1+1' in Calculator'' & 8 & UI elements & \{``id'': ``com.google.android.calculator:id/formula'', ``text'': ``1+1''\} \\
    \midrule
    Calculator & ``input `3×5' in Calculator'' & 8 & UI elements & \{``id'': ``com.google.android.calculator:id/formula'', ``text'': ``3×5''\} \\
    \midrule
    Calculator & ``input square root of 25 in Calculator'' & 8 & UI elements & \{``id'': ``com.google.android.calculator:id/formula'', ``text'': ``$\sqrt{25}$''\} \\
    \midrule
    Calculator & ``input `cos(60)' in Calculator'' & 9 & UI elements & \{``id'': ``com.google.android.calculator:id/formula'', ``text'': ``c60''\} \\
    \midrule
    Calculator & ``compute 50\% of 28 ('50\%28') in Calculator'' & 9 & UI elements & \{``id'': ``com.google.android.calculator:id/formula'',\newline ``text'': ``50\%28''\} \\
    \midrule
    Calculator & ``input `17×23' in Calculator'' & 10 & UI elements & \{``id'': ``com.google.android.calculator:id/formula'',\newline ``text'': ``17×23''\} \\
    \midrule
    Calculator & ``input `2+24÷3' in Calculator'' & 10 & UI elements & \{``id'': ``com.google.android.calculator:id/formula'',\newline ``text'': ``2+24÷3''\} \\
    \midrule
    Calculator & ``input `cos(180)' in Calculator'' & 10 & UI elements & \{``id'': ``com.google.android.calculator:id/formula'',\newline ``text'': ``c180''\} \\
    \midrule
    Calculator & ``input `ln(1234)' in Calculator'' & 10 & UI elements & \{``id'': ``com.google.android.calculator:id/formula'',\newline ``text'': ``l1234''\} \\
    \midrule
    Calculator & ``input the formula for computing sum of the first 5 Fibonacci numbers in Calculator'' & 13 & UI elements & \{``id'': ``com.google.android.calculator:id/formula'',\newline ``text'': ``0+1+1+2+3'' or ``1+1+2+3+5''\} \\
    \midrule
    Calculator & ``input the formula for converting 45 degrees to radians ('45$\times \pi \div$180') in Calculator'' & 13 & UI elements & \{``id'': ``com.google.android.calculator:id/formula'',\newline ``text'': ``45$\times \pi \div$180''\} \\
    \midrule
    Calculator & ``input the formula for computing sum of the first 5 prime numbers in Calculator'' & 14 & UI elements & \{``id'': ``com.google.android.calculator:id/formula'',\newline ``text'': ``2+3+5+7+11''\} \\
    \midrule
    Calculator & ``input `5!÷(2!x3!)' in Calculator'' & 15 & UI elements & \{``id'': ``com.google.android.calculator:id/formula'',\newline ``text'': ``5!÷(2!×3!)''\} \\
    \midrule
    Calculator & ``input `10!÷(2!x8!)' in Calculator'' & 15 & UI elements & \{``id'': ``com.google.android.calculator:id/formula'',\newline ``text'': ``10!÷(2!×8!)''\} \\
    \midrule
    Calculator & ``compute the harmonic mean of 4 and 5 in Calculator'' & 18 & UI elements & \{``id'': [``com.google.android.calculator:id/result\_preview'', ``com.google.android.calculator:id/result\_final''],\newline ``text'': starts with ``4.44''\} \\
    \midrule
    Calculator & ``compute the geometric mean of 3, 4, and 5 in Calculator'' & 18 & UI elements & \{``id'': [``com.google.android.calculator:id/result\_preview'', ``com.google.android.calculator:id/result\_final''],\newline ``text'': starts with ``3.91''\} \\
    \midrule
    Calendar & ``open the calendar app'' & 4 & System log & [ActivityTaskManager:I] \newline ``\symbol{94}(.*?)START(.*?)com.android.calendar'' \\
    \midrule
    Camera & ``open the camera app'' & 4 & System log & [ActivityTaskManager:I] \newline ``\symbol{94}(.*?)Start proc(.*?)com.android.camera'' \\
    \midrule
    Chrome & ``open Chrome'' & 4 & System log & [ActivityTaskManager:I] \newline ``\symbol{94}(.*?)START(.*?)com.google.android.apps.chrome'' \\
    \midrule
    Chrome & ``open a new tab in Chrome'' & 5 & UI elements & \{``id'': ``com.android.chrome:id/tab\_switcher\_button'',\newline ``content-desc'': ``.*2.*''\} \\
    \midrule
    Chrome & ``go to search history in Chrome'' & 6 & System log & [ActivityTaskManager:I] \newline ``\symbol{94}(.*?)START(.*?)chrome.browser.history.
    \newline HistoryActivity'' \\
    \toprule
    \end{tabular}
}
\caption{Comprehensive list of tasks.}
\label{tab:task_list_0}
\end{table}

\clearpage 

\begin{table}[!ht]
\centering
\resizebox{\textwidth}{!}{
\small
    \begin{tabular}{>{\centering}p{0.1\textwidth}p{0.25\textwidth}>{\centering}p{0.05\textwidth}>{\centering}p{0.15\textwidth}p{0.45\textwidth}}
    \toprule
    Application & Task instruction  & \multilinecell{Step\\limit} & \multilinecell{Success criteria\\(source)} & Success criteria (detail) \\
    \midrule
    Clock & ``open the clock app'' & 4 & System log & [ActivityTaskManager:I] ``\symbol{94}(.*?)START(.*?)com.android.deskclock'' \\
    \midrule
    Clock & ``go to the stopwatch page in clock'' & 5 & System log & [AlarmClock:D] \symbol{94}(.*?)Events: \texttt{[}Stopwatch\texttt{]} \texttt{[}Show Tab\texttt{]} \texttt{[}Tap\texttt{]} \\
    \midrule
    Clock & ``go to the alarm page in clock'' & 5 & System log & [AlarmClock:D] \newline ``\symbol{94}(.*?)Events: \texttt{[}Alarm\texttt{]} \texttt{[}Show Tab\texttt{]} \texttt{[}Tap\texttt{]}'' \\
    \midrule
    Clock & ``go to the timer page in clock'' & 5 & System log & [AlarmClock:D] \newline ``\symbol{94}(.*?)Events: \texttt{[}Timer\texttt{]} \texttt{[}Show Tab\texttt{]} \texttt{[}Tap\texttt{]}'' \\
    \midrule
    Clock & ``turn on alarm at 9 am'' & 6 & System log & [AlarmClock:D] \newline ``\symbol{94}(.*?)Created new alarm instance'' \\
    \midrule
    Clock & ``start the stopwatch in clock'' & 7 & System log & [AlarmClock:D] ``\symbol{94}(.*?)Start'' \\
    \midrule
    Clock & ``create alarm at 06:30 am'' & 11 & System log & [ConditionProviders.SCP:D] \newline ``\symbol{94}(.*?)nextUserAlarmTime(.*?)06:30:00'' \\
    \midrule
    Clock & ``create alarm at 10:30 am'' & 11 & System log & [ConditionProviders.SCP:D] \newline ``\symbol{94}(.*?)nextUserAlarmTime(.*?)10:30:00'' \\
    \midrule
    Clock & ``create alarm at 13:30 pm'' & 11 & System log & [ConditionProviders.SCP:D] \newline ``\symbol{94}(.*?)nextUserAlarmTime(.*?)13:30:00'' \\
    \midrule
    Clock & ``create alarm at 17:30 pm'' & 11 & System log & [ConditionProviders.SCP:D] \newline ``\symbol{94}(.*?)nextUserAlarmTime(.*?)17:30:00'' \\
    \midrule
    Clock & ``create alarm at 20:30 pm'' & 11 & System log & [ConditionProviders.SCP:D] \newline ``\symbol{94}(.*?)nextUserAlarmTime(.*?)20:30:00'' \\
    \midrule
    Clock & ``create alarm at 23:30 pm'' & 11 & System log & [ConditionProviders.SCP:D] \newline ``\symbol{94}(.*?)nextUserAlarmTime(.*?)23:30:00'' \\
    \midrule
    Clock & ``create alarm at 10:30 am on every weekday'' & 14 & App data (Database) & [/data/user\_de/0/com.google.android.deskclock/
    \newline databases/alarms.db] \{``hour''=10,``minutes''=30, ``daysofweek''=31\} \\
    \midrule
    Clock & ``create alarm at 10:30 am on every midweek'' & 14 & App data (Database) & [/data/user\_de/0/com.google.android.deskclock/
    \newline databases/alarms.db] \{``hour''=10,``minutes''=30, ``daysofweek''=15\} \\
    \midrule
    Clock & ``create alarm at 13:30 pm and another alarm 2 hours before it)'' & 14 & App data (Database) & [/data/user\_de/0/com.google.android.deskclock/
    \newline databases/alarms.db] [\{``hour''=13,``minutes''=30\}, \{``hour''=11,``minutes''=30\}] \\
    \midrule
    Clock & ``create alarm at 13:30 pm on every weekday'' & 14 & App data (Database) & [/data/user\_de/0/com.google.android.deskclock/
    \newline databases/alarms.db] \{``hour''=10,``minutes''=30, ``daysofweek''=31\} \\
    \midrule
    Clock & ``create alarm at 10:30 am on every weekend'' & 15 & App data (Database) & [/data/user\_de/0/com.google.android.deskclock/
    \newline databases/alarms.db] \{``hour''=10,``minutes''=30, ``daysofweek''=96\} \\
    \midrule
    Clock & ``create alarm at 13:30 pm on every weekend'' & 16 & App data (Database) & [/data/user\_de/0/com.google.android.deskclock/
    \newline databases/alarms.db] \{``hour''=10,``minutes''=30, ``daysofweek''=96\} \\
    \midrule
    Clock & ``create alarm at 13:30 pm and another alarm 2 hours after it'' & 18 & App data (Database) & [/data/user\_de/0/com.google.android.deskclock/
    \newline databases/alarms.db] \{``hour''=10,``minutes''=30\} \\
    \toprule
    \end{tabular}
}
\caption{Comprehensive list of tasks.}
\label{tab:task_list_1}
\end{table}

\clearpage 

\begin{table}[!ht]
\centering
\resizebox{\textwidth}{!}{
\small
    \begin{tabular}{>{\centering}p{0.1\textwidth}p{0.25\textwidth}>{\centering}p{0.05\textwidth}>{\centering}p{0.15\textwidth}p{0.45\textwidth}}
    \toprule
    Application & Task instruction  & \multilinecell{Step\\limit} & \multilinecell{Success criteria\\(source)} & Success criteria (detail) \\
    \midrule
    Clock (*) & ``turn on alarm at 9 am in clock and increase alarm volume in setting'' & 12 & App data (Database) + System setting & [/data/user\_de/0/com.google.android.deskclock/
    \newline databases/alarms.db] \{``hour''=10,``minutes''=30, ``daysofweek''=15, ``enabled''=1 \} \\
    \midrule
    Clock (*) & ``create alarm at 13:30 pm in clock and increase alarm volume in setting'' & 15 & App data (Database) + System setting & [/data/user\_de/0/com.google.android.deskclock/
    \newline databases/alarms.db] [\{``hour''=13,``minutes''=30\}, \{``hour''=15,``minutes''=30\}] \\
    \midrule
    Clock (*) & ``create alarm at 13:30 pm on every weekend and increase alarm volume in setting'' & 16 & App data (Database) + System setting & [/data/user\_de/0/com.google.android.deskclock/
    \newline databases/alarms.db] \{``hour''=13,``minutes''=30, ``daysofweek''=31\} \\
    \midrule
    Clock (*) &  ``turn on airplane mode in setting and create alarm at 10:30 am in clock" & 17 & App data (Database) +  System setting & [/data/user\_de/0/com.google.android.deskclock/
    \newline databases/alarms.db] \{``hour''=10,``minutes''=30\} \\
    \midrule
    Clock (*) &  ``turn on airplane mode in setting and create alarm at 13:30 pm in clock" & 17 & App data (Database) +  System setting & [/data/user\_de/0/com.google.android.deskclock/
    \newline databases/alarms.db] \{``hour''=13,``minutes''=30\} \\
    \midrule
    Clock (*) & ``create alarm at 10:30 am in clock and increase alarm volume in setting'' & 18 & App data (Database) + System setting & [/data/user\_de/0/com.google.android.deskclock/
    \newline databases/alarms.db] \{``hour''=10,``minutes''=30\} \\
    \midrule
    Contacts & ``open the contact app'' & 4 & System log & [ActivityManager:I] \newline
    ``\symbol{94}(.*?)Start proc(.*?)com.android.contacts'' \\
    \midrule
    Contacts & ``activate the insert page in contact'' & 5 & System log & [ActivityManager:I] ``\symbol{94}(.*?)START(.*?)INSERT(.*?)
    \newline ContactEditorActivity'' \\
    \midrule
    Files & ``open the file manager app'' & 4 & System log & [ActivityTaskManager:I] \newline \symbol{94}(.*?)START(.*?)files.FilesActivity'' \\
    \midrule
    Files & ``list audio files in file manager'' & 6 & System log & [DirectoryFragment:D] \newline ``\symbol{94}(.*?)Showing directory(.*?)audio(.*?)root'' \\
    \midrule
    Files & ``list image files in file manager'' & 6 & System log & [DirectoryFragment:D] \newline ``\symbol{94}(.*?)Showing directory(.*?)images'' \\
    \midrule
    Files & ``list video files in file manager'' & 6 & System log & [DirectoryFragment:D] \newline ``\symbol{94}(.*?)Showing directory(.*?)videos'' \\
    \midrule
    Files & ``list download files in file manager'' & 6 & System log & [DirectoryFragment:D] \newline
    ``\symbol{94}(.*?)Showing directory(.*?)download'' \\
    \midrule
    Gmail & ``open the Gmail'' & 4 & System log & [ActivityTaskManager:I] \newline ``\symbol{94}(.*?)START(.*?)com.google.android.gm'' \\
    \midrule
    Instagram & ``open Instagram'' & 4 & UI elements & \{``id'': ``com.instagram.android:id/feed\_tab'', ``selected'': ``true''\} \\
    \midrule
    Instagram & ``go to my profile in Instagram'' & 5 & UI elements & \{``id'': ``com.instagram.android:id/profile\_tab'', ``selected'': ``true''\} \\
    \midrule
    Instagram & ``go to reels tab in Instagram'' & 5 & UI elements & \{``id'': ``com.instagram.android:id/clips\_tab'', ``selected'': ``true''\} \\
    \midrule
    Instagram & ``go to search tab in Instagram'' & 5 & UI elements & \{``id'': ``com.instagram.android:id/search\_tab'', ``selected'': ``true''\} \\
    \midrule
    Maps & ``open Maps'' & 4 & System log & 
    [ActivityTaskManager:I] ``\symbol{94}(.*?)START
    \newline (.*?)com.google.android.maps.MapsActivity'' \\
    \midrule
    Messages & ``open the message app'' & 4 & System log & [ActivityTaskManager:I] ``\symbol{94}(.*?)START
    \newline (.*?)com.google.android.apps.messaging'' \\
    \midrule
    Messages & ``start chatting in message'' & 5 & System log & [BugleUsageStatistics:I] \newline ``\symbol{94}(.*?)BUGLE CREATE(.*?)DEFAULT'' \\
    \toprule
    \end{tabular}
    
}
\caption{Comprehensive list of tasks.}
\label{tab:task_list_2}
\end{table}

\clearpage 

\begin{table}[!ht]
\centering
\resizebox{\textwidth}{!}{
\small
    \begin{tabular}{>{\centering}p{0.1\textwidth}p{0.25\textwidth}>{\centering}p{0.05\textwidth}>{\centering}p{0.15\textwidth}p{0.45\textwidth}}
    \toprule
    Application & Task instruction  & \multilinecell{Step\\limit} & \multilinecell{Success criteria\\(source)} & Success criteria (detail) \\
    \midrule
    Phone & ``open the phone app'' & 4 & System log & [Dialer:I] ``\symbol{94}(.*?)MainActivity.onCreate'' \\
    \midrule
    Phone & ``call 911'' & 9 & System log & [Telecom:I] ``\symbol{94}(.*?)Emergency number detected'' \\
    \midrule
    Phone & ``call 11489'' & 11 & UI elements & [\{``id'':``com.android.dialer:id/incall\_end\_cal'',
    \newline ``enabled'': ``true''\}, \{``id'':``com.android.
    \newline dialer:id/contactgrid\_contact\_name'', ``text'': ``11489''\}] \\
    \midrule
    Phone & ``call 311311'' & 12 & UI elements & 
    [\{``id'':``com.android.dialer:id/incall\_end\_cal'', 
    \newline ``enabled'': ``true''\},  \{``id'':``com.android.
    \newline dialer:id/contactgrid\_contact\_name'', ``text'': ``311311''\}] \\
    \midrule
    Phone & ``call 123-4578'' & 13 & UI elements & [\{``id'':``com.android.dialer:id/incall\_end\_cal'',
    \newline ``enabled'': ``true''\}, \{``id'':``com.android.
    \newline dialer:id/contactgrid\_contact\_name'', ``text'': ``1234578''\}] \\
    \midrule
    Phone & ``call 223-4458'' & 13 & UI elements & [\{``id'':``com.android.dialer:id/incall\_end\_cal'',
    \newline ``enabled'': ``true''\}, \{``id'':``com.android.
    \newline dialer:id/contactgrid\_contact\_name'', ``text'': ``223-4458''\}] \\
    \midrule
    Phone & ``call 402-7717'' & 13 & UI elements & [\{``id'':``com.android.dialer:id/incall\_end\_cal'',
    \newline ``enabled'': ``true''\}, \{``id'':``com.android.
    \newline dialer:id/contactgrid\_contact\_name'', ``text'': ``402-7717''\}] \\
    \midrule
    Phone & ``call 766-3394'' & 13 & UI elements & [\{``id'':``com.android.dialer:id/incall\_end\_cal'',
    \newline ``enabled'': ``true''\}, \{``id'':``com.android.
    \newline dialer:id/contactgrid\_contact\_name'', ``text'': ``766-3394''\}] \\
    \midrule
    Phone & ``call 987-6654'' & 13 & UI elements & [\{``id'':``com.android.dialer:id/incall\_end\_cal'',
    \newline ``enabled'': ``true''\}, \{``id'':``com.android.
    \newline dialer:id/contactgrid\_contact\_name'', ``text'': ``987-6654''\}] \\
    \midrule
    Phone & ``call 2000-0202'' & 14 & UI elements & [\{``id'':``com.android.dialer:id/incall\_end\_cal'',
    \newline ``enabled'': ``true''\}, \{``id'':``com.android.
    \newline dialer:id/contactgrid\_contact\_name'', ``text'': ``(200)002-02''\}] \\
    \midrule
    Phone & ``call the national weather service (301-713-0622)'' & 14 & UI elements & [\{``id'':``com.android.dialer:id/incall\_end\_cal'',
    \newline ``enabled'': ``true''\}, \{``id'':``com.android.
    \newline dialer:id/contactgrid\_contact\_name'', ``text'': ``(301)713-0622''\}] \\
    \midrule
    Phone & ``call the social security administration (800-772-1213)'' & 14 & UI elements & [\{``id'':``com.android.dialer:id/incall\_end\_cal'',
    \newline ``enabled'': ``true''\}, \{``id'':``com.android.
    \newline dialer:id/contactgrid\_contact\_name'', ``text'': ``(800)772-1213''\}] \\
    \midrule
    Phone & ``call 26-445-1193'' & 15 & UI elements & [\{``id'':``com.android.dialer:id/incall\_end\_cal'',
    \newline ``enabled'': ``true''\}, \{``id'':``com.android.
    \newline dialer:id/contactgrid\_contact\_name'', ``text'': ``(264)451-193''\}] \\
    \midrule
    Phone & ``call the US national contact center (800-333-4636)'' & 16 & UI elements & 
    [\{``id'':``com.android.dialer:id/incall\_end\_cal'',
    \newline ``enabled'': ``true''\}, \{``id'':``com.android.
    \newline dialer:id/contactgrid\_contact\_name'', ``text'': ``(800)333-4636''\}] \\
    \midrule
    Phone & ``call the white house (202-456-1111)'' & 17 & UI elements & 
    [\{``id'':``com.android.dialer:id/incall\_end\_cal'', 
    \newline ``enabled'': ``true''\},  \{``id'':``com.android.
    \newline dialer:id/contactgrid\_contact\_name'', ``text'': ``(202)456-1111''\}] \\
    \midrule
    Photos & ``open the photos app'' & 4 & System log & [ActivityTaskManager:I] \symbol{94}(.*?)START(.*?)com.google.android.apps.photos \\
    \toprule
    \end{tabular}
}
\caption{Comprehensive list of tasks.}
\label{tab:task_list_3}
\end{table}

\clearpage 

\begin{table}[!ht]
\centering
\resizebox{\textwidth}{!}{
\small
    \begin{tabular}{>{\centering}p{0.1\textwidth}p{0.25\textwidth}>{\centering}p{0.05\textwidth}>{\centering}p{0.15\textwidth}p{0.45\textwidth}}
    \toprule
    Application & Task instruction  & \multilinecell{Step\\limit} & \multilinecell{Success criteria\\(source)} & Success criteria (detail) \\
    \midrule
    Settings & ``open the setting app'' & 4 & System log & [ActivityManager:I] \newline
    ``\symbol{94}(*.?)Start proc(.*?)com.android.settings.Settings'' \\
    \midrule
    Settings & ``turn on airplane mode'' & 5 & System log & [PhoneGlobals:I] \newline
    ``\symbol{94}(.*?)Turning radio off(.*?)airplane'' \\
    \midrule
    Settings & ``turn off wifi'' & 5 & System log & [WifiService:I] ``\symbol{94}(.*?)setWifiEnabled
    \newline (.*?)com.android.settings(.*?)enable\texttt{=}false'' \\
    \midrule
    Settings & ``decrease the screen brightness in setting'' & 6 & System log & [DisplayPowerController:V] \newline
    ``\symbol{94}(.*?)Brightness(.*?)changing(.*?)manual'' \\
    \midrule
    Settings & ``go to app info list in setting'' & 6 & System log & [SettingsActivity:D] ``\symbol{94}(.*?)Switching
    \newline (.*?)android.settings(.*?)System log'' \\
    \midrule
    Settings & ``go to bluetooth setting'' & 6 & System log & [PrefCtrlListHelper:D] 
    \newline ``\symbol{94}(.*?)android.settings.bluetooth.BluetoothDevice'' \\ 
    \midrule
    Settings & ``toggle dark theme in setting'' & 6 & System log & [SettingsProvider:V] \newline
    ``\symbol{94}(.*?)content(.*?)settings(.*?)dark(.*?)mode'' \\
    \midrule
    Settings & ``toggle vibrate for calls in setting'' & 6 & System log & [SettingsProvider:V] \newline
    ``\symbol{94}(.*?)vibrate(.*?)when(.*?)ringing'' \\
    \midrule
    Settings & ``increase media volume in setting'' & 6 & System log & [SettingsProvider:V]
    ``\symbol{94}(.*?)MEDIA'' \\
    \midrule
    Settings & ``increase call volume in setting'' & 6 & System log & [SettingsProvider:V]
    ``\symbol{94}(.*?)CALL'' \\
    \midrule
    Settings & ``increase ring volume in setting'' & 6 & System log & [SettingsProvider:V]
    ``\symbol{94}(.*?)MUSIC'' \\
    \midrule
    Settings & ``increase alarm volume in setting'' & 6 & System log & [SettingsProvider:V]
    ``\symbol{94}(.*?)ALARM'' \\
    \midrule
    Settings & ``go to `add a language' page in setting'' & 7 & System log & [ActivityTaskManager:I] \newline
    ``\symbol{94}(.*?)LocalePicker'' \\  
    \toprule
    \end{tabular}
}
\caption{Comprehensive list of tasks.}
\label{tab:task_list_4}
\end{table}

\clearpage

\begin{table}[!ht]
\centering
\resizebox{\textwidth}{!}{
\small
    \begin{tabular}{>{\centering}p{0.1\textwidth}p{0.25\textwidth}>{\centering}p{0.05\textwidth}>{\centering}p{0.15\textwidth}p{0.45\textwidth}}
    \toprule
    Application & Task instruction  & \multilinecell{Step\\limit} & \multilinecell{Success criteria\\(source)} & Success criteria (detail) \\
    \midrule
    Snapseed & ``open Snapseed'' & 4 & UI elements & \{``id'': ``com.niksoftware.snapseed:id/logo\_view'', ``enabled'': ``true''\} \\
    \midrule
    Snapseed & ``open image in Snapseed'' & 6 & UI elements & \{``id'': ``com.niksoftware.snapseed:id/looks\_button'', ``selected'': ``true''\} \\
    \midrule
    Snapseed & ``set dark theme in Snapseed'' & 7 & App data (xml) & [/data/data/com.niksoftware.snapseed/
    \newline  shared\_prefs/Preferences.xml] \newline \{``pref\_appearance\_use\_dark\_theme''=``true''\} \\
    \midrule
    Snapseed & ``open image and apply portrait filter in Snapseed'' & 7 & UI elements & \{``-android uiautomator'': `` new UiSelector().text(``Portrait'')'',``selected'': ``true''\} \\
    \midrule
    Snapseed & ``set format quality to JPG 100\% in Snapseed'' & 9 & App data (xml) & [/data/data/com.niksoftware.snapseed/
    \newline shared\_prefs/Preferences.xml] \newline \{``pref\_export\_setting\_compression''=``100''\} \\
    \midrule
    Snapseed & ``set image sizing to 2000 px in Snapseed'' & 9 & App data (xml) & [/data/data/com.niksoftware.snapseed/
    \newline  shared\_prefs/Preferences.xml] \newline \{``pref\_export\_setting\_long\_edge''=``2000''\} \\
    \midrule
    Snapseed & ``open image and go to tools tab in Snapseed'' & 9 & UI elements & \{``id'': ``com.niksoftware.snapseed:id/tools\_button'', ``selected'': ``true''\} \\
    \midrule
    Snapseed & ``open image and apply noir S03 filter in Snapseed'' & 10 & UI elements & \{``-android uiautomator'': ``new UiSelector().text(``S03'')'', ``selected'': ``true''\} \\
    \midrule
    Snapseed & ``apply noir S03 filter to an image after setting dark theme in Snapseed'' & 13 & App data (xml) + UI elements  & \{“-android uiautomator”: “new UiSelec-
tor().text(“S03”)”, “selected”: “true”\}, [/data/data/com.niksoftware.snapseed/
shared\_prefs/Preferences.xml]
{“pref\_appearance\_use\_dark\_theme”=“true”} \\
    \midrule
    Snapseed & ``apply noir S03 filter to an image after setting format quality to JPG 100\% in Snapseed'' & 14 & App data (xml) + UI elements & \{“-android uiautomator”: “new UiSelec-
tor().text(“S03”)”, “selected”: “true”\}, [/data/data/com.niksoftware.snapseed/
shared\_prefs/Preferences.xml]
{“pref\_export\_setting\_compression”=“100”} \\
    \midrule
    Snapseed & ``apply noir S03 filter to an image after setting image sizing to 2000 px in Snapseed'' & 14 & App data (xml) + UI elements & \{“-android uiautomator”: “new UiSelec-
tor().text(“S03”)”, “selected”: “true”\}, [/data/data/com.niksoftware.snapseed/
shared\_prefs/Preferences.xml]
{“pref\_export\_setting\_long\_edge”=“2000”} \\
    \midrule
    Snapseed (*) & decrease screen brightness in setting and apply portrait filter to an image in Snapseed & 17 & UI elements + System setting & \{“-android uiautomator”: “new UiSelec-
tor().text(“Portrait”)”, “selected”: “true”\} \\
    \midrule
    Snapseed (*) & decrease screen brightness in setting and apply noir S03 filter to an image in Snapseed & 18 & UI elements + System setting & \{“-android uiautomator”: “new UiSelec-
tor().text(“S03”)”, “selected”: “true”\} \\
    \toprule
    \end{tabular}
}
\caption{Comprehensive list of tasks.}
\label{tab:task_list_5}
\end{table}

\clearpage

\begin{table}[!ht]
\centering
\resizebox{\textwidth}{!}{
\small
    \begin{tabular}{>{\centering}p{0.1\textwidth}p{0.25\textwidth}>{\centering}p{0.05\textwidth}>{\centering}p{0.15\textwidth}p{0.45\textwidth}}
    \toprule
    Application & Task instruction  & \multilinecell{Step\\limit} & \multilinecell{Success criteria\\(source)} & Success criteria (detail) \\
    \midrule
    Walmart & ``open Walmart'' & 4 & UI elements & \{``id'': ``com.walmart.android:id/navigation
    \newline \_shop'', ``enabled'': ``true''\} \\
    \midrule
    Walmart & ``go to account tab in Walmart'' & 5 & UI elements & \{``id'': ``com.walmart.android:id/navigation
    \newline \_account'', ``selected'': ``true''\} \\
    \midrule
    Walmart & ``go to my cart in Walmart'' & 5 & UI elements & \{``id'': 
    ``com.walmart.android:id/cart\_fragment
    \newline \_constraint\_layout'', ``selected'': ``true''\} \\
    \midrule
    Walmart & ``go to my items tab in Walmart'' & 5 & UI elements & \{``id'': ``com.walmart.android:id/navigation
    \newline \_my\_items'', ``selected'': ``true''\} \\
    \midrule
    Walmart & ``go to search tab in Walmart'' & 5 & UI elements & \{``id'': ``com.walmart.android:id/navigation
    \newline \_search'', ``selected'': ``true''\} \\
    \midrule
    Walmart & ``go to services tab in Walmart'' & 5 & UI elements & \{``id'': ``com.walmart.android:id/navigation
    \newline \_services'', ``selected'': ``true''\} \\
    \midrule
    Walmart & ``go to grocery category and show subcategories in Walmart'' & 7 & UI elements & \{``id'': ``com.walmart.android:id/category\_
    \newline container\_title'', ``text'': ``Grocery''\} \\
    \midrule
    Walmart & ``go to store map in Walmart'' & 7 & UI elements & \{``id'': ``com.walmart.android:id/instoremaps\_
    \newline webview\_container'', ``displayed'': ``true''\} \\
    \midrule
    Wikipedia & ``open Wikipedia'' & 4 & UI elements & \{``id'': ``org.wikipedia:id/nav\_tab\_explore'', ``selected'': ``true''\} \\
    \midrule
    Wikipedia & ``go to saved tab in Wikipedia'' & 5 & UI elements & \{``id'': ``org.wikipedia:id/nav\_tab\_reading\_lists'', \newline ``selected'': ``true''\} \\
    \midrule
    Wikipedia & ``go to search tab in Wikipedia'' & 5 & UI elements & \{``id'': ``org.wikipedia:id/nav\_tab\_search'', ``selected'': ``true''\} \\
    \midrule
    Wikipedia & ``disable the top 1 and 'randomizer' topics in the feed customization settings on Wikipedia and go back to the feed'' & 10 & App data (xml) + UI elements & 
    [/data/data/org.wikipedia/shared\_prefs/org.
    \newline wikipedia\_preferences.xml] \{``feedCardsEnabled'': 
    \newline ``[false,true,true,true,true,true,false,true,true,true]''\}, 
    \newline \{``id'': ``org.wikipedia:id/nav\_tab\_explore'', ``selected'': ``true''\} \\
    \midrule
    Wikipedia & ``disable the top 2 topics in the feed customization settings on Wikipedia and go back to the feed'' & 10 & App data (xml) + UI elements & 
    [/data/data/org.wikipedia/shared\_prefs/org.
    \newline wikipedia\_preferences.xml] \{``feedCardsEnabled'':
    \newline ``[false,false,true,true,true,true,true,true,true,true]''\}, 
    \newline \{``id'': ``org.wikipedia:id/nav\_tab\_explore'', ``selected'': ``true''\} \\
    \midrule
    Wikipedia & ``disable the top 2 and 'randomizer' topics in the feed customization settings on Wikipedia and go back to the feed'' & 11 & App data (xml) + UI elements & [/data/data/org.wikipedia/shared\_prefs/org.
    \newline wikipedia\_preferences.xml] \{``feedCardsEnabled'': 
    \newline ``[false,false,true,true,true,true,false,true,true,true]''\}, 
    \newline \{``id'': ``org.wikipedia:id/nav\_tab\_explore'', ``selected'': ``true''\} \\
    \midrule
    Wikipedia & ``disable the `show link previews', `top read' feed settings, and return to the feed on Wikipedia'' & 11 & App data (xml) + UI elements & [/data/data/org.wikipedia/shared\_prefs/org.
    \newline wikipedia\_preferences.xml] \{``feedCardsEnabled'': [false,true,true,true,true,true,true,true,true,true]\}  
    \newline \{``id'': ``org.wikipedia:id/nav\_tab\_explore'', ``selected'': ``true''\} \\
    \midrule
    Wikipedia & ``decrease the text size to 50\% in Wikipedia'' & 12 & App data (xml) & [/data/data/org.wikipedia/shared\_prefs/org.
    \newline wikipedia\_preferences.xml] \{"textSizeMultiplier": -5\} \\
    \midrule
    Wikipedia & ``disable the topics that are related to `history' in the feed customization settings on Wikipedia and go back to the feed'' & 12 & App data (xml) + UI elements & [/data/data/org.wikipedia/shared\_prefs/org.
    \newline wikipedia\_preferences.xml] \{``feedCardsEnabled'': [true,true,true,false,true,false,true,true,true,true]\} 
    \newline \{``id'': ``org.wikipedia:id/nav\_tab\_explore'', ``selected'': ``true''\}\\
    \midrule
    Wikipedia & ``disable the topics that include `day' in their names in the feed customization settings on Wikipedia and go back to the feed'' & 13 & App data (xml) + UI elements & [/data/data/org.wikipedia/shared\_prefs/org.
    \newline wikipedia\_preferences.xml] \{``feedCardsEnabled'': [true,true,false,true,true,false,true,true,true,true]\} 
    \newline \{``id'': ``org.wikipedia:id/nav\_tab\_explore'', ``selected'': ``true''\}\\
    \toprule
    \end{tabular}
}
\caption{Comprehensive list of \tasknum tasks.}
\label{tab:task_list_6}
\end{table}

\clearpage 

\begin{table}[!ht]
\centering
\resizebox{\textwidth}{!}{
\small
    \begin{tabular}{>{\centering}p{0.1\textwidth}p{0.25\textwidth}>{\centering}p{0.05\textwidth}>{\centering}p{0.15\textwidth}p{0.45\textwidth}}
    \toprule
    Application & Task instruction  & \multilinecell{Step\\limit} & \multilinecell{Success criteria\\(source)} & Success criteria (detail) \\
    \midrule
    Wikipedia & ``disable the topics with odd-numbered indices in the feed customization settings on Wikipedia and go back to the feed'' & 14 & App data (xml) + UI elements &
    [/data/data/org.wikipedia/shared\_prefs/org.
    \newline wikipedia\_preferences.xml] \{``feedCardsEnabled'': 
    \newline ``[false,true,false,true,false,true,false,true,true,true]''\}, 
    \newline \{``id'': ``org.wikipedia:id/nav\_tab\_explore'', ``selected'': ``true''\} \\
    \midrule
    Wikipedia & ``disable the topics with even-numbered indices in the feed customization settings on Wikipedia and go back to the feed'' & 14 & App data (xml) + UI elements & [/data/data/org.wikipedia/shared\_prefs/org.
    \newline wikipedia\_preferences.xml] \{``feedCardsEnabled'': [true,false,true,false,true,false,true,false,true,true]\} 
    \newline \{``id'': ``org.wikipedia:id/nav\_tab\_explore'', ``selected'': ``true''\}\\
    \midrule
    Wikipedia & ``disable the topics with prime-numbered indices in the feed customization settings on Wikipedia and go back to the feed'' & 14 & App data (xml) + UI elements & [/data/data/org.wikipedia/shared\_prefs/org.
    \newline wikipedia\_preferences.xml] \{``feedCardsEnabled'': [true,false,false,true,false,true,false,true,true,true]\} 
    \newline \{``id'': ``org.wikipedia:id/nav\_tab\_explore'', ``selected'': ``true''\}\\
    \midrule
    Wikipedia & ``increase the text size to 180\% in Wikipedia'' & 15 & App data (xml) & [/data/data/org.wikipedia/shared\_prefs/org.
    \newline wikipedia\_preferences.xml] \{"textSizeMultiplier": -5\} \\
    \midrule
    Wikipedia & ``disable featured article feed, decrease the text size to 50\%, and return to the feed on Wikipedia'' & 17 & App data (xml) + UI elements & [/data/data/org.wikipedia/shared\_prefs/org.
    \newline wikipedia\_preferences.xml] 
    \newline \{``textSizeMultiplier'': -5, ``feedCardsEnabled'': [false,true,true,true,true,true,true,true,true,true]\}  
    \newline \{``id'': ``org.wikipedia:id/nav\_tab\_explore'', ``selected'': ``true''\}\\
    \midrule
    Wikipedia & ``disable featured article feed, increase the text size to 180\%, and return to the feed on Wikipedia'' & 19 & App data (xml) + UI elements & [/data/data/org.wikipedia/shared\_prefs/org.
    \newline wikipedia\_preferences.xml] 
    \newline \{``textSizeMultiplier'': 8, ``feedCardsEnabled'': [false,true,true,true,true,true,true,true,true,true]\} 
    \newline \{``id'': ``org.wikipedia:id/nav\_tab\_explore'', ``selected'': ``true''\}\\
    \midrule
    Youtube & ``open Youtube'' & 4 & System log & [ActivityTaskManager:I] \newline
    ``\symbol{94}(.*?)START(.*?)com.google.android.youtube'' \\
    \toprule
    \end{tabular}
}
\caption{Comprehensive list of \tasknum tasks.}
\label{tab:task_list_7}
\end{table}

\clearpage

\subsection{Exemplary demonstrations on representative tasks} \label{app:task_demonstration_example}

\begin{figure}[ht!]
\centering
    \begin{minipage}{.95\textwidth}
        \centering
        \subfigure[{\tt Calculator}]{
            \includegraphics[width=\linewidth]{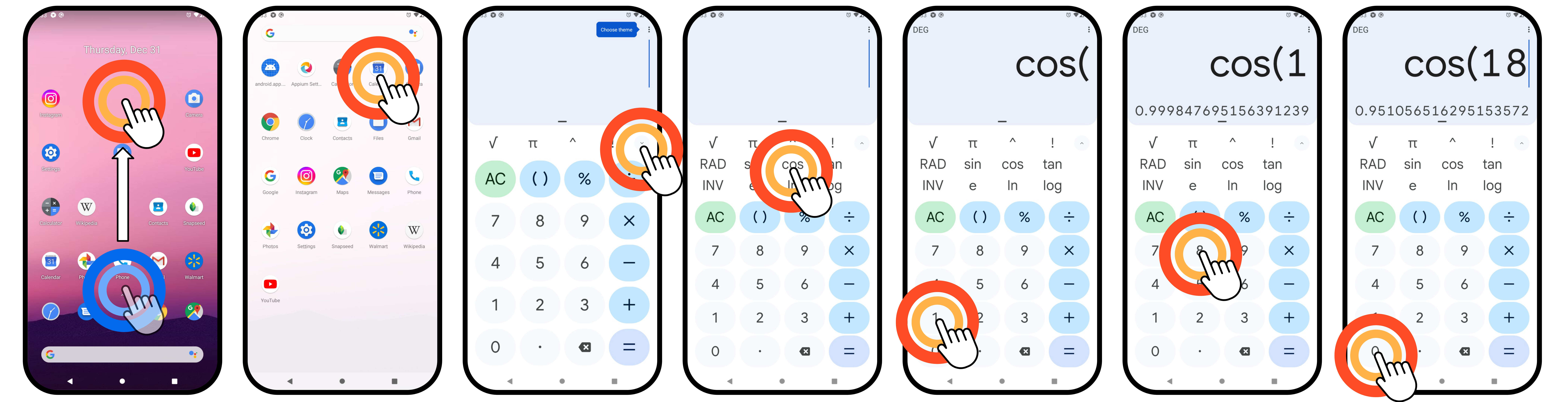}
        }
    \end{minipage}
    \begin{minipage}{.95\textwidth}
        \centering
        \includegraphics[width=\linewidth]{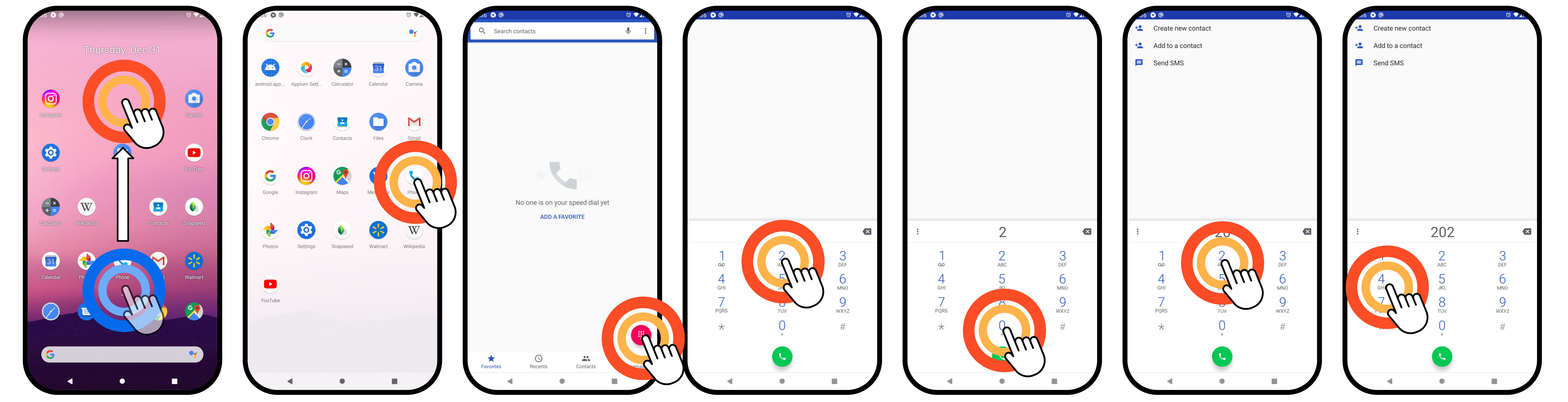}
        \subfigure[{\tt Call}]{
            \includegraphics[width=\linewidth]{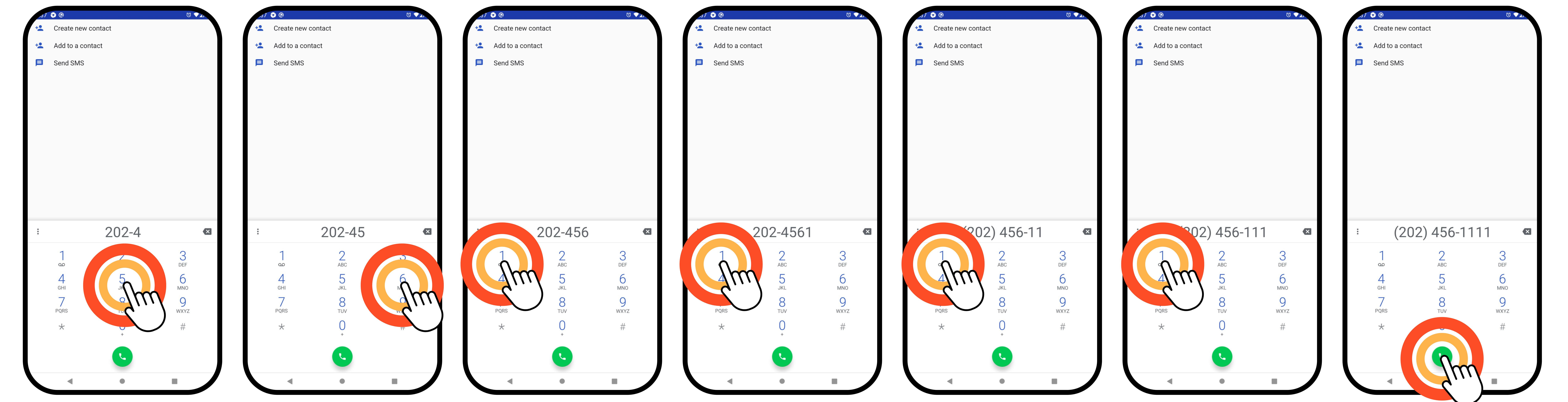}
        }
    \end{minipage}
    \begin{minipage}{.95\textwidth}
        \centering
        \includegraphics[width=\linewidth]{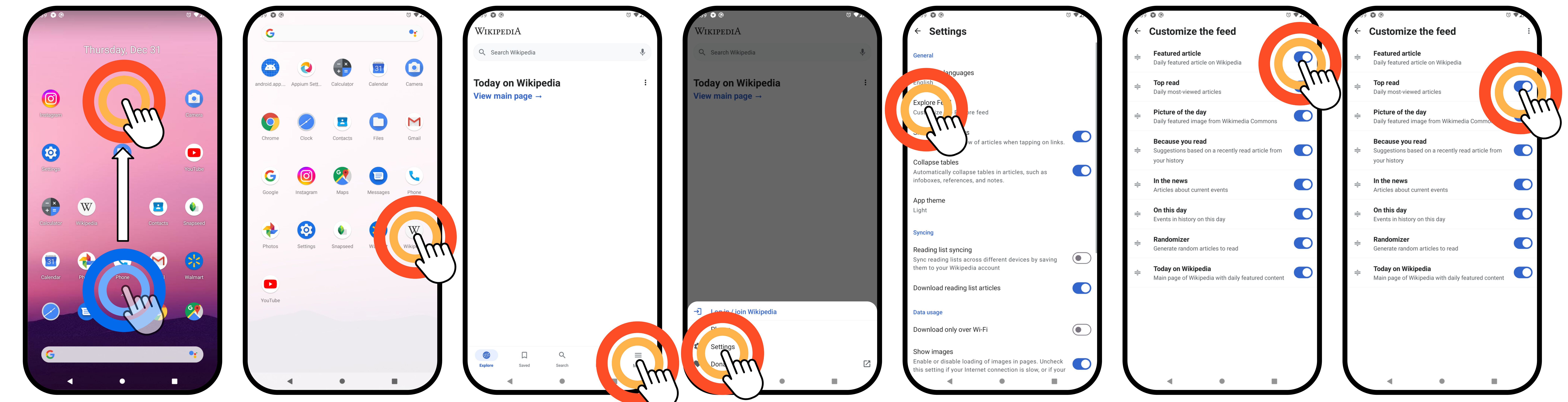}
        \subfigure[{\tt Wikipedia}]{
            \includegraphics[width=\linewidth]{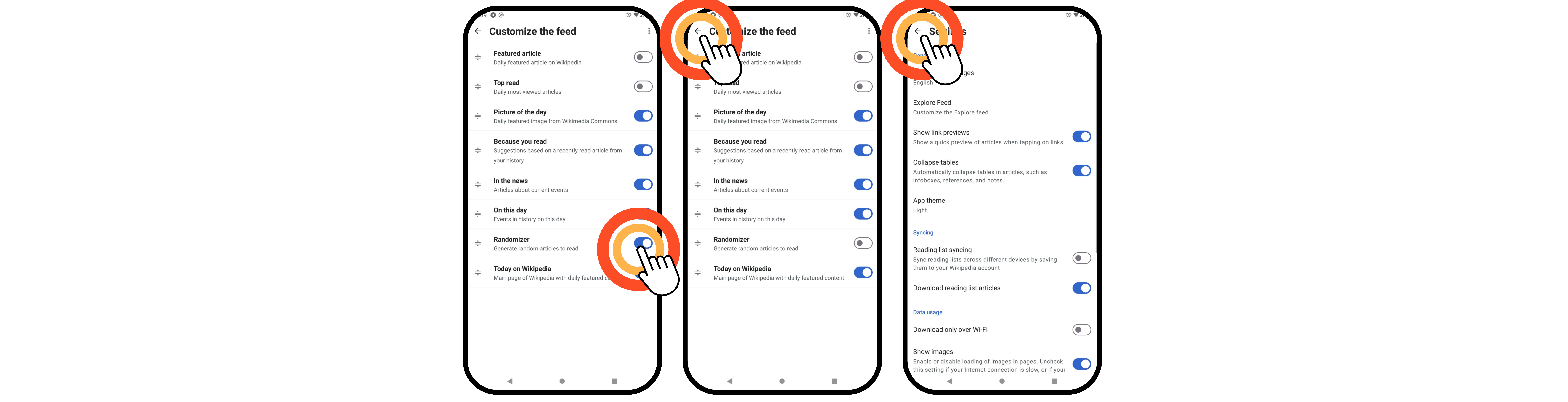}
        }
    \end{minipage}
    \caption{Examples of human expert demonstrations of six representative tasks. The blue and red cursors linked with a white arrow identify the \texttt{swiping} action, while the red cursor alone identifies the \texttt{tapping} action.}
    \label{fig:demonstration_example}
\end{figure}

In our experiments, we select six representative tasks. 
The tasks are selected to cover various functionalities, such as navigating pages (e.g., tab in the clock application or different setting pages in the setting application) and manipulating various UI elements (e.g., checkbox, dial pad, time pickers, etc.). 
On each task, we display the successful demonstration in \autoref{fig:demonstration_example}.

\begin{figure}[ht!]
\centering
    \begin{minipage}{.95\textwidth}
        \centering
        \subfigure[{\tt Alarm(Simple)}]{
            \includegraphics[width=\linewidth]{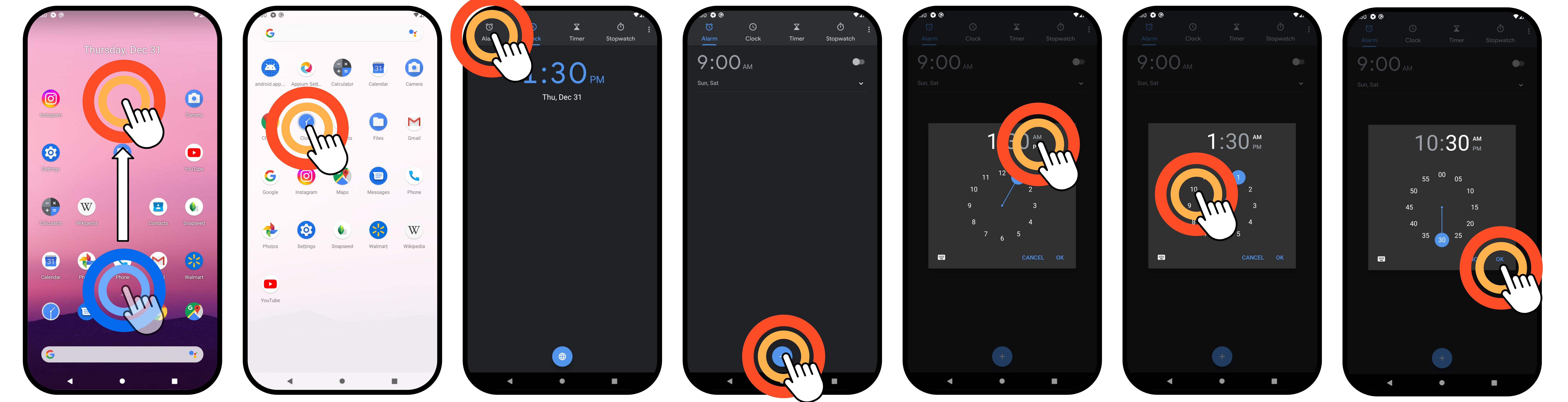}
        }
    \end{minipage}
    \begin{minipage}{.95\textwidth}
    \centering
    \includegraphics[width=\linewidth]{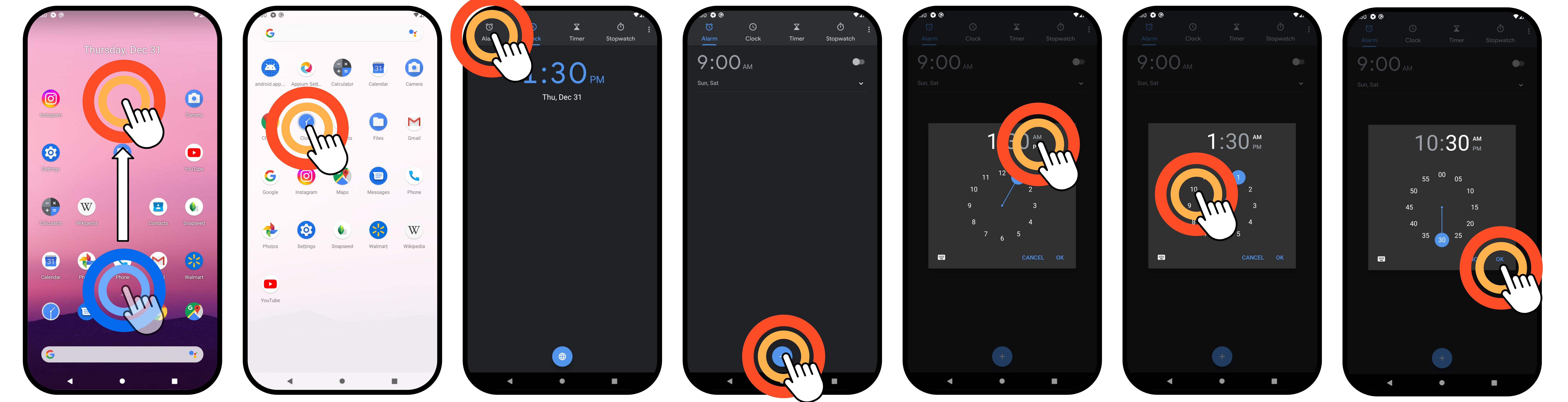}
    \subfigure[{\tt Alarm(Complex)}]{
        \includegraphics[width=\linewidth]{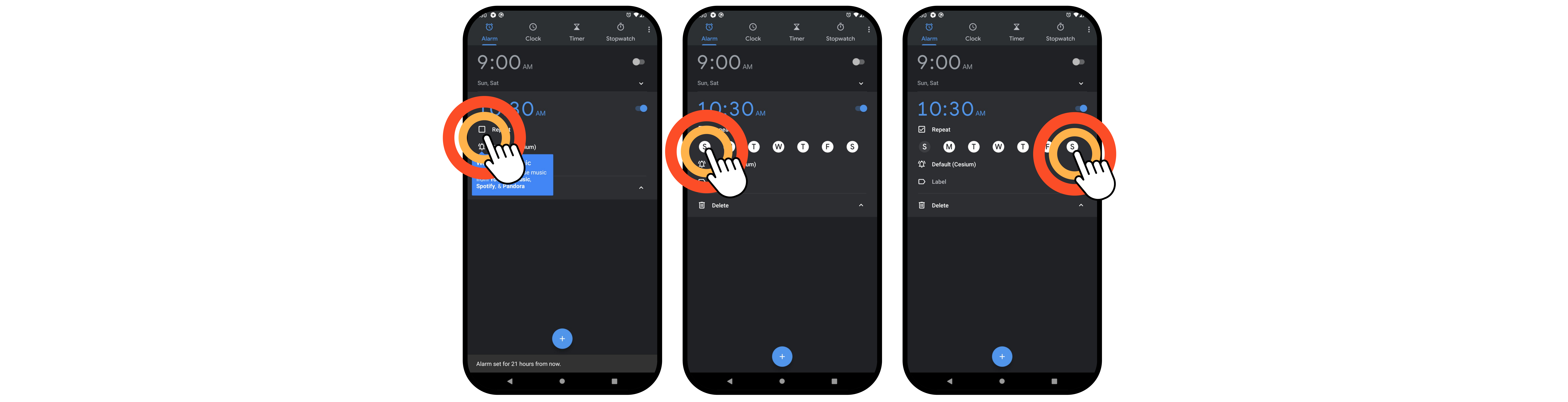}
    }
    \end{minipage}
    \begin{minipage}{.95\textwidth}
        \centering
        \subfigure[{\tt Language}]{
            \includegraphics[width=\linewidth]{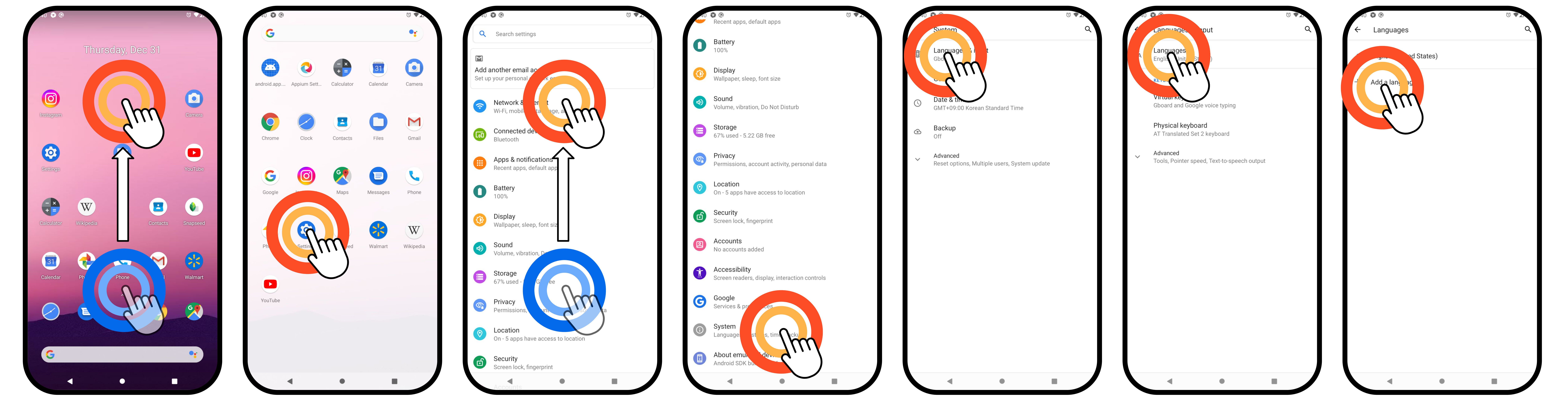}
        }
    \end{minipage}
\end{figure}

\clearpage

%% file: sections/appendix/continual_learning.tex
\clearpage
{\section{Guidelines of using \metabbr for continual learning}\label{app:continual_learning}

Among many topics related to continual learning, \metabbr has high relevancy on the topics below:
\begin{itemize}[leftmargin=6mm,itemsep=0mm]
    \item Open-world learning, open-ended learning
    \item Knowledge transfer (multi-task learning, domain adaptation)
    \item Continual learning paradigms (task incremental, domain incremental)
    \item Continual reinforcement learning (skill discovery)
    \item Continual learning in LLMs (in-context learning, fine-tuning)
\end{itemize}
Specifically, our environment simulates an open-ended environment by, for example, incorporating access to the web (inducing dynamicity in the tasks using Chrome, Wikipedia, Walmart, and Instagram), challenging the agent to face continuously changing contexts. 
This also allows \metabbr to be a useful testbed for open-ended learning.
Within the open digital world, we examine the adaptabilities of agents in environments with different configurations, including the transferability of knowledge of agents trained with multiple tasks.

Furthermore, a user can design various scenarios to test the agents associated with the topics above. These scenarios would include where agents encounter tasks with gradually increasing complexities, enabling the study of abilities in efficient in-context continual learning~\citep{momeni2024context} and skill extraction abilities~\citep{liu2024skillact}.
For example, we provide an example of using the subset of tasks in B-MoCA to challenge continual learning with a list of 6 different tasks with increasing difficulties (i.e., step limit):
\begin{itemize}[leftmargin=6mm,itemsep=0mm]
    \item 1. ``open the clock app'' (Step limit 4)
    \item 2. ``increase alarm volume in setting'' (Step limit 6)
    \item 3. ``create alarm at 10:30 am'' (Step limit 11)
    \item 4. ``create alarm at 10:30 am on every weekday'' (Step limit 13)
    \item 5. ``create alarm at 10:30 am in clock and increase alarm volume in setting'' (Step limit 18)
    \item 6. ``turn on airplane mode in setting and create alarm at 10:30 am in clock'' (Step limit 19)
\end{itemize}
To illustrate, we test Gemini-1.5-pro agents' in-context continual learning given this subset of tasks. In detail, we include the experiences of solving tasks 1, 2, and 3 in the prompt to test Gemini-1.5-pro agents on tasks 4, 5, and 6 (`With experience' in \autoref{tab:continual_learning}), and compare its performance to that of Gemini-1.5-pro agents without any experience included in the prompt (`Without experience'). As \autoref{tab:continual_learning} shows, we observe that the agent demonstrates meaningful knowledge transfer ability on tasks 4 and 5, but still struggles with the more complex scenario in task 6, indicating further room for improvement in its in-context continual learning capabilities. We believe this exemplary experiment suggests the potential of our benchmark to evaluate diverse interesting aspects of lifelong learning.

\begin{table}[h]
\centering
\caption{Performance comparison with and without experience. We use the human expert demonstration collected from a training environment with ID 000 and test the agent at a test environment with ID 100 across three different seeds.}
\scalebox{0.8}{
\begin{tabular}{p{11cm}cc}
\toprule
\textbf{Task} & \textbf{With experience} & \textbf{Without experience} \\
\midrule
4. ``Create alarm at 10:30 am on every weekday'' & 0.66 & 0.00 \\
5. ``Create alarm at 10:30 am in clock and increase alarm volume in setting'' & 1.00 & 0.00 \\
6. ``Turn on airplane mode in setting and create alarm at 10:30 am in clock'' & 0.33 & 0.00 \\
\bottomrule
\end{tabular}
}
\label{tab:continual_learning}
\end{table}

Hoping for \metabbr to be used for continual learning in more diverse ways, we present some example subsets from other applications, as illustrated below:
\begin{itemize}[leftmargin=6mm,itemsep=0mm]
    \item Calculator: ``open Calculator'' - ``input 1 in Calculator'' - ``input `17x23' in Calculator'' - ``compute the harmonic mean of 4 and 5 in Calculator''
    \item Phone: ``open the phone app'' - ``call 911'' - ``call 311311'' - ``call the white house (202-456-1111)''
    \item Settings: ``open the setting app'' - ``turn on airplane mode'' - ``increase media volume in setting''
    \item Wikipedia: ``open Wikipedia'' - ``go to search tap in Wikipedia'' - ``disable the top 1 and `randomizer' topics in the feed customization settings on Wikipeida and go back to the feed'' - ``disable the topics that include `day' in their names in the feed customization settings on Wikipedia and go back to the feed'' - ``disable featured article feed, increase the text size to 180\%, and return to the feed on Wikipedia''
\end{itemize}

%% file: sections/appendix/agent_details.tex
\clearpage
\section{Agent details}

\subsection{Overview of agents employing foundation models}\label{app:agent_llm_overview}

\begin{figure}[!h]
  \centering
  \includegraphics[width=0.5\textwidth]{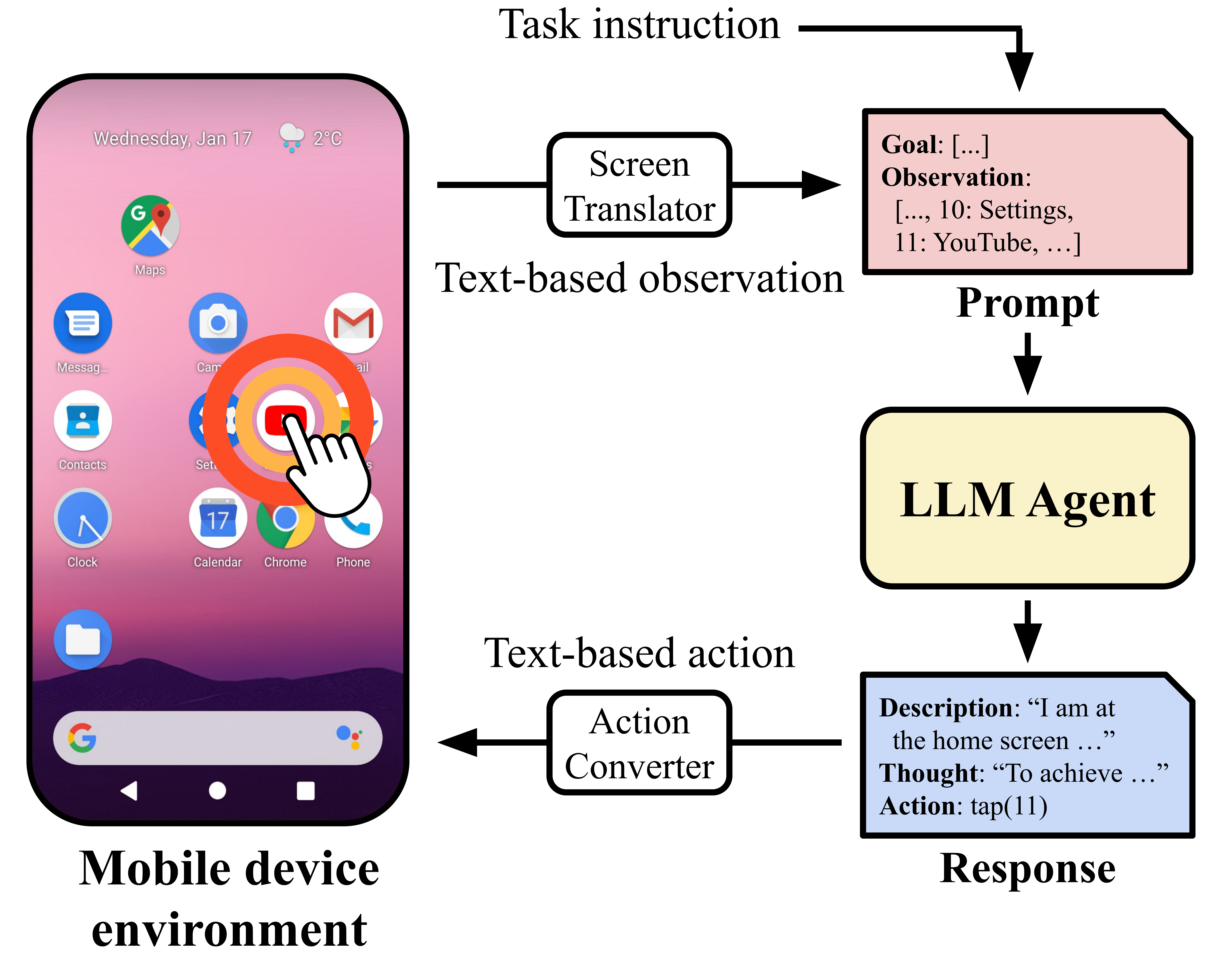} 
  \caption{Illustration of the LLM agent controlling mobile devices.
  LLM agents interact with the environments through an additional screen translator and action converter, to obtain text-based observations and control with text-based actions.}
  \label{fig:llm_overview}
\end{figure}

We provide an overall workflow (illustrated in \autoref{fig:llm_overview}) of agents employing foundation models, especially with LLM.
To employ LLM to develop the agents, we build a prompt based on task instruction and text-based observation.
The text-based observation describes the screen layout information. 
We provide the prompt to the LLM and obtain the response.
The response includes the text-based action chosen from a set of action options (which is provided in the prompt) by the LLM.
The action converter extracts the legal action from the text-based action.
The environment which is realized as an Android emulator, then, processes the provided action.

\subsection{Prompt details for agents with foundation models}\label{app:agent_llm_prompt}

For the agents employing LLMs or MLLMs, we use prompts composed of general instruction and task-specific information.
A specific example of the prompt format used in our experiment is described in \autoref{tab:llm_general_prompt}.

General instruction includes a description of the agent's role, an explanation of the response format, and a definition of action space.
The role description informs the agents about the problem, i.e., the device control. 
The output format is designed to integrate the Chain-of-Thought (CoT) technique~\citep{wei2022chain}. 
The possible actions are provided as callable functions: options of tapping an element in the list, swiping the screen, and pressing the three navigation buttons over the screen with action press(``BACK''), action press(``HOME''), and action press(``OVERVIEW'').

Task-specific information includes instruction on the goal, observation, history of actions agents have taken, and (optionally) few-shot examples of the given task.
For the text-based observations, we provide a numeric tag and detailed information for each UI element, including resource ID information, class information, content description information, text information, and additional information such as checked information or bounding box location information. 
The resource ID and class information capture the coarse-grained information on the UI element, such as the type of UI element.
The content description and text information serve as fine-grained information on the UI element, such as the specific name of the application.
The checked information and bounding box location information supply additional information on the UI element.
While we observe that the agents using GPT-4o, Gemini-1.5-pro, and Llama-3-70B can accurately generate a description of the screen layout, rationale on the current situation and next step, and decide an action option based on the selected set of attributes, we allow the users to easily modify the set of attributes constituting the text-based observations for further investigation.

For few-shot examples, we include a pre-defined number of examples hinting at the correct actions to take.
Specifically, to build a prompt with few-shot examples, the \textcolor{red}{\lbrack few shot prompt\rbrack} part in \autoref{tab:llm_general_prompt} is replaced with the text illustrating the human demonstration.
\autoref{tab:llm_few_shot_prompt} shows an exemplary few show prompts, with one trajectory of the human expert demonstration.
For agents employing GPT-4o or Gemini-1.5-pro, we provide a whole trajectory performed by the human expert on solving each task as a few-shot example by sampling from the set of trajectories performed in various training environments.

For agents employing Llama-3, we note that we use different versions of prompts in our experiments.
These different versions slightly vary in expressions for the instructions, while the overall structure and intent remain consistent across all of them.
For LLM agents employing Llama-3-70B, for example, we avoid terming each few-shot example as a full trajectory, because we sample several observation-action pairs from the trajectory due to the limited context length.
Also, for the \custom agents employing fine-tuned Llama-3-8B, we use a different version of prompt format, which requires them not to generate descriptions or thoughts in their responses, because our training dataset does not contain thought processes.
Please refer to the supplementary codes for the exact prompt format on these baselines.

\subsection{Architecture design for \custom agents trained with VLM encoder}\label{app:agent_bc_architecture}

The network architecture for \custom agents is composed of three components: encoder, attention module, and action head. 
Given the task instruction $c$ and the screen image $o_t \in \mathbb{R}^{3 \times 256 \times 128}$ at each timestep $t$, \custom agents generate a screen-touching action $a_t \in \mathbb{R}^{385}$.
Among the screen-touching action $a_t$, the first 378 values correspond to tapping pre-defined locations (14×27 bins) on the screen, four values to swiping directions (up, down, right, left), and the last three values to pressing buttons (back, home, overview). 

These \custom agents use visual and text encoders to represent screen images $o_t$ and task instruction $c$, respectively.
The visual encoder embeds visual feature $e_{o_{t}} \in \mathbb{R}^{d}$ from the observation $o_t$, and the text encoder extracts features $e_{c} \in \mathbb{R}^{d}$ from the task instruction $c$. 
For the visual encoder, we use EfficientNet-b0~\citep{tan2020efficientnet} pre-trained with ImageNet followed by an adaption layer using a fully connected layer to adapt the output channel to hidden dimension $d$~\citep{liu2023llava}. 
For the text encoder, we use a pre-trained text encoder of Auto-UI~\citep{zhan2023you} which is trained with Android-in-the-wild~\citep{rawles2023android}, which is a dataset composed of demonstrations for solving Android device control problems.
The text encoder is kept frozen during the training process.
The hidden dimension $d$ is set to value of $768$ for both visual embedding and text embedding.

The attention module, then, fuses the visual feature $e_{o_{t}}$ and text feature $e_{c}$ into a single vision-language embedding $e_{\text{fused}} \in \mathbb{R}^{d}$. 
Especially, we use multi-head attention layer~\citep{vaswani2017attention} for cross-attention, with $e_{c}$ given as query and $e_{o_{t}}$ given as key and value.
Given the fused feature $e_{\text{fused}}$, the action heads predict the action $a_t$, by applying the three fully connected (FC) layers with the size of (1024, 1024, 385), respectively.

\clearpage 

\begin{table}[ht!]
\centering
\begin{center}
\resizebox{0.85\textwidth}{!}{%
\begin{tabular}[h!]{@{}l@{}}
\footnotesize
\tcbox[colback=white,boxrule=1pt,arc=3mm]{
\begin{tblr}{
colspec = {@{}X@{}},
rowsep=1pt,
}
You are an agent that is trained to perform daily tasks on digital devices, such as smartphones. \\
You are given a goal task instruction to accomplish, an observation from the environment, and previous actions you have taken (up to 4 past steps). \\
The observation is a screen description parsed from the Android view hierarchy which contains the numeric tag and relevant information (e.g., text description) of each UI element. \\
\\[-1.8em]
For the response, you need to think and call the function needed to achieve the goal task instruction. \\
Your output should include three parts in the given format: \\
- Description: <Describe what you observe in the input>\\[-0.3em]
- Thought: <Provide a rationale on the next action you should take to complete the task>\\[-0.3em]
- Action: <Select an action option in the format of a function call with the correct parameters. You cannot output anything else except a function call.>\\
\\[-1.5em]
For the action, you need to select an action option by calling one of the following functions to control the digital device:\\
$\;$1. dual-gesture(touch y: float, touch x: float, lift y: float, lift x: float): This function is used to operate a dual-gesture action. A dual-gesture comprises four floating-point numeric values between 0 and 1, indicating a normalized location of the screen in each of the x-y coordinates. A dual-gesture action is interpreted as touching the screen at the location of (touch y, touch x) and lifting at the location of (lift y, lift x). The dual-gesture action indicates a tapping action if the touch and lift locations are identical, but a swiping action if they differ. A simple use case is dual-gesture(0.5, 0.5, 0.5, 0.5) to tap the center of the screen.\\
$\;$2. tap(numeric tag: int): This function is used to tap a UI element shown on the digital device screen. "numeric tag" is a tag assigned to a UI element shown on the digital device screen. A simple use case can be tap(5), which taps the UI element labeled with the number 5.\\
$\;$3. swipe(direction: str): This function is used to swipe on the digital device screen. "direction" is a string that represents one of the four directions: up, down, left, right. "direction" must be wrapped in double quotation marks. A simple use case is swipe("up"), which can be used to open the app list on the home screen.\\
$\;$4. press("HOME"): This function is used to press the home button.\\
$\;$5. press("BACK"): This function is used to press the back button.\\
$\;$6. press("OVERVIEW"): This function is used to press the overview button.\\
You can only take one action at a time, so please directly call the function. \\
Please never take action besides the options provided. \\
\\[-1.8em]
Goal: \textcolor{blue}{\lbrack task instruction\rbrack}.\\
\\[-1.8em]
\textcolor{red}{$\{$few shot prompt$\}$} \\
\\[-1.8em]
Previous actions: \textcolor{blue}{\lbrack previous actions\rbrack}\\[-0.3em]
Current observation: \textcolor{blue}{\lbrack current observation\rbrack}\\[-0.3em]
Answer:
\end{tblr}}
\end{tabular}}%
\caption{Prompts used for LLM agents and MLLM agents. 
Parts for \textcolor{blue}{\lbrack...\rbrack} are filled for different environments (including the tasks).
Parts for \textcolor{red}{$\{$...$\}$} are filled in according to different experiments, as few-shot examples are optional.}
\label{tab:llm_general_prompt}
\end{center}
\end{table}

\begin{table}[ht!]
\centering
\begin{center}
\resizebox{0.85\textwidth}{!}{%
\begin{tabular}[h!]{@{}l@{}}
\footnotesize
\tcbox[colback=white,boxrule=1pt,arc=3mm]{
    \begin{tblr}{
      colspec = {@{}X@{}},
      rowsep=1pt,
    }
    Below is an example of a human expert. \\
    Each example is a full trajectory from the beginning to the end of the task completion. \\
    Each observation from the environment and corresponding action taken by the expert is described as:\\
    - Observation: <An observation from the environment>\\[-0.3em]
    - Action: <An action taken by the human expert>\\[-0.3em]
    \\[-1.7em]
    Demonstration Example:\\
    - Observation: $[$\{`numeric\_tag': 0, `resource-id': `', `class': `View', `description': `Appslist', ..., `checked': False\},
    \textcolor{blue}{\lbrack...\rbrack},
    \{`numeric\_tag': 38, `resource-id': `', `class': `FrameLayout', `description': `', ..., `checked':False\}$]$ \\[-0.3em]
    - Action: swipe("up") \\
    \textcolor{red}{$\{$...$\}$}\\
    - Observation: $[$\{`numeric\_tag': 0, `resource-id': `', `class': `FrameLayout' ‘description’: ‘’, ..., `checked': False\},
    \textcolor{blue}{\lbrack...\rbrack},
    \{`numeric\_tag': 83, `resource-id': `', `class': `LinearLayout', ‘description’: ‘’, ..., `checked': False\}$]$ \\[-0.3em]
    - Action: tap(74) \\
    \end{tblr}}
\end{tabular}}%
\caption{An exemplary few-shot prompt with one trajectory of human expert demonstration. 
For the attributes of UI elements in the observation, the value is set to be `' if it is unavailable (e.g., `resource-id' of the first element in the first observation).
Parts for \textcolor{blue}{\lbrack...\rbrack} are filled with a list of descriptions for UI elements. 
Parts for \textcolor{red}{$\{$...$\}$} are filled with a list of intermediate steps in the expert demonstration.}
\label{tab:llm_few_shot_prompt}
\end{center}
\end{table}

\clearpage

%% file: sections/appendix/experiment_setup_details.tex
\clearpage
\section{Experiment setup details}

\subsection{Demonstration details}\label{app:exp_dataset_collection}

\paragraph{Demonstration collection procedure}
We collect human expert demonstrations for the few-shot examples provided to the LLM and MLLM agents as well as training of \custom agents.
The demonstrations are collected by human experts (the authors).
Before collecting the expert demonstrations, the demonstrators are allowed to be accustomed to the environment interfaces by letting themselves interact with the environments (i.e., Android emulators).
In this period, the demonstrators as asked to manipulate the emulators without specifying any certain task or to solve several random daily tasks.
Then, the demonstrators are instructed to perform demonstrations on the target representative tasks.

In detail, the collectors are asked to complete the six representative tasks in each \emph{training} environment.
The instruction for the target tasks is the same as the task instruction provided to the agents.
The demonstrators try to perform the task as optimally as possible and consistently along different device configurations as much as possible.
The definitions of action space for the collected demonstration are in two modes: the action space defined with action options (for agents using foundation models and \custom agent using fine-tuned Llama-3) and the action space defined in Section~\ref{sec:custom_agent} (for \custom agents using VLM encoder).
For few-shot examples of agents employing foundation models, demonstrators use discrete actions, except when the task is impossible to solve without using continuous actions (i.e., dual-gesture actions).
The end of each episode is determined by the success detector we implement.
We exclude the demonstrations with failures.

\paragraph{Training environments employed for experiments}
We explain the details of the environments used for demonstration collection.
For few-shot examples used for LLM agents and MLLM agents in experiments in Section~\ref{sec:exp_main}, we exploit training environments with ID numbers from 000 to 034.
Hence, a total number of 210 trajectories of demonstrations are prepared.
For training \custom agents in Section~\ref{sec:exp_main}, the training environments with ID numbers from 000 to 034 are used.
Hence, a total number of 210 trajectories of demonstrations are employed.
Each triplet (task instruction, observation, action) in the trajectories is used as a data point for composing the training batch.
For \custom agents using fine-tuned Llama-3, we use text-based observation (without screen layout image) and text-based actions for the training.
Also, we do not provide few-shot examples for these agents.
For \custom agents using VLM encoder, the observation is screen layout image and the action is either screen-touching action, swiping, or pressing the button.

For the experiments on the effect of data diversity on \custom agents using VLM encoder (Section~\ref{sec:exp_analysis}), we leverage varying numbers of training environments 7 and 35 where the corresponding ID numbers of the environments are from 000 to 006 and from 000 to 034, respectively.
Hence, a total number of 42 and 210 trajectories of demonstrations are exploited, respectively.

\subsection{Inference configuration details for agents with foundation models}\label{app:exp_llm_configuration}

For the experiments on agents with foundation models, we set the configurations for the foundation models.
We set the temperature to 0.0 and top-p to 1.0 for GPT-4o. 
We set the temperature to 0.0 and top-p to 1.0 for Gemini-1.5-Pro.
We set the temperature to 0.1 and top-p to 1.0 for Llama-3-70B.
The maximum output tokens for all models are set to 256.
The values of the parameters unspecified are set to default.
We experiment with three different runs.
We note that we fix seeds for consistent output generation across the runs when employing the closed-source LLMs, while they do not provide deterministic generation by default.

We explain details on the few-shot examples leveraged for the LLM agents and MLLM agents.
For LLM and MLLM agents using closed-source foundation models (i.e., GPT-4o and Gemini-1.5-pro), a whole trajectory by the human expert on solving a target task is sampled as a few-shot example.
We exploit three few-shot examples for each step for these agents.
For LLM agents employing open-source LLM (i.e., Llama-3), a pair of observation and action is provided randomly sampled from the exemplary trajectories as few-shot examples, similar to several prior methods~\citep{zhang2023mobile,rawles2023android}, owing to the limited context length.
We employ three few-shot examples on the {\tt Alarm(simple)}, {\tt Alarm(complex)}, and {\tt Language} tasks.
We use one few-shot example on the {\tt Calculator}, {\tt Call}, and {\tt Wikipedia} tasks.

\subsection{Training configuration details for \custom agents}\label{app:exp_custom_training_detail}

\paragraph{Custom agents using fine-tuned LLM}
For the experiments on \custom agents using fine-tuned Llama-3, we fine-tune the pre-trained \texttt{Llama-3-8B-Instruct} model for 15 epochs, using a batch size of 8 with gradients accumulation over 8 steps, sampled from a collection of 210 human demonstrations. 
We use the AdamW optimizer~\citep{loshchilov2017decoupled} with a learning rate of 1e-6 with a cosine annealing scheduler (setting a warmup ratio of 0.01 and a weight decay of 1.0).
In particular, we adopt the LoRA~\citep{hu2021lora} technique, setting a rank value to 8 and the alpha to 16. 
We train with the setting of completion only, focusing on the action prediction part only.
Each training is conducted on 8 NVIDIA RTX 3090 GPUs and takes approximately three hours.

\paragraph{Custom agents with VLM encoder}
For the experiments on \custom agents using VLM encoder, we train each multi-task policy for 10K steps with a batch size of 32, sampled from a collection of 210 human demonstrations. 
We use the Adam optimizer~\citep{kingma2017adam} with a learning rate of 3e-4 and adopt a cosine annealing learning rate schedulers
Each training is conducted on a single NVIDIA RTX A6000 GPU and takes approximately one hour. 

%% file: sections/appendix/experiment_result_details.tex
\clearpage
\section{Experiment result details}

\subsection{Analysis of failure cases of LLM agents}\label{app:exp_failure_cases}

We provide several failure cases with a display of the corresponding representative trajectories.
We show six cases demonstrating various aspects of the failing behaviors of the agents.

\begin{figure}[ht!]
\centering
    \includegraphics[height=3.4cm]{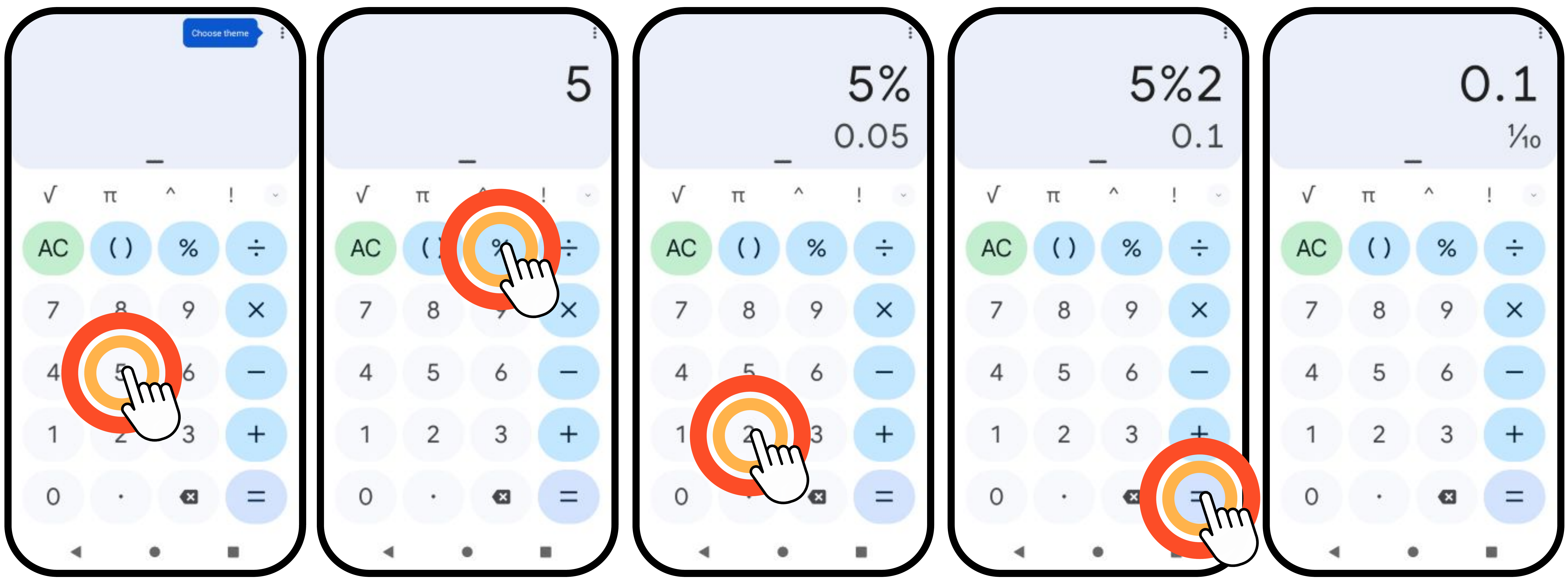}
    \caption{An exemplary trajectory from LLM agent employing GPT-4o in a task with instruction ``compute 50\% of 28 ('50\%28') in Calculator". The agent fails to tap UI elements appropriately in a sequence.}
    \label{fig:failure_calculator}
\end{figure}

The first case is with a task requiring the agents to input a mathematical equation.
The task instruction is to ``compute 50\% of 28 (`50\%28') in Calculator".
As shown in \autoref{fig:failure_calculator}, the LLM agent employing GPT-4o fails to manipulate appropriate UI elements in a sequence.
In the agent's detailed response, shown in \autoref{tab:failure_calculator}, we observe that the agent fails to build an appropriate plan.
We believe that developing an agentic workflow to check the validity of the generated plan iteratively before making the final decision~\citep{madaan2024self} in UI element manipulation can improve task performance.

\begin{table}[!ht]
\begin{center}
\resizebox{0.8\textwidth}{!}{
\begin{tabular}[h]{@{}l@{}}
\fontsize{9}{11}\selectfont 
\centering
\tcbox[colback=white,boxrule=1pt,arc=3mm]{
    \begin{tblr}{colspec = {@{}X@{}}, rowsep=1pt,}
    - Description: The current screen shows the calculator app with the number 5 already entered. There are various buttons available, including digits, operations, and special functions.\\
    - Thought: To compute 50\% of 28, I need to enter the percentage symbol, then the number 2, and finally the number 8. After that, I will press the equals button to get the result.\\
    - Action: tap(18)
    \end{tblr}}
\end{tabular}}
\caption{An exemplary reponse from LLM agent employing GPT-4o in a task with instruction ``compute 50\% of 28 ('50\%28') in Calculator". An inappropriate plan is observable.
}
\label{tab:failure_calculator}
\end{center}
\end{table}

The second case is with a task requiring the agents to perform tasks in two different applications.
The instruction is ``turn on airplane mode in setting and create alarm at 10:30 am in clock".
After enabling airplane mode, the agent mistakenly opens the Gmail application instead of the clock application.
Then, it comes back to the home screen and repeats to enter the settings application to enable airplane mode.
We expect that this is mainly due to a limited understanding of the historical context.
We believe that developing agents with longer historical contexts can mitigate this issue, while developing a more efficient method is desirable for reducing the length of the input.

\begin{figure}[ht!]
\centering
    \includegraphics[height=3.4cm]{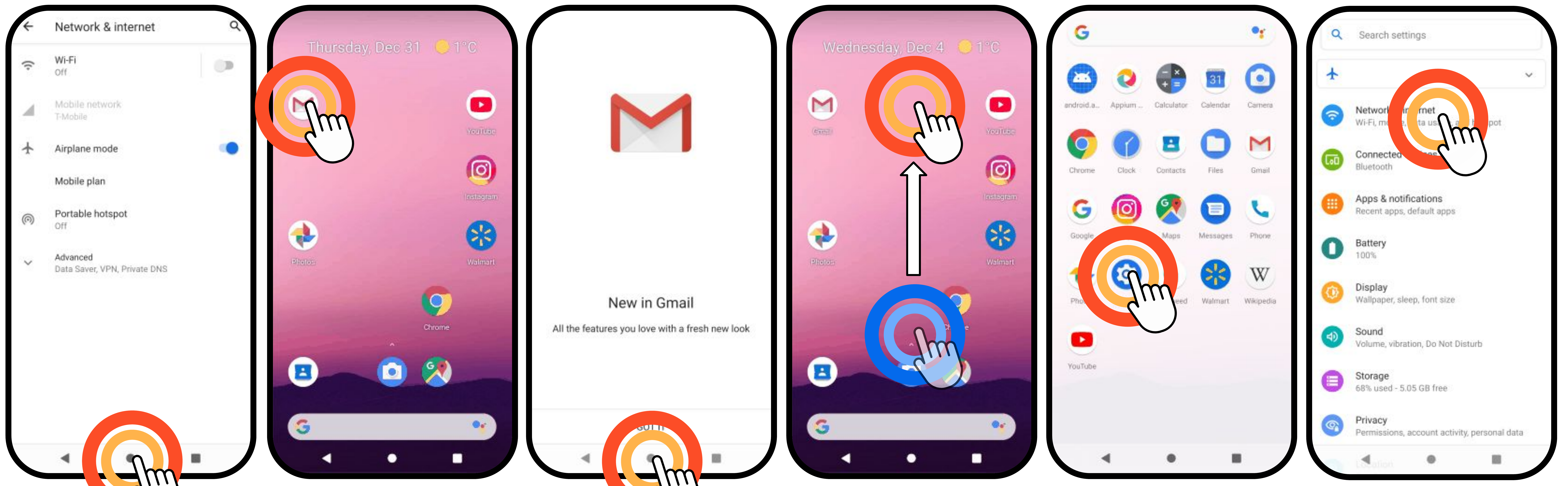}
    \caption{An exemplary trajectory from LLM agent employing GPT-4o in a task with instruction ``turn on airplane mode in setting and create alarm at 10:30 am in clock". The agent forgets that it has already enabled the airplane mode.}
    \label{fig:failure_cross}
\end{figure}

\begin{figure}[ht!]
\centering
    \includegraphics[height=3.4cm]{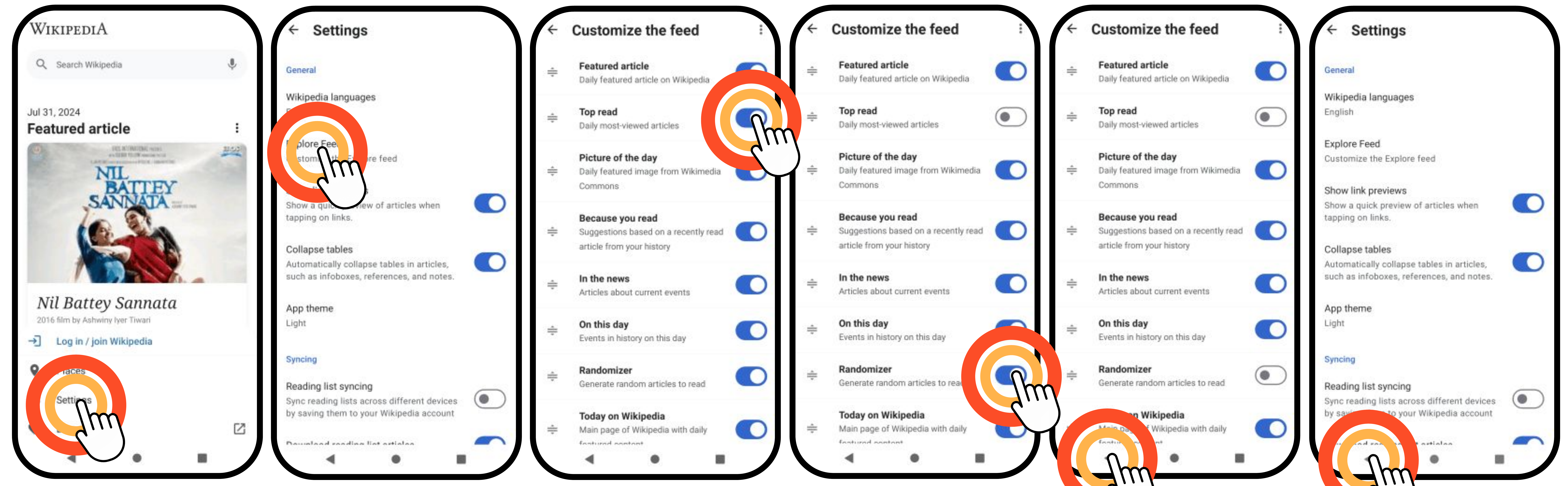}
    \caption{An exemplary trajectory from LLM agent employing GPT-4o in a task with instruction ``disable the top 2 and 'randomizer' topics in the feed customization settings on Wikipedia and go back to the feed". The agent fails to disable the first topic.}
    \label{fig:failure_wikipedia}
\end{figure}

\clearpage
We show another failure case where the agent fails by generating an inaccurate plan.
The task instruction is ``disable the top 2 and `randomizer' topics in the feed customization settings on Wikipedia and go back to the feed".
As shown in \autoref{fig:failure_wikipedia}, the GPT-4o agent often fails to disable the first top topic, even with few-shot examples composed of trajectories of the human expert disabling three topics.
In the response, displayed in \autoref{tab:failure_wikipedia}, the agent generates a plan to disable the second and `randomizer' topics and return to the feed, disregarding the first topic.
We further examine the understanding of the agent on the task in a relevant question, shown in \autoref{tab:question_wikipedia}.

\begin{table}[!ht]
\begin{center}
\resizebox{0.9\textwidth}{!}{
\begin{tabular}[h]{@{}l@{}}
\fontsize{9}{11}\selectfont 
\centering
\tcbox[colback=white,boxrule=1pt,arc=3mm]{
    \begin{tblr}{colspec = {@{}X@{}}, rowsep=1pt,}
    - Description: I see the ``customize\_the\_feed" screen with various feed content types listed. The ``top\_read" and ``randomizer" switches are now off, indicating they have been disabled.\\
    - Thought: I have successfully disabled the ``top\_read" and ``randomizer" topics. Now, I need to navigate back to the feed.\\
    - Action: press("BACK")
    \end{tblr}}
\end{tabular}}
    \caption{An exemplary trajectory from LLM agent employing GPT-4o in a task with instruction ``disable the top 2 and 'randomizer' topics in the feed customization settings on Wikipedia and go back to the feed". The agent fails to generate an accurate plan.}
\label{tab:failure_wikipedia}
\end{center}
\end{table}

\begin{table}[!ht]
\begin{center}
\resizebox{0.9\textwidth}{!}{
\begin{tabular}[h]{@{}l@{}}
\fontsize{9}{11}\selectfont 
\centering
\tcbox[colback=white,boxrule=1pt,arc=3mm]{
    \begin{tblr}{colspec = {@{}X@{}}, rowsep=1pt,}
    Question:\\
    How many topics should be disabled, given the task instruction ``disable the top 2 and `randomizer' topics in the feed customization settings on Wikipedia and go back to the feed"? List each topic.\\
    \\[-1.8em]
    Answer: \\
Based on the task instruction, you need to disable a total of 3 topics in the feed customization settings on Wikipedia. These topics are:\\
1. The top 1 topic\\
2. The top 2 topic\\
3. The `randomizer' topic
    \end{tblr}}
\end{tabular}}
\caption{A question related to the task and a response from the GPT-4o model (not agent). The agent answers that three topics should be disabled. We also find that a similar answer (of disabling three topics) is observable by asking questions differently.
}
\label{tab:question_wikipedia}
\end{center}
\end{table}

\begin{figure}[ht!]
\centering
    \includegraphics[height=3.4cm]{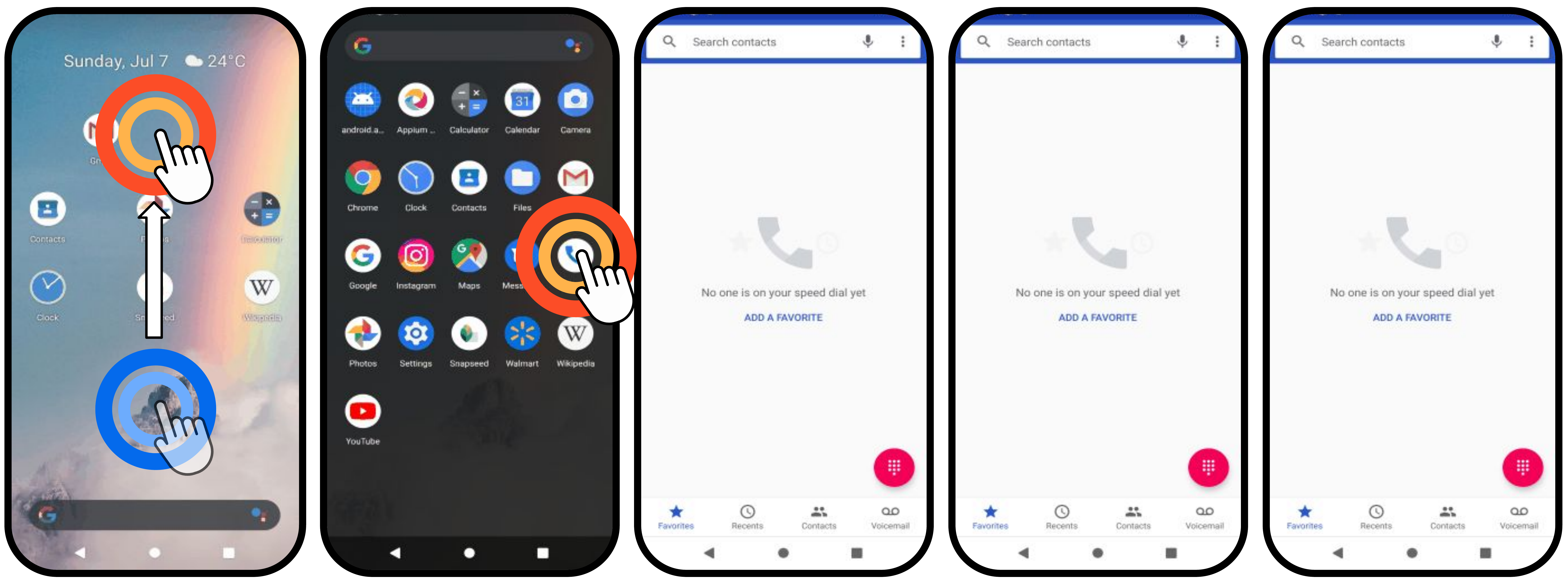}
    \caption{An exemplary trajectory from LLM agent employing GPT-4o in a task with instruction ``call the white house (202-456-1111)". The agent repeats an action that does not induce any change on the device, by tapping a non-existing UI element provided in the few-shot example.}
    \label{fig:failure_phone}
\end{figure}

\clearpage
The failure in another case stems from the hallucination of LLM agents.
The task instruction is to ``call the white house (202-456-1111)".
In this example, the LLM agent employing GPT-4o hallucinates during the task completion, as detailed in \autoref{fig:failure_phone} and \autoref{tab:failure_phone}.
The agent naively copies the last action from the few-shot examples, although the target UI element is not present in the actual current screen.
We believe that this implies the limitation of providing few-shot examples naively and necessitates more efficient methods when incorporating long experiences.

\begin{table}[!ht]
\begin{center}
\resizebox{0.9\textwidth}{!}{
\begin{tabular}[h]{@{}l@{}}
\fontsize{9}{11}\selectfont 
\centering
\tcbox[colback=white,boxrule=1pt,arc=3mm]{
    \begin{tblr}{colspec = {@{}X@{}}, rowsep=1pt,}
    - Description: The screen shows the dial pad interface with a floating action button labeled "dial" and a numeric keypad.\\
    - Thought: To complete the task of calling the White House, I need to tap the dial button to initiate the call.\\
    - Action: tap(74)
    \end{tblr}}
\end{tabular}}
\caption{An exemplary response from LLM agent employing GPT-4o in a task with instruction ``call the white house (202-456-1111)". The agent describes that a keypad appears on the screen and tries tapping a UI element with a numeric tag 74, despite their absence in the current observation.
}
\label{tab:failure_phone}
\end{center}
\end{table}

\begin{figure}[ht!]
\centering
    \includegraphics[height=3.4cm]{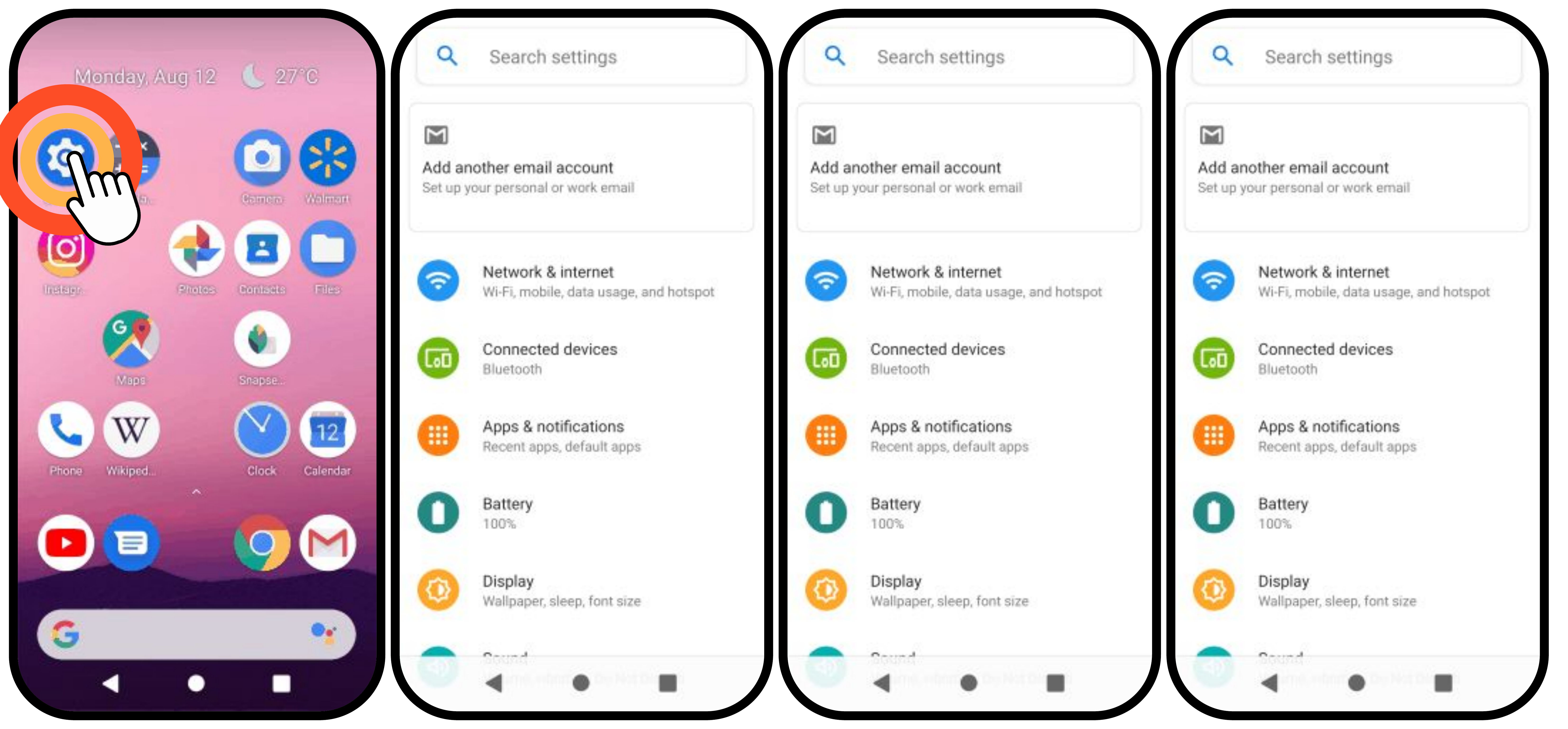}
    \caption{An exemplary trajectory of an LLM agent employing GPT-4o in a task with instruction ``increase call volume in setting" in a test environment having the DPI setting of value 550. 
    The agent repeats tapping the item "Sound" at the bottom of the list, occluded by the navigation bar.}
    \label{fig:failure_settings}
\end{figure}

Furthermore, we show a failure case where the task requires the agents to navigate to a target page by tapping a UI element in the middle of the list.
The task instruction is to ``increase call volume in setting".
As in an exemplary trajectory shown as \autoref{fig:failure_settings}, the LLM agents try tapping the UI element of item "Sound" in the list repeatedly, although the action does not result in entering the page.
To describe, this is mainly because the tapping location is occluded by another UI element, i.e., the navigation bar.
Similar to LLM agents, we also observe that MLLM agents also try repeatedly tapping the UI element, although they can clearly observe the occluded UI elements.
We believe that developing agents to have a better understanding of the effect of their actions on the device can potentially address the issue~\citep{shinn2023reflexion}.
We also acknowledge that building a more agent-friendly text-based observation parsing method (such as by excluding occluded UI elements) and allowing more adept action options can better help the agents, which can be easily modifiable with our open-source code materials.

\clearpage
Finally, we demonstrate a failure case due to the response in the wrong syntax.
We observe that the agents often generate action in the wrong format, especially from the agents employing Gemini-1.5-pro and Llama-3-70B in zero-shot, but rarely with few-shot examples or GPT-4o (even in zero-shot).
For example, as detailed in \autoref{tab:failure_format}, the Gemini-1.5 agent often decides to perform the action of tapping with a parameter that is different from the explanation and example in the prompt shown in \autoref{tab:llm_general_prompt}.
We assume that this is because the agents are hallucinated by the format of previous actions provided in the input.
Specifically, the format for the history of actions is designed to include all the detailed information of UI elements, in particular for the tapping action, being the same as the description in the text-based observation instead of including the numeric tag only.
The agents follow the format in the previous action instead of the format described in the earlier part of the action space in the prompt.
This implies a potential room for improvements, even in the zero-shot, by further engineering the interface or developing prompt formats for each agent with advanced design techniques~\citep {shin2020autoprompt}.

\begin{table}[!ht]
\begin{center}
\resizebox{0.9\textwidth}{!}{
\begin{tabular}[h]{@{}l@{}}
\fontsize{9}{11}\selectfont 
\centering
\tcbox[colback=white,boxrule=1pt,arc=3mm]{
    \begin{tblr}{colspec = {@{}X@{}}, rowsep=1pt,}
    Previous actions: ["tap({`numeric\_tag': 7, `resource\_id': `', `class': 'TextView', `content\_description': `Calculator', `text': `Calculator', `checked': 'false'})", 
    "tap({`numeric\_tag': 28, `resource\_id': `digit\_1', `class': `ImageButton', `content\_description': `1', `text': `', `checked': `false'})"]\\
    Current observation: [{`numeric\_tag': 0, `resource\_id': `', `class': `FrameLayout', `content\_description': `', `text': `', `checked': `false'}, $\cdots$, {`numeric\_tag': 41, `resource\_id': `', `class': `FrameLayout', `content\_description': `', `text': `', `checked': `false'}]\\
    \\[-1.8em]
    Answer: \\
        - Description: The Calculator app is open and the input is currently "1".\\
        - Thought: I need to input "+" next.\\
        - Action: tap({`numeric\_tag': 31, `resource\_id': `op\_add', `class': `ImageButton', `content\_description': `plus', `text': `', `checked': `false'}) 
    \end{tblr}}
\end{tabular}}
\caption{An exemplary response from LLM agent employing Gemini-1.5-pro in a task with instruction ``input `1+1' in Calculator", with the input of current observation (abbreviated) and previous actions. 
The agent generates an action with a parameter different from the guidelines specified in the input prompt, resulting in an error.
}
\label{tab:failure_format}
\end{center}
\end{table}

\clearpage
\subsection{Performances of LLM agents using Llama-3-70B on all tasks}\label{app:exp_llm_agent_all_tasks}
\autoref{tab:all_tasks_llama3} reports the evaluation results on all tasks in three representative test environments of LLM agents using Llama-3~({\tt meta-llama/Meta-Llama-3-70B-Instruct}) in zero-shot without fine-tuning.
We observe that Llama-3 agents exhibit reasonable performances across all tasks in our benchmark in zero-shot, compared to their performances on the six selected representative challenging tasks.
The result in all tasks demonstrates that they are capable of understanding and reasoning based on their pre-trained knowledge of the text-based observation of mobile devices.
However, we observe that Llama-3 agents often provide less detailed descriptions during task performance, compared to GPT-4o agents, as shown in \autoref{tab:llama3_vs_gpt4o_response}.
Also, regarding different test environments, a similar trend to other agents exhibiting degradation of performance as more randomizations of environmental features are applied is observable.

\begin{table}[h!]
\centering
\resizebox{0.55\textwidth}{!}{
\small
    \begin{tabular}{c c c c}
        \toprule
            Test Env ID
            & \multilinecell{LLM Agent\\{(GPT-4o)}}
            & \multilinecell{LLM Agent\\{(Gemini-1.5-pro)}}
            & \multilinecell{LLM Agent\\{(Llama-3)}}
            \\
        \midrule 
            100
            & $53.69$\stdv{0.75}
            & $36.13$\stdv{0.75}
            & $37.40$\stdv{0.72}
            \\
            101
            & $47.07$\stdv{0.91}
            & $26.97$\stdv{0.55}
            & $28.24$\stdv{0.95}
            \\
            105
            & $42.24$\stdv{1.04}
            & $25.45$\stdv{1.81}
            & $24.17$\stdv{0.42}
            \\
        \toprule
    \end{tabular}
}
\caption{Success rates of LLM agents using Llama-3 on all tasks with three representative test environments.
For easy comparison, we also show the performances of LLM agents using closed-source models (i.e., GPT-4o and Gemini-1.5-pro).
}
\label{tab:all_tasks_llama3}
\end{table}

\begin{table}[!ht]
\centering
\resizebox{0.9\textwidth}{!}{
\small
    \begin{tabular}{p{0.12\textwidth} p{0.8\textwidth}}
    \toprule
GPT-4o agent & I see the home screen with various app icons, such as Settings, Calculator, Camera, Walmart, Instagram, Photos, Contacts, Files, Maps, Snapseed, Phone, Wikipedia, Clock, Calendar, YouTube, Messages, Chrome, and Gmail.\\
Llama-3 agent & The current observation shows the home screen of the Android device, displaying various app icons and widgets.\\
   \bottomrule
    \end{tabular}
}
\caption{Exemplary `Description' parts of responses from the Llama-3 agent and GPT-4o agent, while performing a task with instruction ``call the white house (202-456-1111)".
The responses are from both agents, describing the home screen of the test environment with ID 100 at the first step.}
\label{tab:llama3_vs_gpt4o_response}
\end{table}

\subsection{Performances of \custom agents using Llama-3 in interactive environments} \label{app:exp_agent_llama}

\begin{figure}[h!]
    \centering
    \includegraphics[width=0.7\textwidth]{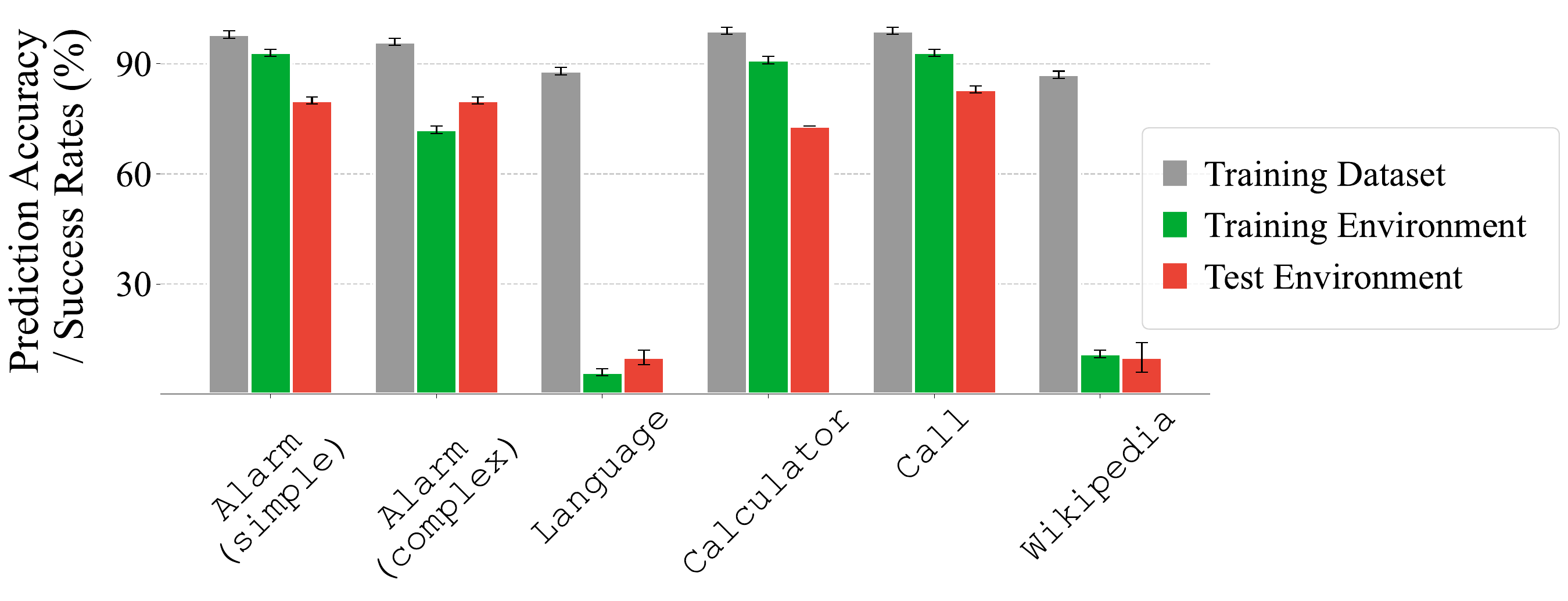}
    \captionof{figure}{
         The accuracy of predictions of \custom agents using fine-tuned Llama-3 on the training dataset and their performances in the interactive environments (both training and test). 
        Unlike their high accuracy in the training datasets, they show limited success rates in the real-system evaluation.
    }
\label{fig:llama3_interactive_env}
\end{figure}

\autoref{fig:llama3_interactive_env} displays evaluation results on six representative challenging tasks in all test environments of \custom agents using fine-tuned Llama-3~({\tt meta-llama/Meta-Llama-3-8B-Instruct}).
We also display the success rates in the training environments and action accuracies in the training dataset. 
Unlike their high accuracy in the training dataset after fine-tuning with BC, they show limited success rates in interactive environments.
We observe that the agents fail to recover from their errors in the environments.
We believe this reveals the limitation of the naive adaptation of BC and the necessity of real-system evaluation for rigorous assessment.

\clearpage

\subsection{Performances of \custom agents using VLM encoder in training environments}\label{app:exp_bc_training_env_performance}

\begin{figure}[h!]
    \centering
    \includegraphics[width=0.55\textwidth]{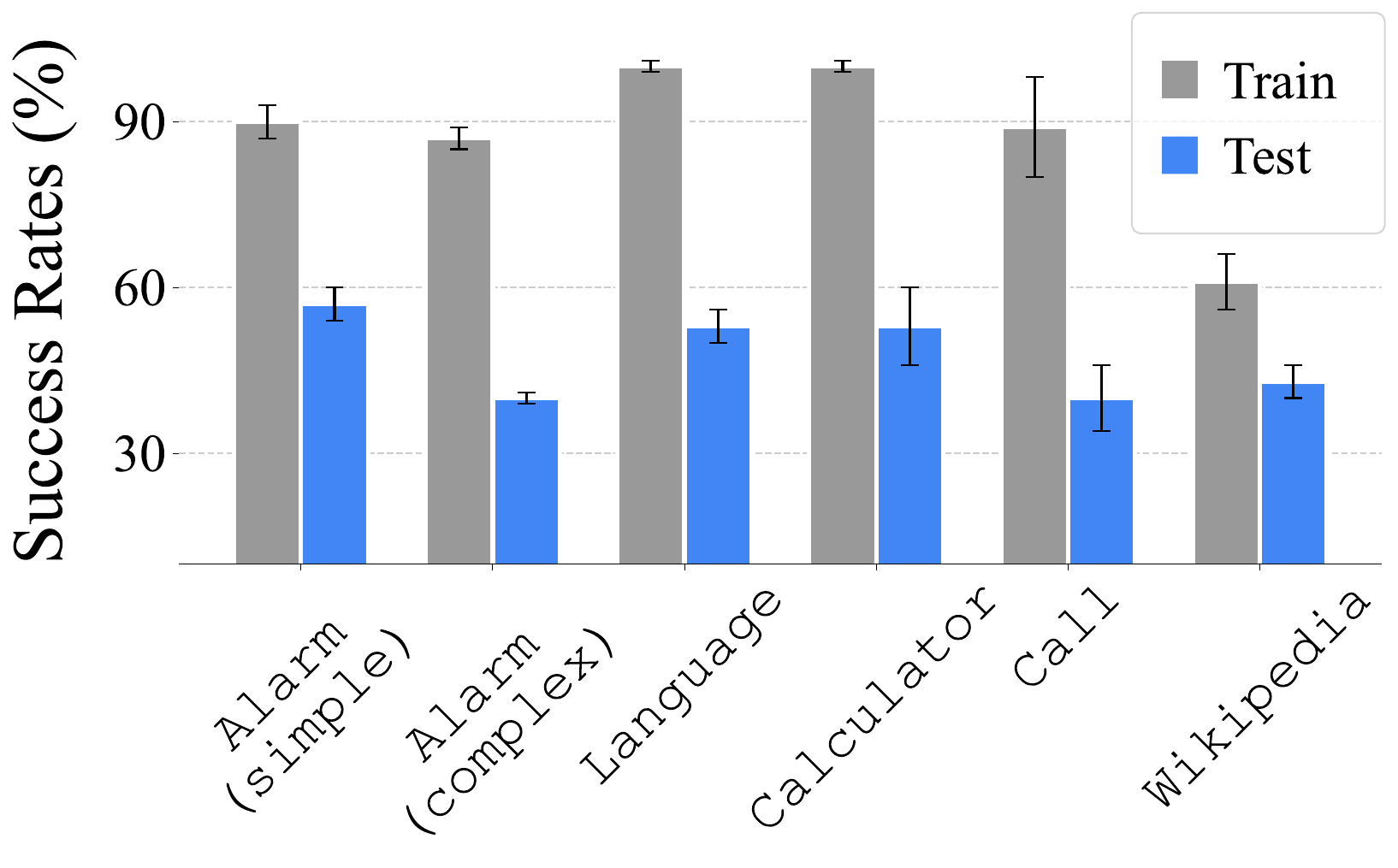}
    \captionof{figure}{
         Success rates of \custom agents with VLM encoder on training and test environments. 
         The differences between the success rates demonstrate the headroom for the generalization ability.
    }
\label{fig:vlui_generalization}
\end{figure}

\autoref{fig:vlui_generalization} displays the differences in the success rates of \custom agents in training and test environments. 
The challenges with diverse device configurations degenerate the performances of the \custom agents.
For example, on the {\tt Language} task, the success rates decrease from higher than 90\% in the training environments to less than 60\% in the test environments. 
The differences between success rates in the training and test environments demonstrate the headroom for the generalization ability.

\subsection{Performances of MLLM agents with and without few-shot examples}\label{app:exp_mllm_few_shot}

\autoref{tab:exp_mllm_few_shot} shows the performances of MLLM agents using GPT-4o with and without few-shot examples (with the performances of LLM agents using GPT-4o for easy comparison).
We observe a similar trend between LLM agents and MLLM agents.
To be specific, the few-shot examples help MLLM agents on several tasks (e.g., {\tt Alarm(complex)}), while they are not beneficial in some cases and even decrease the performances (e.g., {\tt Alarm(Simple)}).

\begin{table}[h!]
\centering
\resizebox{0.7\textwidth}{!}{
\small
    \begin{tabular}{c c c c c}
        \toprule
            & \multilinecell{LLM Agent\\{(zero-shot)}}
            & \multilinecell{LLM Agent\\{(few-shot)}}
            & \multilinecell{MLLM Agent\\{(zero-shot)}}
            & \multilinecell{MLLM Agent\\{(few-shot)}}
            \\
        \midrule 
            {\tt Alarm(simple)}
            & $87$\stdv{07}
            & $83$\stdv{05}
            & $80$\stdv{05}
            & $63$\stdv{07}
            \\
            {\tt Alarm(complex)}
            & $33$\stdv{05}
            & $67$\stdv{03}
            & $23$\stdv{03}
            & $43$\stdv{11}
            \\
            {\tt Language}
            & $97$\stdv{03}
            & $83$\stdv{03}
            & $90$\stdv{05}
            & $83$\stdv{05}
            \\
            {\tt Calculator}
            & $17$\stdv{03}
            & $17$\stdv{07}
            & $10$\stdv{05}
            & $23$\stdv{11}
            \\
            {\tt Call}
            & $00$\stdv{00}
            & $30$\stdv{08}
            & $00$\stdv{00}
            & $07$\stdv{05}
            \\
            {\tt Wikipedia}
            & $30$\stdv{05}
            & $20$\stdv{00}
            & $30$\stdv{05}
            & $20$\stdv{05}
            \\
        \midrule
            Average
            & $44$\stdv{03}
            & $50$\stdv{04}
            & $39$\stdv{01}
            & $40$\stdv{03}
            \\
        \bottomrule
    \end{tabular}
}
\caption{Success rates of MLLM agents and LLM agents using GPT-4o with and without few-shot examples.}
\label{tab:exp_mllm_few_shot}
\end{table}

%% file: sections/appendix/rl_details.tex
\clearpage 

\section{Agents trained with reinforcement learning}\label{app:exp_rl_discussions}
In this section, we explain further analysis with the agents employing policies trained with reinforcement learning. 
We, first, train the policies using the success signal as a reward.
We, then, further study the efficacy of reward shaping for improving efficiency.

\subsection{Experimental setup}\label{app:agent_rl_algorithm}

\paragraph{Algorithm}
To train custom agents trained with RL, we use double deep Q-network (DDQN)~\citep{van2016deep} and proximal policy optimization (PPO)~\cite{ppo}.
DDQN trains the two optimal Q functions $Q_1$ and $Q_2$ by minimizing the TD loss for the agent data sampled from the replay buffer $D$. As an example, we write the objective for $Q_1$(and its parameterization $Q_{\theta_1}$) as follows: 
\begin{equation*}
L_{\text{TD}}(Q_1)=\mathbb{E}_{(s,a,r,s') \sim D}[(r + \gamma Q_{\Bar{\theta_1}}(s',a^*) - Q_{\theta_1}(s,a)^2].
\end{equation*}
where $a^*=\argmax_{a'}Q_{\Bar{\theta_2}}(s,a)$ and $\Bar{\theta}$ denotes the target network for $\theta$, which in practice is replaced by the moving average of $\theta$. 
Especially, when calculating the TD target, we swap the maximizing actions of the next state action values between $Q_1$ and $Q_2$ to prevent the overestimation of the values as presented in~\citep{hasselt2010double}. 
To balance the exploration and the exploitation during the training, we employ the $\epsilon$-greedy technique to sample actions for collecting the training data. 
For the critic network architecture, we adopt the same visual encoder and text encoder backbone as the \custom agent (using VLM encoder) and extract the fused feature $e_\text{fused}$. 
Finally, we add three FC layers comprising the dimensions of (1024, 1024, 385) to output the Q values for each action.

PPO~\citep{ppo} builds upon the policy gradient method~\citep{sutton2018reinforcement} and uses a clipped surrogate objective. 
To be detailed, the objective is given by
\begin{equation*}
L_{\text{PPO}}(\theta)=E_{\pi_{\text{old}}} \left[ \min \left( \text{clip}(r(\theta), 1-\epsilon, 1+\epsilon \right)A_t, r(\theta)A_t)\right],
\end{equation*}
where $A_t$ denotes the estimator for the future expected return and $r(\theta)=\pi_\theta(a|s) / \pi_{\text{old}}(a|s)$ corresponds to an importance sampling ratio between the current and the previous policy.
Following ~\citep{ppo}, we use generalized advantage estimator (GAE)~\citep{schulman2015high} in place of $A_t$.
We use the same network architecture as the \custom agent (using VLM encoder) for implementing $\pi_\theta$. 
For the value network $V$, we apply three FC layers that follow the size of (1024, 1024, 1) using the feature $e_{\text{fused}}$ produced by the pre-trained backbone.

\paragraph{Task and environment}
We experiment RL with a sparse success signal using a task with the instruction ``open Gmail'' (denoted as {\tt Gmail}) and {\tt Language} task. We train at 10 training environments, whose IDs are 000, 001, 002, 003, 004, 005, 007, 008, 021, 022 and then evaluate under 10 test environments. We also experiment with dense reward setup using the task with the instruction ``call 911'' (denoted as {\tt Call 911}) task under a training environment with id 000.

\paragraph{Training procedure}
For DDQN, we update the critic network by sampling 1/4 transitions from the replay buffer that stores the successful history and the last from the failure buffer in every 5 episodes. For PPO, we apply one update using the on-policy samples for every 5 episodes and use the Adam optimizer with a learning rate of 2e-4. We use the hyperparameters of $\epsilon=0.1$, $\gamma=0.9$, and $\lambda=0.9$ for both actor and critic unless otherwise specified.
Each training is conducted on a single NVIDIA RTX A6000 GPU and Intel(R) Xeon(R) Gold 6426Y CPU.
We note that the training procedure can be accelerated by vectorizing the training environments.

\subsection{Results with RL training} \label{app:exp_rl_experiments_sparse}

\begin{figure}[!ht]
    \centering
    \includegraphics[width=0.7\textwidth]{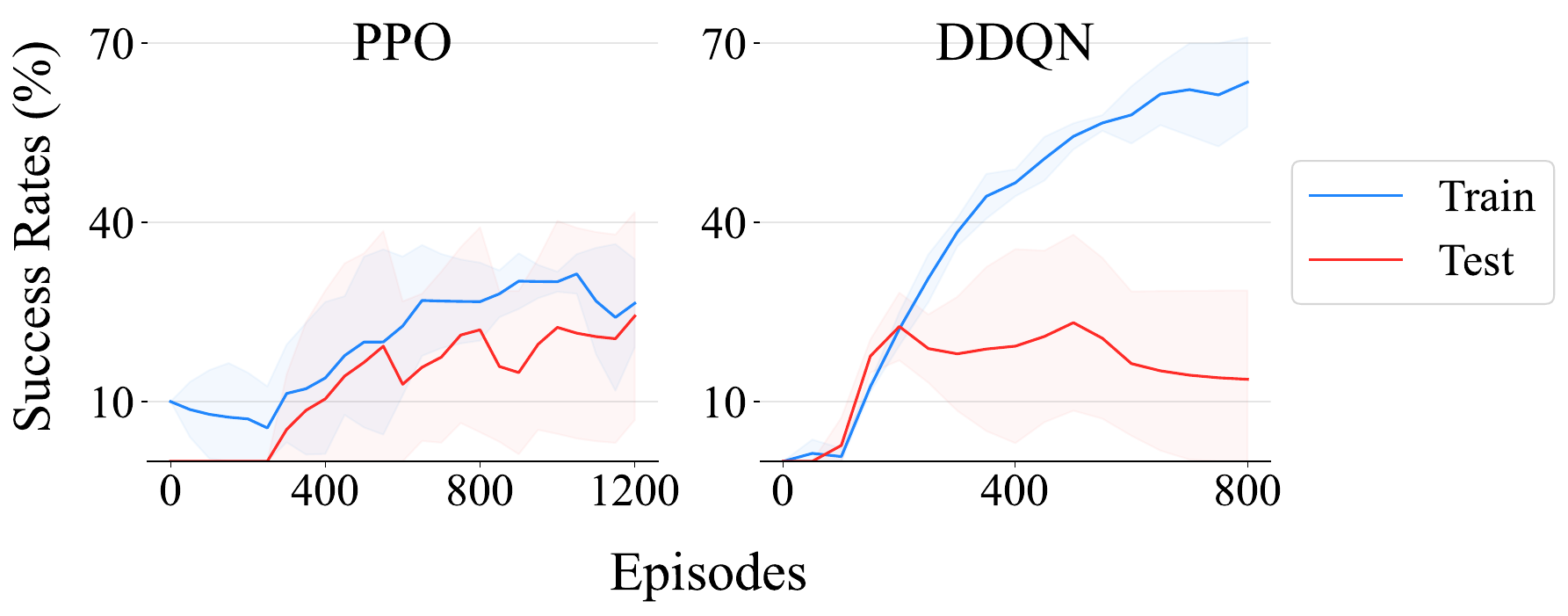}
    \caption{Training and test curves of DDQN and PPO algorithms on the ``open Gmail'' task.}
\label{fig:exp_rl_ppo_dqn}
\end{figure}

We first train custom agents using RL on a {\tt Gmail} task. We show the success rate in 10 training and test environments with three different seeds in \autoref{fig:exp_rl_ppo_dqn}.
As shown in \autoref{fig:exp_rl_ppo_dqn}, RL agents learn to solve the task via interactions in training environments resulting in their test performance increases. Still, we remark that RL agents fail to complete more complex tasks that require longer interactions for success. On the {\tt Language} task, we find that these agents fail to complete the task even in training environments. While the agents enter the target application (i.e., the setting application) during exploration, they fail to further navigate the appropriate pages. This is mainly because the sparse success signal is only provided after completing the task, unable to guide the exploration from scratch.

\subsection{Effect of reward shaping}  \label{app:exp_rl_experiments_dense}

Instead of directly using sparse success signals for reward, reward shaping can be a helpful direction to guide agents in the long-horizon task. 
To further study, we design a dense reward function in a {\tt Call 911} task having an episode length of 6. 
The reward function is defined as a value of $i/N$ for accomplishing the $i$-th step of the total $N$ step task, and -1 otherwise.
The value for $N$ and the completion criteria for each step are determined based on a human expert demonstration.
To clarify, we created an additional reward function (i.e., evaluator) checking 5 subgoals we designed based on the optimal trajectory of task completion, which is composed of 6 steps as follows:
\begin{itemize}[leftmargin=6mm]
    \item Step 1. Opening the Phone application
    \item Step 2. Entering the call page
    \item Step 3-5. Typing 9-1-1 sequentially
    \item Step 6. Clicking the ‘Call’ button
\end{itemize}
Each subgoal 1-5 corresponds to executing steps 1-5 precisely, and the additional evaluator examines the completion of subgoals by examining the attributes of UI elements (e.g., for Step 1, the evaluator examines if the agents has accessed to the Phone application, and, for Step 3, it checks if any UI element containing text value `9' exists). When the agent achieves the $i$-th subgoal, the evaluator provides the reward with a numeric value of $i/6$.
We train the agent using DDQN with $\gamma = 0.1$ for 3500 episodes.
We use a single training environment (with environment id 000).

\begin{figure}[!ht]
    \centering
    \includegraphics[width=0.4\textwidth]{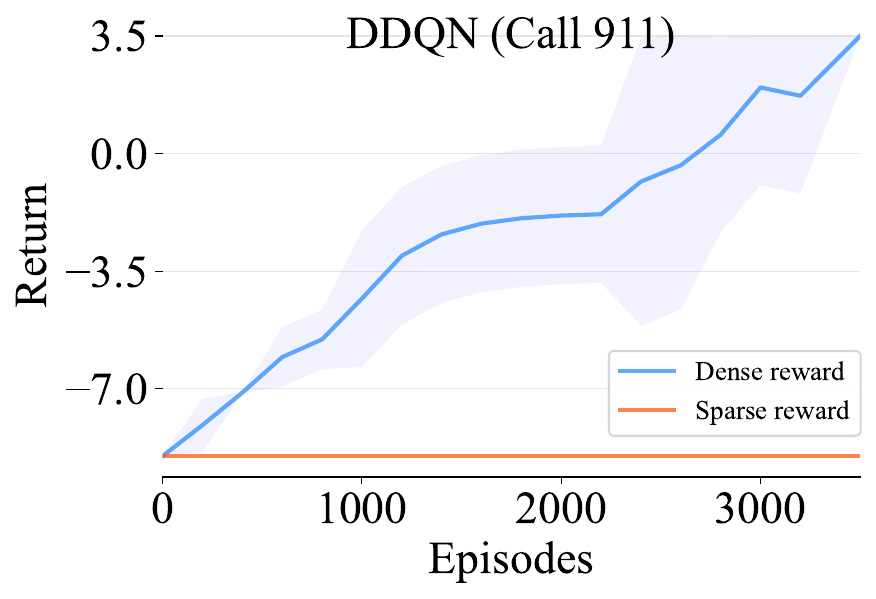}
    \caption{Training curves of DDQN algorithm on the ``call 911'' task with reward shaping.}
\label{fig:ddqn_dense_reward}
\end{figure}

Fig \ref{fig:ddqn_dense_reward} displays the training curve with return values.
The sum of the reward keeps increasing as the agent learns to progress each step. 
The return achieves the maximal value of $1/6 + \dots + 6/6$, which indicates the success of the task in a training environment, while it fails in a slightly different test environment with id 100. 
We note that designing dense reward functions is an effortful process since it requires an understanding of UI elements per task step. 
The automation of the reward-shaping technique to many other tasks is highly expected to contribute to the success of reinforcement learning from scratch, suggesting a future direction toward more scalable reward function design techniques. 
We also expect that initializing the agent with a pre-trained policy (such as with a VLM encoder or LLM) can also improve the efficiency of RL.

%% file: sections/appendix/related_work.tex
\clearpage
\section{Additional related work}\label{app:related_work}

We provide additional studies related to our work.

\paragraph{Foundation models for decision-making system}

Inspired by the strong emergent properties of foundation models~\citep{brown2020language,wei2022chain}, many researches have adopted LLMs to develop decision-making system~\citep{yao2022react,shinn2023reflexion}.
In robot learning, for example, LLMs have been widely equipped for reasoning, planning, manipulation, and navigation~\citep{driess2023palm,liang2023code,huang2023visual}.
Furthermore, agents with LLMs have shown capabilities of performing interesting tasks in numerous simulated worlds, including game environments~\citep{wang2023voyager,tan2024towards} and virtual reality~\citep{qian2023communicative,yang2024v}.
In recent days, focusing on practicalness, solving computer tasks with foundation models has also been actively explored~\citep{nakano2021webgpt,furuta2023instructionfinetuned}.
We further study the abilities of foundation models to control mobile devices toward assistive agents in real life.

\paragraph{Developing assistive agent for device control}

For agents that effectively understand and manipulate the UI elements, a large body of work has leveraged structural information, such as document object model in HTML or Android view hierarchy~\citep{branavan2010reading,gur2019learning}.
In addition, methods for equipping agents with the ability to understand information-rich screen images have been widely investigated, mainly with vision-based reinforcement learning~\citep{liu2018reinforcement,humphreys2022data,shaw2023pixels}.
Recently, diverse strategies to build device control agents with foundation models are introduced, including prompting methods~\citep{wen2023empowering,kim2023language}, instruction-tuning~\citep{furuta2023instructionfinetuned}, fine-tuning with images~\citep{zhan2023you,hong2023cogagent}, and visual prompting~\citep{yan2023gpt,yang2023appagent}. 
Here, we present an elaborate analysis of the main methods for building mobile device control agents.



%% file: sections/appendix/limitation.tex
\section{Limitations and future directions}\label{app:limitation}

We discuss several limitations and promising future directions of this work:

\begin{itemize}[leftmargin=6mm]

\item{\em Tasks with text typing:} 
While agents can input text by touching the soft keyboard on the screen in the current interface, it demands excessively long interactions.
We plan to include tasks requiring text typing, such as web search or e-mail sending, with advanced interfaces in the future.

\item {\em Open-ended tasks and reward modeling:}
Since the ADB-based success detector does not capture the semantics of agent behaviors, tasks with ambiguous success criteria are hard to evaluate.
Alternatively, we believe employing the reward model learned from demonstrations~\citep{fan2022minedojo} can be used for integrating open-ended tasks.

\item {\em More on agents with foundation models:}
LLM agents can be employed in different ways.
For example, they can be high-level planners operating a set of pre-defined APIs~\citep{chen2024octopus} or low-level actors implemented with neural network policies~\citep{ahn2022can}.
Also, we note that employing different agentic workflows or experimenting with engineered prompts are highly valuable approaches that remain unexplored within our benchmark.

\end{itemize}

We remark that we open-source \metabbr to allow easy modification and improvements on interfaces and experiments.